\documentclass[a4paper,12pt]{report}

\usepackage{LaTexPkg}

\title{Lecture Notes on High Dimensional Linear Regression}
\author{Alberto Quaini\footnote{
  Department of Econometrics, Erasmus University of Rotterdam,
  P.O. Box 1738, 3000 DR Rotterdam, The Netherlands.
  }\footnote{
  Please, let me know if you find typos or mistakes at 
  \href{mailto:quaini@ese.eur.nl}{\texttt{quaini@ese.eur.nl}}.
}}
\date{\today\ (first version: October 17, 2024)}

\begin{document}

\maketitle \thispagestyle{empty}

\section*{Introduction}

These lecture notes were developed for 
a Master's course in {\it advanced machine learning}
at Erasmus University of Rotterdam. 
The course is designed for graduate students in mathematics, 
statistics and econometrics. 
The content follows a proposition-proof structure, 
making it suitable for students seeking a formal 
and rigorous understanding of the statistical theory 
underlying machine learning methods.

At present, the notes focus on linear regression, 
with an in-depth exploration of the existence, uniqueness, relations, 
computation, and non-asymptotic properties of 
prominent estimators in this setting: 
least squares, ridgeless, ridge, and lasso.

\subsubsection*{Background}
It is assumed that readers have a solid background in mathematical analysis, 
linear algebra, convex analysis, and probability theory.
Some definitions and results from these fields, relevant for the 
course, are provided in the Appendix for reference.

\subsubsection*{Book-length references}
The content of these lecture notes is inspired by a wide range of existing literature, but the presentation of topics follows my own interpretation and logical structure. Although most of the content can be traced back to established sources, certain sections reflect my perspective, and some material is original to this course. For those interested in more comprehensive, book-length discussions of related topics, 
the following key references are recommended:
\citet{hastie2009elements},
\citet{buhlmann2011statistics},
\citet{hastie2015statistical}, and
\citet{wainwright2019high}.

\subsubsection*{Disclaimer}
Please note that despite my efforts, these lecture notes may contain errors. I welcome any feedback, corrections, or suggestions you may have. If you spot any mistakes or have ideas for improvement, feel free to contact me via email at \href{mailto:quaini@ese.eur.nl}{\texttt{quaini@ese.eur.nl}}.

\tableofcontents
\newpage

\addcontentsline{toc}{section}{Notation}
\section*{Notation}

\begin{itemize}
\item All random variables are defined on complete probability space
$(\Omega, \mathcal F, \P)$ and take values in a real Euclidean space.
\item For a random variable (vector) [matrix] $x$ ($\bm x$) [$\bm X$], 
the notation $x\in\R$ ($\bm x\in\R^n$) [$\bm X\in\R^{n\times p}$]
means that $x$ ($\bm x$) [$\bm X$] takes values in 
$\R$ ($\R^n$) [$\R^{n\times p}$].
\item The symbols $\overset{\P}{\to}$ and $\overset{d}{\to}$ denote
convergence in probability and in distribution, respectively.
\item Given a random variable $x$, its expectation is denoted $\E[x]$
and its variance $\Var[x]$.
\item For a vector $\bm x\in\R^n$, the $i-$th element is denoted $x_i$
for $i=1,\ldots,n$.
\item For a matrix $\bm A\in\R^{n\times p}$, 
the $i,j-$th element is denoted $A_{i,j}$,
the $j-$th column is denoted $\bm A_j$ and
the $i-$th row is denoted $A_{(i)}$,
for $i=1,\ldots,n$ and $j=1,\ldots,p$.
\item The transpose of a matrix $\bm A\in\R^{n\times p}$ is
denoted $\bm A'$.
\item The Moore-Penrose inverse of a matrix $\bm A\in\R^{n\times p}$ is
denoted $\bm A^+$.
\item The {\it $l_p-$norm}\index{l$_p-$norm} $\norm{\cdot}_p$ on $\R^n$
    is defined for all $\bm v\in\R^n$ as
    $\norm{\bm v}_p:=\left(\sum_{i=1}^{n}|v_i|^p\right)^{1/p}$ when $p\in[1,+\infty)$,
    and $\norm{\bm v}_p:=\max_{i=1}^{n}\vert v_i\vert$ when $p=+\infty$.
\item Given vector $\bm\theta\in\R^p$, the $l_0-$norm (which is not a norm!) $\norm{\bm x}_0$ counts the number of nonzero elements of $\bm\theta$.
\item $\argmin_{x\in X}f(x)$ denotes the set of minimizers of $f$ over set $X$.
\item $\diag(\bm A)$ denotes the diagonal elements of a square 
matrix $\bm A\in\R^{n\times n}$
\item $\diag(a_1,\ldots,a_n)$ denotes a square matrix in $\R^{n\times n}$
that has diagonal elements given by $a_1,\ldots,a_n\in\R$ and that has zero elsewhere.
\item Given a matrix $\bm A\in\R^{n\times p}$,
its rank is $\Rank(\bm A)$, its range 
is $\Range(\bm A)$, its kernel is $\Ker(\bm A)$,
its trace is denoted $\Trace(\bm A)$.
\item We denote $\Proj_S$ the orthogonal projection onto
set $S\subset\R^{n\times p}$.
\item Given a vector space $V$, 
the sum of two subsets $A,B\subset V$ is defined as 
$A+B:=\{a+b\ :\ a\in A, b\in B\}$.
The sum of a set $A\in V$ and a vector $b\in V$
is defined as 
$A+b:=\{a+b\ :\ a\in A\}$.
\item The symbol $\partial$ indicates the subdifferential.
\end{itemize}

\chapter{Linear Regression}

Linear regression is a supervised learning technique
aimed at predicting
a {\it target} random variable $y$
using a linear combination 
$$\bm x'\bm\theta=x_1\theta_1+\ldots+x_p\theta_p$$ 
of 
{\it explanatory variables} $\bm x=[x_1,\ldots,x_p]'$,
where $\bm\theta\in\R^p$ and $p\in\N$.\footnote{
  An intercept in $f(\bm x;\bm\theta)$ can be introduced 
  by adding a constant term to the predictors.
}
The target variable $y$ is also referred to as {\it dependent} or {\it output} variable, while the explanatory variables $\bm x$ are also 
known as {\it independent variables}, 
{\it predictors} or {\it input} variables.

\begin{figure}[H]
    \centering
    \includegraphics[scale=0.3]{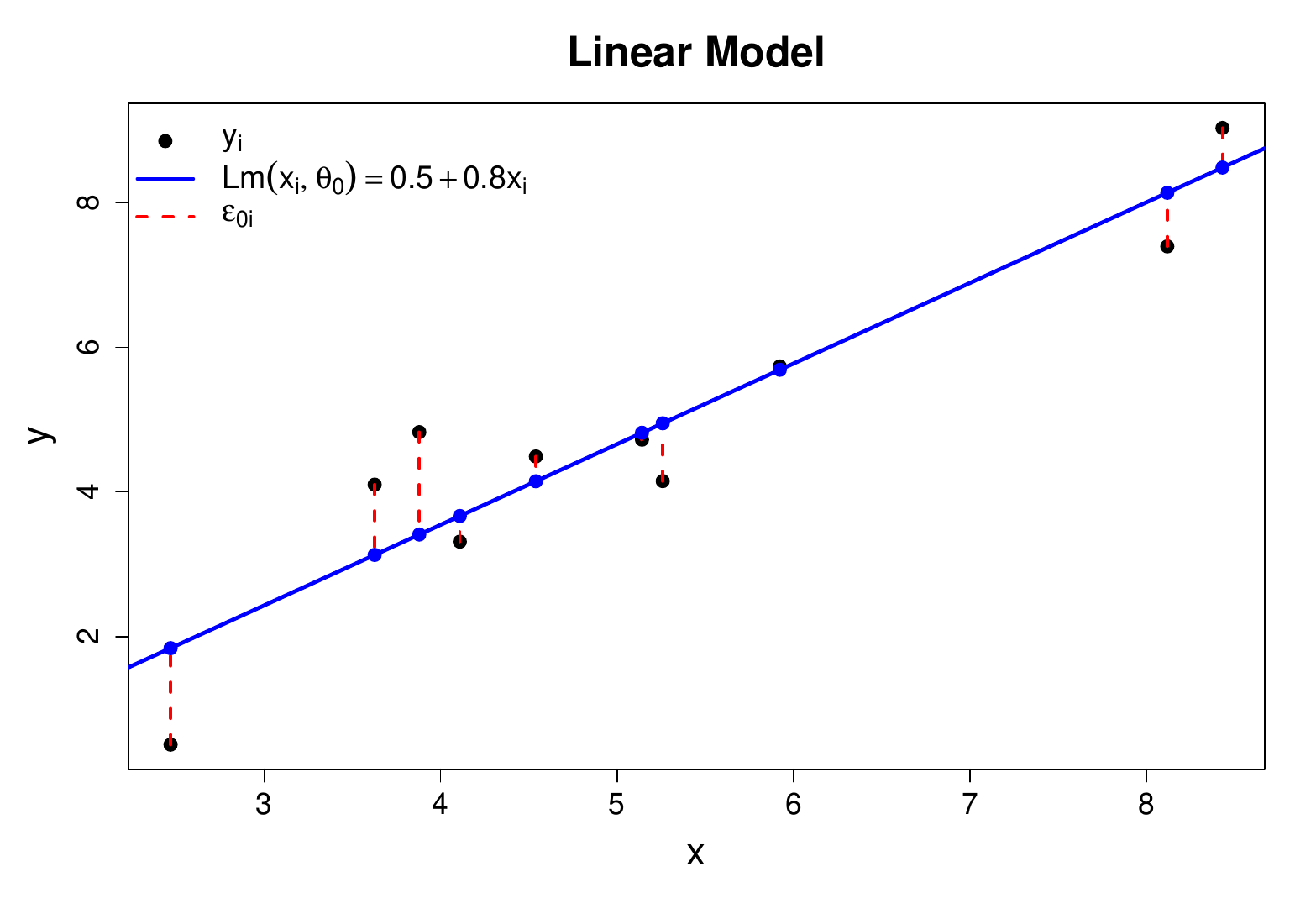}
    \caption{Statistical linear model $y_i=\tilde{\bm x}_i'\bm\theta_0+\varepsilon_{0i}$ where
    $\tilde{\bm x}_i=[1,x_i]'$ and $\bm\theta_0=[0.5, 0.8]'$.}
    \label{fig:linear model}
\end{figure}

In typical applications, we observe only a sample of size $n\in\N$ 
of these random variables, 
represented by the pairs $(\bm x_i, y_i)_{i=1}^n$,
where $\bm x_i\in\R^p$ and $y_i\in\R$ for each $i$.
Given a regression coefficient $\bm\theta_0\in\R^p$, 
a {\it statistical linear model}\index{Linear model}, or simply linear model, is
expressed as
\begin{equation}\label{linear model}
  y_i=\bm x_i'\bm\theta_0+\varepsilon_{0i},\quad i=1,\ldots,n,
\end{equation}
where $\varepsilon_{0i}$ 
are real-valued {\it residual} random variables.
Figure \ref{fig:linear model} depicts a linear model for
$i=1,\ldots,n$ observations $y_i$,
with two predictors $\tilde{\bm x}_i=[1,x_i]'$ consisting of a unit constant
and a variable $x_i$,  
a coefficient $\bm\theta_0\in\R^2$ and
the associated error terms $\varepsilon_{0i}$.
The {\it Data Generating Process} (DGP), 
i.e., the joint distribution of the predictors $\bm x$ and 
the real-valued {\it residual} random variables 
$\bm\varepsilon_0=[\varepsilon_{0i}]_{i=1}^n$, 
is subject to certain restrictions. 
Depending on the type of restrictions imposed on the DGP, 
different types of linear models are obtained. 
The two general forms of linear models are fixed and random design models,
which are defined as follows.
\begin{definition}[Fixed design model]
  In a {\it fixed design model}\index{Fixed design}, the sequence $(\bm x_i)_{i=1}^n$ is fixed. 
  The residuals $\varepsilon_{0i}$ are 
independent and identically distributed.
\end{definition}
\begin{definition}[Random design model]
  In a {\it random design model}\index{Random design}, the pair 
  $(\bm x_i,y_i)_{i=1}^n$ is a sequence of
  independent and identically distributed random variables.
\end{definition}
The fixed design model is particularly suitable when the predictors 
are controlled by the analyst, such as the dose of medication
administered 
to patients in the treatment group in a clinical trial. 
Conversely, the random design model is appropriate when 
the explanatory variables are stochastic, such as the wind speed 
observed at a specific time and location.

We organize the observed values of the target variable 
in the vector $\bm y=[y_1,\ldots,y_n]'\in\R^n$, 
and the observations on the predictors in the {\it design matrix}\index{Design matrix} 
$$\bm X:=\begin{bmatrix}
    x_{11}&\ldots&x_{1p}\\
    \vdots&\ddots&\vdots\\
    x_{n1}&\ldots&x_{np}
\end{bmatrix}\in\R^{n\times p}.$$
With this notation, the linear model (\ref{linear model}) can be expressed as:
$$\bm y=\bm X\bm\theta_0+\bm\varepsilon_0,$$
where $\bm\varepsilon_0=[\varepsilon_{01},\ldots,\varepsilon_{0n}]'\in\R^n$.

\begin{example}
A classic example of linear regression is found in the work of \citet{nerlove1961returns}, which examines returns to scale in the U.S. electricity power supply industry.
In this study, the total cost $y_i$ for firm $i$
is predicted using a linear model based on the firm's 
output production $x_{i1}$, the wage rate $x_{i2}$, 
the price of fuel $x_{i3}$, and the rental price of capital $x_{i4}$,
with data from $n=145$ electric utility companies. 
\end{example}

The rest of the chapter is organized as follows.
First, we study the most basic linear regression approach, 
the method of least squares projection, 
and some of its recent machine learning extensions.
Our study focuses on their existence, uniqueness, connections,
geometric interpretation, and computation.
Then, we study we cover both their finite- or small-sample properties, 
that are valid for any given sample size,
and their asymptotic properties, 
that are useful approximations when the sample size is large enough.

\subsection{Estimators and their properties}

\begin{definition}[Estimand]
An {\it estimand}\index{Estimand} is a feature, or parameter, of interest of the population.
\end{definition}
\begin{definition}[estimator and estimate]
    An {\it estimator}\index{Estimator} is a function taking as input the data,
    and possibly other auxiliary variables,
    and outputting an {\it estimate}\index{Estimate}, 
    which is a specific value assigned to the estimand.
\end{definition}
For instance, 
in the context of the linear model (\ref{linear model}),
the coefficient $\bm\theta_0\in\R^p$ represents the estimand.
An estimator is a function 
$\hat{\bm\theta}_n:\R^n\times\R^{n\times p}\to\R^p$ that takes as inputs the
data $(\bm y, \bm X)\in\R^n\times\R^{n\times p}$ 
and produces an estimate $\hat{\bm\theta}_n(\bm y, \bm X)\in\R^p$.
For simplicity, we use the same notation, $\hat{\bm\theta}_n$, 
to refer to both the estimator and the resulting estimate,
although formally the estimate should be written as 
$\hat{\bm\theta}_n(\bm y, \bm X)$. 
We may also write $\hat{\bm\theta}_n\in\R^p$ to indicate that an estimator
outputs values in $\R^p$.
\begin{definition}[Linear prediction]\label{linear prediction}
  The quantity $\Lm(\bm X,\bm\theta):=\bm X\bm\theta$ denotes the {\it linear prediction}\index{Linear prediction} associated to coefficient vector $\bm\theta\in\R^p$ 
  and predictors $\bm X\in\R^{n\times p}$.  
\end{definition}
Let $\bm M^+$ denote the Moore-Penrose inverse of a generic real-valued matrix $\bm M$.
We make extensive use of the following useful projections.\footnote{
  We use the notation $\Range$ ($\Ker$) for the range (kernel) of a matrix.
  Details on these sets, the Moore-Penrose inverse and orthogonal projections
  are given in Appendix Section \ref{app:sec:linear algebra}.
}
\begin{definition}[Useful projections]
    Given a fixed matrix $\bm X\in\R^{n\times p}$:
    \begin{itemize}
        \item $\bm\Proj_{\Range(\bm X')}:=\bm X^+\bm X$ is the orthogonal projector onto $\Range(\bm X')$;
        \item $\bm\Proj_{\Ker(\bm X)}:=\bm I-\bm X^+\bm X$ is the orthogonal projector onto $\Ker(\bm X)$;
        \item $\bm\Proj_{\Range(\bm X)}:=\bm X\bm X^+$ is the orthogonal projector onto $\Range(\bm X)$;
        \item $\bm\Proj_{\Ker(\bm X')}:=\bm I-\bm X\bm X^+$ is the orthogonal projector onto $\Ker(\bm X')$.
    \end{itemize}  
\end{definition}
The next proposition demonstrates that, if we fix the design matrix $\bm X$, 
we can focus on regression coefficients in $\Range(\bm X')$.
Indeed, \[\R^p=\Ker(\bm X) + \Range(\bm X')\] and coefficients in \(\Ker(\bm X)\) are immaterial for linear prediction,
while coefficients in \(\Range(\bm X')\) span all possible linear predictions that
can be achieved through $\bm X$.
\begin{proposition}\label{prop:prediction in range X'}
    Given a matrix $\bm X\in\R^{n\times p}$, for any $\bm\theta\in\R^p$:
    \begin{equation*}
        \Lm(\bm X,\bm\theta)=\Lm(\bm X,\bm\Proj_{\Range(\bm X')}\bm\theta).
    \end{equation*}
\end{proposition}
\begin{proof}
    Using the identity $\bm I=\bm\Proj_{\Range(\bm X')}+\bm\Proj_{\Ker(\bm X)}$, we have for any $\bm\theta\in\R^p$:
    \begin{align*}
        \bm X\bm\theta=\bm X(\bm\Proj_{\Range(\bm X')}+\bm\Proj_{\Ker(\bm X)})\bm\theta=
        \bm X\bm\Proj_{\Range(\bm X')}\bm\theta.
    \end{align*}
\end{proof}

\subsection*{Finite-sample properties}

Since an estimator is derived from data, it is a random variable.
Intuitively, when comparing two estimators of the same estimand, 
we prefer the one whose probability distribution is 
"more concentrated" around the true value of the estimand. 
Formally, estimators are compared using several key properties.\footnote{
  Some of these properties are defined by means of the $l_2-$norm.
  Note that this choice is typical, but arbitrary.
}
\begin{definition}[Bias]
  The {\it bias}\index{Bias} of an estimator 
  $\hat{\bm\theta}_n$ of $\bm\theta_0$ is the difference 
  between the expected value of the estimator and the estimand: 
  \begin{equation*}
      \Bias(\hat{\bm\theta}_n,\bm\theta_0):=\E[\hat{\bm\theta}_n] - \bm\theta_0.
  \end{equation*} 
\end{definition}
\begin{definition}[Estimation risk]
  The {\it estimation risk}\index{Estimation risk} of an estimator 
  $\hat{\bm\theta}_n$ of $\bm\theta_0$ measures the difference 
  between the estimator and the estimand as: 
  \begin{equation*}\label{estimation risk}
      \ER(\hat{\bm\theta}_n,\bm\theta_0):=\lVert\hat{\bm\theta}_n - \bm\theta_0\rVert_2^2.
  \end{equation*} 
\end{definition}
\begin{definition}[MSE]\label{def:MSE}
  The {\it Mean Squared Error}\index{Mean squared error} (MSE) of an estimator 
  $\hat{\bm\theta}_n$ of $\bm\theta_0$ is the expected 
  estimation risk of the estimator:
  \begin{equation*}
      \MSE(\hat{\bm\theta}_n,\bm\theta_0):=
      \E[\lVert\hat{\bm\theta}_n - \bm\theta_0\rVert_2^2].
  \end{equation*}
\end{definition}
\begin{definition}[Predictive risk]\label{def:predictive risk}
  The {\it predictive risk}\index{Predictive risk} of an estimator
  $\hat{\bm\theta}_n$ of $\bm\theta_0$ measures the difference 
  between the linear predictions of $\hat{\bm\theta}_n$
and those of $\bm\theta_0$:
\begin{equation*}
      \PR(\hat{\bm\theta}_n,\bm\theta_0):=
      \lVert\Lm(\bm X,\hat{\bm\theta}_n) - 
      \Lm(\bm X,\bm\theta_0)\rVert_2^2/n.
  \end{equation*}
\end{definition}
\begin{definition}[Mean predictive risk]\label{def:mean predictive risk}
  The {\it Mean Predictive Risk}\index{Mean predictive risk} (MPR) of an estimator
  $\hat{\bm\theta}_n$ of $\bm\theta_0$ is the expected 
  predictive risk of the estimator:
\begin{equation*}
      \MPR(\hat{\bm\theta}_n,\bm\theta_0):=
      \E[\lVert\Lm(\bm X,\hat{\bm\theta}_n) - 
      \Lm(\bm X,\bm\theta_0)\rVert_2^2/n].
  \end{equation*}
\end{definition}
While the MSE and and the MPR are deterministic quantities,
the estimation risk and the predictive risk are stochastic,
unless we treat the design matrix as fixed.

As a corollary to Proposition \ref{prop:prediction in range X'},
the predictive risk of an estimator is unchanged if both the
estimator and the estimand are projected onto $\Range(\bm X')$.
\begin{corollary}
    Given a matrix $\bm X\in\R^{n\times p}$, for any estimator $\hat{\bm\theta}_n$
    of $\bm\theta_0\in\R^p$:
    \begin{equation*}
        \PR(\hat{\bm\theta}_n,\bm\theta_0)=\PR(\bm\Proj_{\Range(\bm X')}\hat{\bm\theta}_n,\bm\Proj_{\Range(\bm X')}\bm\theta_0).
    \end{equation*}
\end{corollary}
\begin{proof}
    The result follows from Proposition \ref{prop:prediction in range X'}.
\end{proof}
The next proposition justifies the definition of mean predictive risk
given in Definition \ref{def:mean predictive risk}.
\begin{proposition}\label{prop:MPR justification}
    Assume that the linear model (\ref{linear model}) holds with $\E[\bm x\varepsilon_0]=\bm 0$.
    Then, for any $\bm\theta\in\R^p$:
    $$\E[\lVert\bm y-\bm X\bm\theta\rVert_2^2/n]=\MPR(\bm\theta,\bm\theta_0)+\E[\norm{\bm\varepsilon_0}_2^2/n].$$
\end{proposition}
\begin{proof}
Since $\bm y=\bm X\bm\theta_0+\bm\varepsilon_0$, we have
  $$\E[\lVert\bm y-\bm X\bm\theta\rVert_2^2/n]=\E[\lVert\bm X(\bm\theta_0-\bm\theta)\rVert_2^2/n]+
  \E[\norm{\bm\varepsilon_0}_2^2/n]+2(\bm\theta_0-\bm\theta)\E[\bm X'\bm\varepsilon_0].$$
  Then, the result follows since  
  $\E[\bm X'\bm\varepsilon_0]=\sum_{i=1}^n\E[\bm X_i\varepsilon_{0i}]=\bm 0$,
  where $\bm X_i$ denotes the $i-$th row of $\bm X$.
\end{proof}
If our primary goal is to accurately predict the target variable, we seek a estimators $\hat{\bm\theta}$ with a low mean prediction risk
$\E[\lVert\bm y-\bm X\hat{\bm\theta}\rVert_2^2/n]$.
Since we cannot control the error term $\varepsilon_0$,
Proposition \ref{prop:MPR justification} suggests that
we should focus on estimators with a low mean predictive risk.

On the other hand, if our interest lies in understanding which predictors influence the target variable and how they do so, the true coefficient $\bm\theta_0$
becomes our focus.
In this case, we might prefer unbiased estimators -- those with zero bias -- over biased ones. However, estimators with lower mean squared error (MSE) are generally favored, even if they feature some bias. 
The following proposition demonstrates that the MSE can be decomposed into
a bias and a variance term.
\begin{proposition}[Bias-variance decomposition of MSE]\label{prop:MSE bias-var decomp}
  Given an estimator $\hat{\bm\theta}_n\in\R^p$ for $\bm\theta_0\in\R^p$,
  the MSE can be decomposed as follows:
  $$\MSE(\hat{\bm\theta}_n,\bm\theta_0)=
  \lVert\Bias(\hat{\bm\theta}_n,\bm\theta_0)\rVert_2^2 + 
  \Trace(\Var[\hat{\bm\theta}_n]).$$
\end{proposition}
\begin{proof}
  The result follows from \(\E[\hat{\bm\theta}_n - \E[\hat{\bm\theta}_n]]=0\) and
  \begin{align*}
      \MSE(\hat{\bm\theta}_n,\bm\theta_0)&=
      \E[(\hat{\bm\theta}_n - \bm\theta_0)'(\hat{\bm\theta}_n - \bm\theta_0)]\\
      &=\E[\Trace\{(\hat{\bm\theta}_n - \bm\theta_0)(\hat{\bm\theta}_n - \bm\theta_0)'\}]\\
      &=\Trace(\E[(\hat{\bm\theta}_n - \bm\theta_0)(\hat{\bm\theta}_n - \bm\theta_0)'])\\
      &=\Trace(\E[(\hat{\bm\theta}_n - \E[\hat{\bm\theta}_n] + \Bias(\hat{\bm\theta}_n,\bm\theta_0))(\hat{\bm\theta}_n - \E[\hat{\bm\theta}_n] + \Bias(\hat{\bm\theta}_n,\bm\theta_0))'])\\
      &=\Trace(\Var[\hat{\bm\theta}_n] + \Bias(\hat{\bm\theta}_n,\bm\theta_0)\Bias(\hat{\bm\theta}_n,\bm\theta_0)'
      +\\ 
      &\qquad\qquad\
      \E[\hat{\bm\theta}_n-\E[\hat{\bm\theta}_n]]
    \Bias(\hat{\bm\theta}_n,\bm\theta_0)'+\Bias(\hat{\bm\theta}_n,\bm\theta_0)\E[
    \hat{\bm\theta}_n-\E[\hat{\bm\theta}_n]]')\\
    &=\Trace(\Var[\hat{\bm\theta}_n] + \Bias(\hat{\bm\theta}_n,\bm\theta_0)\Bias(\hat{\bm\theta}_n,\bm\theta_0)')\\
    &=\lVert\Bias(\hat{\bm\theta}_n,\bm\theta_0)\rVert_2^2 + 
  \Trace(\Var[\hat{\bm\theta}_n]).
  \end{align*}
\end{proof}
Loosely speaking, the bias and the variance of an estimator 
are linked to the estimator's "complexity".
Estimators with higher complexity often fit the data better, 
resulting in lower bias, but they are more sensitive to data variations, 
leading to higher variance. 
Conversely, estimators with lower complexity 
tend to have lower variance but higher bias,
a phenomenon known as the {\it bias-variance tradeoff}\index{Bias-variance tradeoff}.

Apart from simple cases, computing the finite-sample properties of 
estimators, such as their MSE or predictive risk, 
is infeasible or overly complicated. 
This is because they require computations under the DGP of complex transformations of the data. 
When direct computation is not possible, 
we can rely on {\it concentration inequalities} or {\it asymptotic approximations}.

Concentration inequalities are inequalities that bound the probability 
that a random variable deviates from a particular value, typically its expectation. 
In this chapter, we focus on inequalities that control the MSE 
or predictive risk of an estimator, such as:
$$\P[d(\hat{\bm\theta}_n,\bm\theta_0)\le h(\bm y, \bm X,n,p)]\ge1-\delta,$$
or 
$$\P[d(\Lm(\bm X,\hat{\bm\theta}_n),\Lm(\bm X,\bm\theta_0))
\le h(\bm y, \bm X,n,p)]\ge1-\delta,$$
where $\delta\in(0,1)$ is the {\it level of confidence}, 
$d:\mathbb R^p\times\mathbb R^p\to[0,+\infty)$
is a distance, and $h$ is a real-valued function of
the data, the sample size, and the number of predictors.

\subsection*{Large-sample properties}

Large-sample or asymptotic theory
provides an alternative approach to study and analyse estimators.
Classically, this framework develops approximations of the finite-sample properties of estimators,
such as their distribution, MSE or predictive risk,
by letting the sample size $n\to\infty$.
Consequently, these approximations work well when the sample size $n$ is much larger than 
the number of predictors $p$.
More recently, asymptotic approximations are also developed by letting $p\to\infty$,
or having both $n,p\to\infty$ at some rate.
Note that, given a sample of size $n$ and number of variables $p$, 
there is no general indication on how to choose the appropriate asymptotic regime for $n$ and $p$, as the
goodness of fit of the corresponding asymptotic approximations should be assessed on a case by case basis.
In this chapter, we work with two notions from large-sample theory: consistency and asymptotic distribution.
\begin{definition}[Consistency] 
  Estimator $\hat{\bm\theta}_n$ of $\bm\theta_0$ is {\it consistent}\index{Consistency},
  written $\hat{\bm\theta}_n\overset{\mathbb P}{\to}\bm\theta_0$ as $n\to\infty$,
  if for all $\varepsilon>0$,
  $$\lim_{n\to\infty}\P[\lvert\hat{\bm\theta}_n-\bm\theta_0\rvert>\varepsilon]=0.$$
\end{definition}
\begin{definition}[Asymptotic distribution] 
  Given a deterministic real-valued sequence $r_{n,p}\to\infty$, 
  let $F_{n,p}$ be the probability distribution of 
  $r_{n,p}(\hat{\bm\theta}_n-\bm\theta_0)$ and
  $F$ a non-degenerate probability distribution.
  Estimator $\hat{\bm\theta}_n$ of $\bm\theta_0$ has {\it asymptotic distribution}\index{Asymptotic distribution}
  $F$ with rate of convergence $r_{n,p}$
  if $F_{n,p}(\bm z)\to F(\bm z)$ as $r_{n,p}\to\infty$ for all $\bm z$ at which $F(\bm z)$ is continuous.
  Equivalent short-hand notations are $r_{n,p}(\hat{\bm\theta}_n-\bm\theta_0)\overset{d}{\to} F$ and
  $r_{n,p}(\hat{\bm\theta}_n-\bm\theta_0)\overset{d}{\to}\bm\eta\sim F$, as $r_{n,p}\to\infty$.
\end{definition}

\section{Least Squares and Penalized Least Squares}
\label{sec:least squares}

In this chapter, we study the most widely used methods in linear regression analysis:
the method of least squares and some of its penalized variants.
The method of least squares was first introduced 
by \citet{legendre1805nouvelles} and \citet{gauss1809theoria},
and it consists in minimizing the squared $l_2-$distance between 
the target values $\bm y$ and the
linear prediction $\Lm(\bm X,\bm\theta)=\bm X\bm\theta$
in the coefficient vector $\bm\theta\in\R^p$.
\begin{figure}[H]
    \centering
    \includegraphics[width=0.2\textwidth]{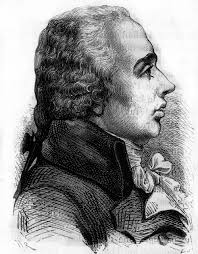} \includegraphics[width=0.225\textwidth]{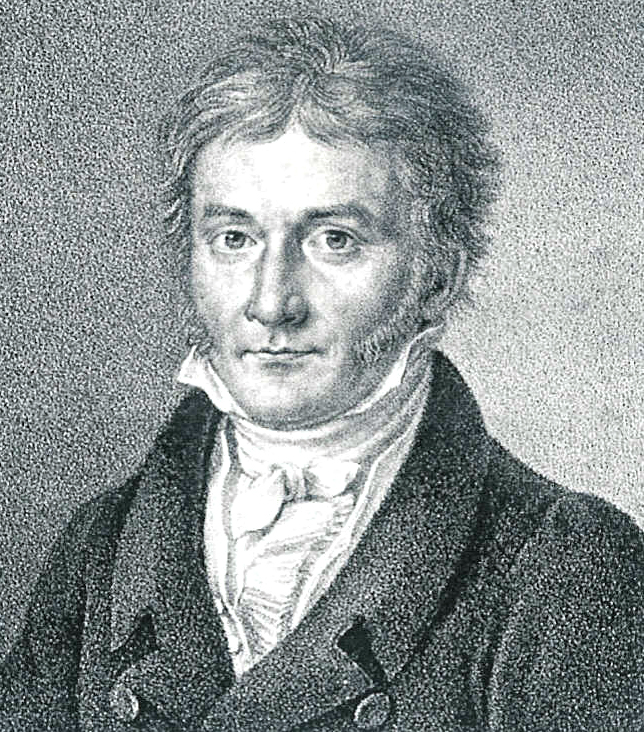}
    \caption{Adrien-Marie Legendre (1752--1833) and Johann Carl Friedrich Gauss (1977--1855).}
\end{figure}
\begin{definition}[Least squares estimator]
  The {\it Least Squares Estimator}\index{Least squares} (LSE) is defined as:
  \begin{equation}\label{least squares problem}
    \hat{\bm\theta}_n^{ls}\in\argmin_{\bm\theta\in\R^p}
    \frac{1}{2}\sum_{i=1}^n(y_i-\bm\theta'\bm x_i)^2=
    \argmin_{\bm\theta\in\R^p}
\frac{1}{2}\norm{\bm y-\bm X\bm\theta}_2^2.
\end{equation}
\end{definition}
In addition to the LSE, we consider the following variants.
\begin{definition}[Ridgeless estimator]
  The {\it ridgeless} estimator\index{Ridgeless}
  is defined as:
\begin{equation}\label{ridgeless estimator}
  \hat{\bm \theta}_n^{rl}=\argmin_{\hat{\bm\theta}\in\R^p}
  \left\{\lVert\hat{\bm\theta}\rVert_2^2\ :\ 
  \hat{\bm\theta}\in\argmin_{\bm\theta\in\R^p}
\frac{1}{2}\norm{\bm y-\bm X\bm\theta}_2^2\right\}.
\end{equation}
\end{definition}
\begin{definition}[Ridge estimator]
  The {\it ridge estimator}\index{Ridge} is defined for $\lambda>0$ as:
\begin{equation}\label{ridge estimator}
  \hat{\bm \theta}_n^r(\lambda)=\argmin_{\bm\theta\in\R^p}
  \frac{1}{2}\norm{\bm y-\bm X\bm\theta}_2^2 +\frac{\lambda}{2}\norm{\bm\theta}_2^2.
\end{equation}
\end{definition}
\begin{definition}[Lasso estimator]
  The {\it lasso estimator}\index{Lasso} is defined for $\lambda>0$ as:
\begin{equation}\label{lasso estimator}
  \hat{\bm \theta}_n^l(\lambda)\in\argmin_{\bm\theta\in\R^p}
  \frac{1}{2}\norm{\bm y-\bm X\bm\theta}_2^2 +\lambda\norm{\bm\theta}_1.
\end{equation}
\end{definition}
Here is a brief overview of the results that are discussed in detail in the rest of this chapter.
A solution to the least squares problem 
\eqref{least squares problem} always exists. 
However, when the predictors (i.e., the columns of $\bm X$)
are linearly dependent, there are infinitely many solutions.\footnote{This situation always arises when $p>n$, and it may arise even when $n\le p$.}
In such cases, the LSE typically considered is the ridgeless estimator, 
which is always unique.

The ridge and lasso estimators
are penalized or regularized versions of the LSE,
with penalty term $\lambda\norm{\bm\theta}_2^2$ and $\lambda\norm{\bm\theta}_1$, respectively.
The penalty parameter $\lambda>0$ controls the strength of the penalty.
The ridge estimator, introduced by \citet{hoerl1970ridge}, was developed 
to address certain shortcomings of the LSE, 
particularly in scenarios involving collinear or multicollinear 
designs -- where the predictors in $\bm X$ are linearly dependent 
or nearly-linearly dependent. 
The ridge estimator is uniquely defined 
and often exhibits better statistical properties compared to those of the LSE 
in settings with multicollinear or many predictors.
On the other hand, the lasso estimator, popularized by \citet{tibshirani1996regression}, 
offers an approximation of the 
$l_0$ estimator, which is defined for some $R>0$:
\begin{equation}\label{l0 estimator}
    \hat{\bm\theta}_n^{l_0}(\lambda)\in\argmin_{\bm\theta\in\R^p}\left\{
  \frac{1}{2}\norm{\bm y-\bm X\bm\theta}_2^2\ :\ \norm{\bm\theta}_0\le R\right\},
\end{equation}
where $\norm{\bm\theta}_0$ is the number of nonzero elements in $\bm\theta$.
A key feature of this estimator is its ability 
to produce sparse solutions, i.e.,
to set some coefficients exactly to zero.
Consequently, the $l_0$ estimator can be used to perform 
parameter estimation and variable selection simultaneously.
However, it is the solution of a non-convex problem, and,
in general, computing it can be an "NP-hard" problem. 
The lasso instead shares the ability to produce sparse solutions
and it can be easily computed even for large datasets.
\begin{remark}[Data standardization]\label{remark:standardization}
    For computational stability, it is recommended to compute linear regression estimators
    with a least squares loss after having standardized 
    the predictors $\bm X$ so that $\bar{\bm x}:=\bm X'\bm 1/n=\bm 0$ and $\bm X_j'\bm X_j=1$ for each $j=1\ldots,p$. 
    Without standardization, the solutions would depend on the units used to measure the predictors.
    Moreover, we may also center the target variable $\bm y$, meaning $\bar{y}:=\bm y'\bm 1/n=0$.
    These centering conditions are convenient, since they mean that we can omit the intercept term.
    Given an optimal solution $\hat{\bm\theta}$ on the centered data, we can recover the
    optimal solutions for the uncentered data: $\hat{\bm\theta}$ is the same and the intercept is given by 
    $\bar{y}-\bar{\bm x}'\hat{\bm\theta}$.
\end{remark}

\subsection{Existence and uniqueness}
\label{sec:ls existence uniqueness}

From here on, we make extensive use of the spectral decomposition of $\bm X$.
\begin{definition}[Spectral decomposition of $\bm X$]\label{def:spectral X}
  The spectral decomposition of $\bm X$ is $$\bm X=\bm U\bm S\bm V',$$ where
$\bm U\in\R^{n\times n}$ and $\bm V\in\R^{p\times p}$ are orthogonal matrices
and, for $r:=\Rank(\bm X)\le\min\{n,p\},$
$$\bm S=\begin{bmatrix}\diag(s_{1},\ldots,s_{r})&\bm 0\\ 
\bm 0&\bm 0\end{bmatrix}\in\R^{n\times p}.$$
\end{definition}

We establish the following key results: 
the existence of the LSE, the ridgeless, 
the ridge and the lasso estimators; 
the closed-form expression of the LSE, ridgeless, and ridge; 
the uniqueness of the ridgeless and ridge; 
the uniqueness of the LSE when $\Rank(\bm X)=p$, i.e., when 
the predictors in $\bm X$ are linearly independent.
Notice that this rank condition cannot hold if $n<p$.
\begin{theorem}[Existence and uniqueness of LSE, ridgeless, ridge and lasso]\label{prop:existence uniqueness ls}
\begin{enumerate}[label=(\roman*)]
The following statements hold:
    \item
    The set of solutions to the least squares problem 
(\ref{least squares problem}) is non-empty and given by
$$\argmin_{\bm\theta\in\R^p}
\norm{\bm y-\bm X\bm\theta}_2^2/2=\bm X^+\bm y+\Ker(\bm X).
$$
\item
The ridgeless estimator exists, is an element of $\Range(\bm X')$, 
and is uniquely given in closed form by
\begin{equation}\label{ridgeless closed form}
  \hat{\bm \theta}_n^{rl}=\bm X^+\bm y.
\end{equation}
\item 
If $\Rank(\bm X)=p$, 
    then the LSE and the ridgeless estimator 
    are uniquely given in closed form by:
    \begin{equation}\label{least squares estimator}
      \hat{\bm\theta}_n^{ls}=\hat{\bm \theta}_n^{rl}
      =(\bm X'\bm X)^{-1}\bm X'\bm y.
    \end{equation}
\item
The ridge estimator with penalty parameter $\lambda>0$ exists, is an element of $\Range(\bm X')$,
and is uniquely given in closed form by:
  \begin{equation}\label{ridge closed form}
    \hat{\bm\theta}_n^r(\lambda)=(\bm X'\bm X+\lambda\bm I)^{-1}\bm X'\bm y.
  \end{equation}
\item
The lasso estimator exists and, in general, it is not unique.
\end{enumerate}
\end{theorem}
\begin{proof}
\begin{enumerate}[label=(\roman*)]
    \item The least squares problem (\ref{least squares problem}) is an 
  unconstrained optimization problem with the objective function 
  $$f:\mathbb R^p\to[0,+\infty);\
  \bm\theta\mapsto \norm{\bm y-\bm X\bm\theta}_2^2/2.$$
  By Theorem \ref{thm: fermat's rule},
  the set of least squares minimizers is given by 
  $$S:=\{\hat{\bm\theta}\in\R^p : \bm X'\bm X\hat{\bm\theta}=\bm X'\bm y\}.$$
  Since $\bm X'\bm y\in\Range(\bm X')$ and 
  $\Range(\bm X')=\Range(\bm X'\bm X)$,
  set $S$ is not empty.
  Consider a vector $\hat{\bm\theta}\in \bm X^+\bm y+\Ker(\bm X)$.
  Using that $\bm X=\bm\Proj_{\Range(\bm X)}\bm X$ which implies
  $\bm X'=\bm X'\bm\Proj_{\Range(\bm X)}$,
  we obtain
  $$\bm X'\bm X\hat{\bm\theta}=\bm X'\bm X\bm X^+\bm y=\bm X'\bm\Proj_{\Range(\bm X)}\bm y=\bm X'\bm y.$$
  Therefore, $\bm X^+\bm y+\Ker(\bm X)\subset S$.
  Now consider a vector $\bm v\in\R^p$ not in 
  set $\bm X^+\bm y+\Ker(\bm X)$.
  That is, $\bm v=\hat{\bm\theta}+\bm u$ with 
  $\hat{\bm\theta}\in \bm X^+\bm y+\Ker(\bm X)$ and
  $\bm u\in\Range(\bm X')$.
  Since $\bm X\bm u\not=\bm 0$, 
  $$\bm X'\bm X\bm v=\bm X'\bm y+\bm X'\bm X\bm u\not=\bm X'\bm y.$$
  We conclude that $\bm X^+\bm y+\Ker(\bm X)= S$. 
  \item The minimum norm least squares problem in 
  (\ref{ridgeless estimator})
  has a strictly convex and coercive objective function
  $$f:\R^p\to\R;\bm\theta\mapsto\norm{\bm\theta}_2^2,$$
  and a closed convex feasible set $\bm X^+\bm y+\Ker(\bm X)\subset\R^p$.
  It follows that a solution exists and it is unique; see Propositions \ref{app:prop:existence minimizers} and
  \ref{app:prop:uniqueness minimizers}.
  Since $\bm X^+\bm y \in \Range(\bm X')$ and 
  $\bm v \in \Ker(\bm X) = \Range(\bm X')^\perp$ are orthogonal, we have
  \[
  \|\bm X^+\bm y\|_2^2 \le \|\bm X^+\bm y\|_2^2 + \|\bm v\|_2^2 \le 
    \|\bm X^+\bm y + \bm v\|_2^2,
    \]
    and therefore
    \[
    \|\bm X^+\bm y\|_2\le\|\bm X^+\bm y + \bm v\|_2.
    \]
  Thus, the ridgeless estimator can be expressed in closed form as 
  $\hat{\bm \theta}_n^{rl}=\bm X^+\bm y$,
  which is an element of $\Range(\bm X')$ since
  $\bm X^+=\bm\Proj_{\Range(\bm X')}\bm X^+$.
  \item If $\Rank(\bm X)=p$, then $\Ker(\bm X)=\{\bm 0\}$.
  Moreover, $\bm X'\bm X$ is invertible and we can use the identity
  $\bm X^+=(\bm X'\bm X)^{-1}\bm X'$ to conclude that 
  the LSE and the ridgeless estimator are uniquely given by 
  (\ref{least squares estimator}). 
  \item The ridge problem in (\ref{ridge estimator}) is
  an unconstrained optimization problem with the strictly convex, coercive and 
  continuously differentiable objective function 
  $$f:\mathbb R^p\to[0,+\infty);\
  \bm\theta\mapsto \norm{\bm y-\bm X'\bm\theta}_2^2/2 + \lambda/2\norm{\bm\theta}_2^2.$$
  It follows that a solution $\hat{\bm\theta}^r\in\R^n$ exists 
  and it is unique; see, again, Propositions \ref{app:prop:existence minimizers} and
  \ref{app:prop:uniqueness minimizers}.
  Theorem \ref{thm: fermat's rule} implies
  $$(\bm X'\bm X + \lambda\bm I)\hat{\bm\theta}^r(\lambda)
  =\bm X'\bm y.$$
  Consider the spectral decomposition $\bm X=\bm U\bm S\bm V'$ in Definition \ref{def:spectral X}.
Then, 
\begin{align*}
    \bm X'\bm X + \lambda\bm I=\bm V\bm S'\bm S\bm V'+ \lambda\bm V\bm V'
    =\bm V\begin{bmatrix}
        \diag(s_1^2,\ldots,s_r^2)+\lambda&\bm 0\\ 
\bm 0&\lambda\bm I
    \end{bmatrix}\bm V',
\end{align*}
which is positive definite, and thus
$\hat{\bm\theta}_n^r(\lambda)=(\bm X'\bm X + \lambda\bm I)^{-1}\bm X'\bm y$ is the solution to the FOCs.
Finally, to prove that $\hat{\bm \theta}_n^r(\lambda)\in\Range(\bm X')$,
notice that $\bm\Proj_{\Range(\bm X')}=\bm V\bm S^+\bm S\bm V'$, 
where
$$\bm S^+\bm S=\begin{bmatrix}
    \bm I&\bm 0\\ 
\bm 0&\bm 0
\end{bmatrix}.$$
Thus,
\begin{align*}
\bm\Proj_{\Range(\bm X')}\hat{\bm \theta}_n^r(\lambda)=&\bm V\bm S^+\bm S\bm V'\bm V(\bm S'\bm S+\lambda\bm I)^{-1}\bm V'\bm V\bm S'\bm U'\bm y\\
=&\bm V(\bm S'\bm S+\lambda\bm I)^{-1}\bm S'\bm U'\bm y=\hat{\bm \theta}_n^r(\lambda).
\end{align*}
We conclude that 
$\bm\Proj_{\Range(\bm X')}\hat{\bm \theta}_n^r(\lambda)=\hat{\bm \theta}_n^r(\lambda)$,
i.e., $\hat{\bm \theta}_n^r(\lambda)\in\Range(\bm X')$.
  \item The lasso problem in (\ref{lasso estimator}) is an unconstrained optimization problem with convex and coercive objective function
    \[
    f:\mathbb{R}^p \to [0,+\infty), 
    \qquad 
    \bm\theta \mapsto \tfrac12\|\bm y-\bm X\bm\theta\|_2^2 + \lambda\|\bm\theta\|_1.
    \]
    Hence a solution $\hat{\bm\theta}^l(\lambda)\in\mathbb{R}^p$ always exists; see Proposition~\ref{app:prop:existence minimizers}.  
    To see that such a solution is not necessarily unique, consider the case $(\bm y,\bm X)\in\mathbb{R}^n\times\mathbb{R}^{n\times 2}$ where the two predictors coincide, i.e.\ $\bm x_1=\bm x_2=\bm x\in\mathbb{R}^n$.  
    For any $(\theta_1,\theta_2)\in\mathbb{R}^2$, the lasso objective reduces to
    \[
    f(\theta_1,\theta_2)
    = \tfrac12\|\bm y-\bm x(\theta_1+\theta_2)\|_2^2 + \lambda(|\theta_1|+|\theta_2|).
    \]
    Writing $s=\theta_1+\theta_2$, we have $|\theta_1|+|\theta_2|\ge |s|$ with equality if $\theta_1$ and $\theta_2$ have the same sign. Hence the minimization problem reduces to the univariate lasso in $s$:
    \[
    \min_{s\in\mathbb{R}} \;\tfrac12\|\bm y-\bm x s\|_2^2 + \lambda|s|.
    \]
    Let $s^\star$ denote the optimal solution of this problem. Then any coefficient vector of the form $[s^\star,0]'$ or $[0,s^\star]'$ (and more generally, any decomposition $[\theta_1,\theta_2]'$ with $\theta_1+\theta_2=s^\star$ and $|\theta_1|+|\theta_2|=|s^\star|$) achieves the same objective value. These are distinct whenever $s^\star\neq 0$, establishing that the lasso solution is not unique in this example.
\end{enumerate}
\end{proof}
\begin{remark}[Computation ridgeless and ridge]
  The closed form expressions of the LSE, ridgeless and ridge estimators 
  are useful analytical result.
However, for numerical stability,
it is recommended to compute these estimators by solving their corresponding
{\it normal equations}, which are $\bm X'\bm X\hat{\bm\theta}_n=\bm X'\bm y$
for the LSE or ridgless, and
 $(\bm X'\bm X+\lambda\bm I)\hat{\bm\theta}_n^r=\bm X'\bm y$
for the ridge.
\end{remark}
\begin{remark}[Collinearity]
     Using the notation in Definition \ref{def:spectral X},
     the minimum nonzero eigenvalue of $\bm X'\bm X$ is $s_r^2$.
     If $r<p$, then $\bm X'\bm X$ has $p-r$ zero eigenvalues and the predictors 
     are said to be collinear, that is, 
     they are linearly dependent.
     In this case $\Ker(\bm X)$ is not trivial (it contains nonzero elements), hence the LSE is not unique.
     Moreover, if $s_r\approx 0$, then the computation of 
     $$\bm X^+=\bm V\begin{bmatrix}
         \diag(1/s_1,\ldots,1/s_r)&\bm 0\\ \bm 0& \bm 0
     \end{bmatrix}\bm U',$$ and hence of the ridgeless estimator, is unstable.
     The ridge estimator instead may not display these computational hurdles, 
     provided that the penalty parameter $\lambda$ is large enough.
     That is because the minimum eigenvalue of $(\bm X'\bm X+\lambda I)$
     is $s_r^2+\lambda$.
     In Section \ref{sec:properties ridgeless ridge} we show that, if $s_r\approx 0$, 
     the ridgeless (ridge) estimator's MSE and MPR satisfy loose (sharp) concentration inequalities.
\end{remark}
\begin{remark}[Uniqueness of the lasso solution]
  \citet{tibshirani2013lasso} shows that,
  under some conditions, the lasso estimator is unique.
  For instance, if the predictors in $\bm X$ are in 
    {\it general position}, 
    then the lasso solution is unique. Specifically, 
    a set $(\bm a_j)_{j=1}^p$ where $\bm a_j\in\R^n$ for all $j$ 
    is in general position if any affine subspace
    of $\R^n$ of dimension $k<n$ contains at most $k + 1$ 
    elements of the
    set $\{\pm\bm x_1,\pm\bm x_2, . . . \pm \bm x_p\}$, excluding antipodal pairs of points (that is, points
    differing only by a sign flip).
    If the predictors are (non redundant) continuous random variables, 
    they are almost surely in general position, 
    and hence the lasso solution is unique. 
    As a result, non-uniqueness of the lasso solution typically 
    occurs with discrete-valued data, 
    such as those comprising dummy or categorical variables.\end{remark}

Since the LSE, ridgeless, ridge and lasso estimators exist, their linear predictions exist too.
Moreover, the linear predictions of the uniquely defined estimators, like ridgeless and ridge, are trivially unique.
Remarkably, some estimators that may not be unique entail unique linear predictions. 
The next lemma implies that the LSE and lasso are among these estimators.
\begin{lemma}\label{lem:uniquness linear predictions}
    Let $h:\R^p\to(-\infty,+\infty]$ be a proper convex function. Then 
    $\bm X\bm\theta_1=\bm X\bm\theta_2$ and $h(\bm\theta_1)=h(\bm\theta_2)$ for every minimizers
    $\bm\theta_1,\bm\theta_2\in\R^p$ of $$f:\R^p\to(-\infty,+\infty];\bm\theta\mapsto\frac{1}{2}\norm{\bm y-\bm X\bm\theta}_2^2+h(\bm\theta).$$
\end{lemma}
\begin{proof}
    Assume that $\bm X\bm\theta_1\neq \bm X\bm\theta_2$, and
    let $\delta:=\inf_{\bm\theta\in\R^p} f(\bm\theta)$.
    By Proposition \ref{app:prop:argmin convex},
    the set of minimizers of $f$ is convex. 
    Thus, for any $\alpha\in(0,1)$:
    \begin{align*}
        \delta =&f(\alpha\bm\theta_1+(1-\alpha)\bm\theta_2) \\
        =&\frac{1}{2}\norm{\bm y-\bm X[\alpha\bm\theta_1+(1-\alpha)\bm\theta_2]}_2^2+h(\alpha\bm\theta_1+(1-\alpha)\bm\theta_2)\\
        <&\alpha\frac{1}{2}\norm{\bm y-\bm X\bm\theta_1}_2^2+(1-\alpha)\frac{1}{2}\norm{\bm y-\bm X\bm\theta_2}_2^2+h(\alpha\bm\theta_1+(1-\alpha)\bm\theta_2)\\
        \le&\alpha\frac{1}{2}\norm{\bm y-\bm X\bm\theta_1}_2^2+(1-\alpha)\frac{1}{2}\norm{\bm y-\bm X\bm\theta_2}_2^2+\alpha h(\bm\theta_1)+(1-\alpha)h(\bm\theta_2)\\
        =&\alpha f(\bm\theta_1)+(1-\alpha)f(\bm\theta_2)=\delta,
    \end{align*}
    where the strict inequality follows from the strict convexity of $g:\bm v\mapsto\norm{\bm y-\bm v}_2^2$.
    Since the conclusion $\delta<\delta$ is absurd,
    we must have $\bm X\bm\theta_1=\bm X\bm\theta_2$;
    and since $f(\bm\theta_1)=f(\bm\theta_2)$, it follows that $h(\bm\theta_1)=h(\bm\theta_2)$.
\end{proof}
While we make use of this lemma for proving uniqueness of the predictions of lasso,
we can use a more direct approach for the other estimators' predictions, which directly provides their closed form expressions. 
We further show that the LSE's prediction has the geometric interpretation of being the unique vector in the range of $\bm X$ 
that is closest to $\bm y$ in $l_2$ distance, and that the residual vector is orthogonal to the range of $\bm X$.
\begin{theorem}[Uniqueness of linear predictions]
The following statements hold:
\begin{enumerate}[label=(\roman*)]
    \item The linear predictions of the LSE and the ridgeless estimator are
  uniquely given by:
  \begin{equation}\label{least squares prediction}
      \Lm(\bm X, \hat{\bm\theta}_n^{ls})=
      \Lm(\bm X, \hat{\bm\theta}_n^{rl})=\bm\Proj_{\Range(\bm X)}\bm y,
  \end{equation}
    which is the unique vector $\bm v\in\Range(\bm X)$
    such that 
    $$\norm{\bm y-\bm v}_2=\inf\{\norm{\bm y-\bm z}_2\ :\ \bm z\in\Range(\bm X)\}.$$
    Moreover, the residual vector 
    $\bm y-\Lm(\bm X, \hat{\bm\theta}_n^{ls})=\bm y-\Lm(\bm X, \hat{\bm\theta}_n^{rl})$ 
    is orthogonal to $\Range(\bm X)$.
    \item The linear prediction of the ridge estimator is uniquely given, for $\lambda>0$, by
    $$\Lm(\bm X, \hat{\bm\theta}_n^r(\lambda))=\bm X(\bm X'\bm X+\lambda\bm I)^{-1}\bm X'\bm y.$$
    \item The linear prediction of the lasso estimator is unique.
\end{enumerate}
\end{theorem}
\begin{proof}
\begin{enumerate}[label=(\roman*)]
  \item The linear predictions 
  $\Lm(\bm X,\hat{\bm\theta}^{ls}_n)$ and 
  $\Lm(\bm X, \hat{\bm\theta}_n^{rl})$
  are uniquely given by (\ref{least squares prediction})
  because all solutions to the least squares problem
  $\hat{\bm\theta}\in \bm X^+\bm y+\Ker(\bm X)$
  yield the same prediction 
  $$\Lm(\bm X,\hat{\bm\theta})=\bm X\bm X^+\bm y=\bm\Proj_{\Range(\bm X)}\bm y.$$ 
  By the definition of $\hat{\bm\theta}_n^{ls}$ and 
  the fact that $\Range(\bm X)$
  is a closed vector subspace of $\R^n$,
  the remaining claims follow as a direct
  application of the Hilbert projection theorem (Theorem \ref{app:thm:hilber projection theorem}).
  \item This result follows directly from the closed form 
  expression (\ref{ridge closed form}) of the ridge estimator.
  \item Since the $l_1-$norm is convex, the result follows by Lemma \ref{lem:uniquness linear predictions}. 
  \end{enumerate}
\end{proof}

\subsection{Equivalent expressions and relations}

The ridgeless and the ridge, together with their corresponding linear predictions, admit the following simple expressions. 
\begin{proposition}[Spectral expression of ridgeless and ridge]
Given the spectral decomposition $\bm X=\bm U\bm S\bm V'$ in Definition \ref{def:spectral X}:
\begin{enumerate}[label=(\roman*)]
    \item The ridgeless estimator is given by 
    \begin{equation*}\label{ridgeless usv}
        \hat{\bm\theta}^{rl}_n=\left(\sum_{j=1}^r\frac{1}{s_j}\bm v_j\bm u_j'\right)\bm y.
    \end{equation*}
    The corresponding linear prediction is
    \begin{equation}\label{ridgeless usv prediction}
        \Lm(\bm X,\hat{\bm\theta}^{rl}_n)=\left(\sum_{j=1}^r\bm u_j\bm u_j'\right)\bm y.
    \end{equation}
    \item The ridge estimator with $\lambda>0$ is given by
        \begin{equation*}\label{ridge usv}
        \hat{\bm\theta}^r_n(\lambda)=\left(\sum_{j=1}^r\frac{s_j}{s_j^2+\lambda}\bm v_j\bm u_j'\right)\bm y.
    \end{equation*}
    The corresponding linear prediction is
    \begin{equation}\label{ridge usv prediction}
        \Lm(\bm X,\hat{\bm\theta}^r_n)=\left(\sum_{j=1}^r\frac{s_j^2}{s_j^2+\lambda}\bm u_j\bm u_j'\right)\bm y.
    \end{equation}
\end{enumerate}    
\end{proposition}
\begin{proof}
    \begin{enumerate}[label=(\roman*)]
        \item From the closed-form expression of the ridgeless estimator,
        \begin{align*}
            \hat{\bm\theta}^{rl}_n=\bm X^+\bm y=\bm V\bm S^+\bm U'\bm y=\left(\sum_{j=1}^r\frac{1}{s_j}\bm v_j\bm u_j'\right)\bm y.
        \end{align*}
        Therefore,
        \begin{align*}
            \bm X\hat{\bm\theta}^{rl}_n=\bm U\bm S\bm S^+\bm U'\bm y=\left(\sum_{j=1}^r\bm u_j\bm u_j'\right)\bm y.
        \end{align*}
        \item From the closed-form expression of the ridge estimator,
        \begin{align*}
            \hat{\bm\theta}^{r}_n=(\bm X'\bm X+\lambda\bm I)^{-1}\bm X'\bm y=&
            \bm V(\bm S'\bm S+\lambda\bm I)^{-1}\bm S'\bm U'\bm y\\=&\left(\sum_{j=1}^r\frac{s_j}{s_j^2+\lambda}\bm v_j\bm u_j'\right)\bm y.
        \end{align*}
        Therefore,
        \begin{align*}
            \bm X\hat{\bm\theta}^{r}_n=\bm U\bm S(\bm S'\bm S+\lambda\bm I)^{-1}\bm S'\bm U'\bm y=\left(\sum_{j=1}^r\frac{s_j^2}{s_j^2+\lambda}\bm u_j\bm u_j'\right)\bm y.
        \end{align*}
    \end{enumerate}
\end{proof}

\begin{figure}[H]
    \centering
    \includegraphics[scale=0.25]{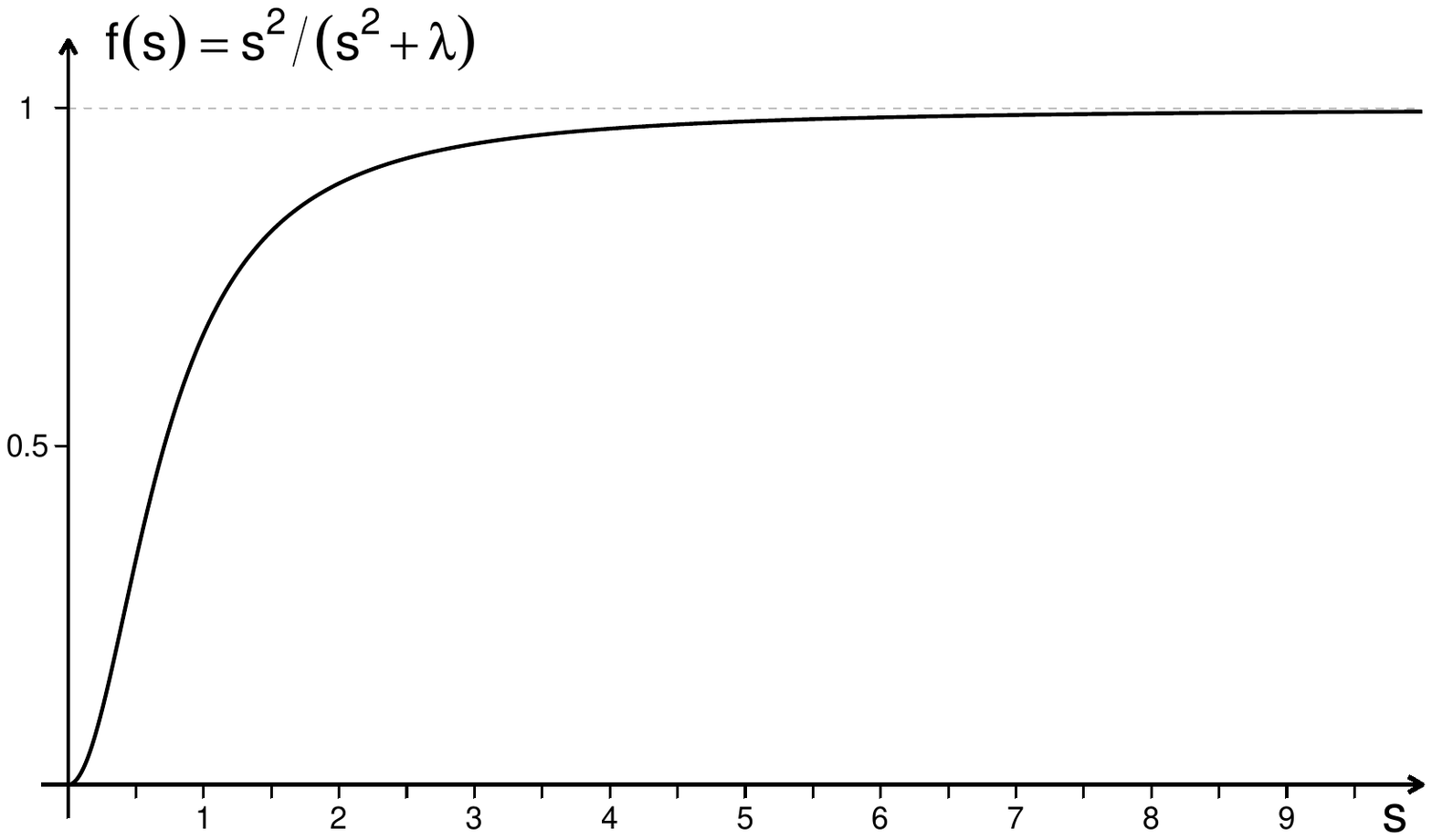}
    \caption{Shrinkage of principal components in the linear prediction of ridge when $\lambda=1/2$.}
    \label{fig:ridge sv shrinkage}
\end{figure}

Using Definition \ref{def:spectral X},
matrix $\bm\Proj_{\Range(\bm X)}=\bm X\bm X^+=\sum_{j=1}^r\bm u_j\bm u_j'$,
where $\{\bm u_1,\ldots,\bm u_r\}$ is an orthogonal basis of $\Range(\bm X)$, i.e., they define orthogonal directions in the observation space $\mathbb R^n$.
From expression (\ref{ridgeless usv prediction}), it follows that the prediction of the ridgeless estimator is the orthogonal projection of $\bm y$
onto the range of $\bm X$.
Expression (\ref{ridge usv prediction}) instead shows that the ridge estimator shrinks this projection, shrinking less
the directions $\bm u_j$ associated to high variance (high $s_j$),
and more the directions $\bm u_j$ associated to low variance (low $s_j$); see Figure \ref{fig:ridge sv shrinkage}.
Indeed, for fixed $\lambda>0$, the weight $s_j^2/(s_j^2+\lambda)\to 0$ as $s_j\to 0$,
and $s_j^2/(s_j^2+\lambda)\to 1$ as $s_j\to\infty$.
 
Each direction $\bm u_j$ corresponds to a latent principal component of the fitted values $\bm X\hat{\bm\theta}$: 
the associated singular value $s_j$ measures how much variation in the data $\bm X$ lies along that direction. 
Hence, directions with large $s_j$ correspond to high-variance directions of $\bm X$, while directions with small $s_j$ correspond to low-variance or nearly collinear directions.

High-variance directions (large $s_j$) are those that $\bm X$ can represent accurately; they correspond to well-identified combinations of regressors, and ridge leaves them nearly unshrunk. 
Low-variance directions (small $s_j$) are those where $\bm X$ has little leverage, so small perturbations in $\bm y$ would cause large changes in the least-squares estimate; ridge regularization dampens those directions heavily by multiplying by $s_j^2/(s_j^2+\lambda)\ll1$.

The ridgeless estimator can also be expressed as a penalized LSE.
\begin{proposition}[Penalized expression of ridgeless]
  The ridgeless estimator is the only solution to the least squares problem (\ref{least squares problem})
  that is in $\Range(\bm X')$, and it can be expressed as:
  \begin{equation}\label{ridgeless penalized problem}
      \hat{\bm \theta}_n^{rl}=\argmin_{\bm\theta\in\R^p}
  \frac{1}{2}\norm{\bm y-\bm X\bm\theta}_2^2
  +\frac{1}{2}\bm\theta'\bm\Proj_{\Ker(\bm X)}\bm\theta.
  \end{equation}
\end{proposition}
\begin{proof}
  From Theorem \ref{prop:existence uniqueness ls},
  the solution set of 
  the least squares problem (\ref{least squares problem}) is
  $\hat{\bm \theta}_n^{rl}+\Ker(\bm X)$, where $\hat{\bm \theta}_n^{rl}=\bm X^+\bm y$ is in $\Range(\bm X')$.
  Since $\Ker(\bm X)\perp\Range(\bm X')$, 
  $\hat{\bm \theta}_n^{rl}$ is the only least squares solution in $\Range(\bm X')$.
  Moreover, penalty \[h:\bm\theta\mapsto\bm\theta'\bm\Proj_{\Ker(\bm X)}\bm\theta=\bm\theta'(I-\bm X^+\bm X)\bm\theta\]
  is zero in $\hat{\bm \theta}_n^{rl}=\bm X^+\bm y$, and strictly positive at any other least squares solution.
  We conclude that $\hat{\bm \theta}_n^{rl}$ minimizes 
  (\ref{ridgeless penalized problem}).
\end{proof}
Following linear transformations relate the ridgeless and the ridge estimators.
\begin{proposition}[Links between ridgeless and ridge]
Following relations between the ridge and the ridgeless estimators hold:
\begin{align}\label{ridge relation ridgeless}
 &\hat{\bm \theta}_n^r(\lambda)= (\bm X'\bm X+\lambda\bm I)^{-1}\bm X'\bm X\hat{\bm \theta}_n^{rl},\\
\label{ridgeless relation ridge}
 &\hat{\bm \theta}_n^{rl}=(\bm X'\bm X)^+(\bm X'\bm X+\lambda\bm I)\hat{\bm \theta}_n^r(\lambda),\\
\label{ridgeless limit ridge}
&\lim_{\lambda\to 0}\hat{\bm \theta}_n^r(\lambda)=\hat{\bm \theta}_n^{rl}.
 \end{align}
\end{proposition}
\begin{proof}
Using $\bm X=\bm\Proj_{\Range(\bm X)}\bm X$ which implies $\bm X'=\bm X'\bm\Proj_{\Range(\bm X)}$, we have
$$(\bm X'\bm X+\lambda\bm I)^{-1}\bm X'\bm y=(\bm X'\bm X+\lambda\bm I)^{-1}\bm X'\bm X\bm X^+\bm y,$$
and thus $$\hat{\bm \theta}_n^r(\lambda)= (\bm X'\bm X+\lambda\bm I)^{-1}\bm X'\bm X\hat{\bm \theta}_n^{rl}.$$
Using $\bm X^+=\bm X^+(\bm X^+)'\bm X'$, we have
$$\bm X^+\bm y=\bm X^+(\bm X^+)'(\bm X'\bm X+\lambda\bm I)(\bm X'\bm X+\lambda\bm I)^{-1}\bm X'\bm y.$$
Moreover, $\bm X^+({\bm X^+})'=(\bm X'\bm X)^+$ implies
$$\hat{\bm \theta}_n^{rl}=(\bm X'\bm X)^+(\bm X'\bm X+\lambda\bm I)\hat{\bm \theta}_n^r(\lambda).$$
Finally, since $\bm X^+=\lim_{\lambda\to 0}(\bm X'\bm X+\lambda\bm I)^{-1}\bm X'$, we have
$\lim_{\lambda\to 0}\hat{\bm \theta}_n^r(\lambda)=\hat{\bm \theta}_n^{rl}$.
\end{proof}
Expression (\ref{ridgeless limit ridge}) explains why estimator (\ref{ridgeless estimator}) is called the ridgeless estimator.
The ridge and lasso estimators can be expressed as 
constrained least squares problems.
\begin{proposition}[Equivalence between penalized and constrained least squares]\label{prop:equivalence penalized constrained}
For $c\ge 0$, $\lambda\ge 0$, and some norm $\norm{\cdot}:\R^p\to\R$, define:
$$\mathcal C(c):=\argmin_{\bm\theta\in\R^p} \left\{
\norm{\bm y-\bm X\bm\theta}_2^2/2\ :\ \norm{\bm\theta}\le c\right\};$$
$$\mathcal P(\lambda):=\argmin_{\bm\theta\in\R^p} \left\{
\norm{\bm y-\bm X\bm\theta}_2^2/2+\lambda\norm{\bm\theta}\right\}.$$
Then, for a given $c> 0$, there exists $\lambda_0\ge 0$ such that 
$\mathcal C(c)\subset\mathcal P(\lambda_0)$.
Conversely, for a given $\lambda> 0$, there exists $c_0\ge 0$ such that 
$\mathcal P(\lambda)\subset\mathcal C(c_0)$.
\end{proposition}
\begin{proof}
The objective function 
$h:\bm\theta\mapsto\norm{\bm y-\bm X\bm\theta}_2^2/2$
is convex and continuous, and the constraint set 
$\{\bm\theta\in\R^p\ :\ \norm{\bm\theta}\le c\}$
is not empty, closed, bounded and convex.
Therefore, a solution exists, i.e., $\mathcal C(c)$ is not empty; see Proposition \ref{app:prop:existence minimizers}.
By the KKT theorem for convex problems,
$\hat{\bm\theta}\in\mathcal C(c)$ for any $c> 0$
if and only if
$\hat{\bm\theta}$ satisfies the KKT conditions, 
for some corresponding $\lambda_0\ge 0$: 
\begin{align*}
    &\bm 0\in\lambda_0\partial\lVert\hat{\bm\theta}\rVert
    +\bm X'\bm X\hat{\bm\theta}-\bm X'\bm y,\\
    &\lVert\hat{\bm\theta}\rVert\le c,\\
    &\lambda_0(\lVert\hat{\bm\theta}\rVert- c)=0.
\end{align*}
By Theorem \ref{thm: fermat's rule}, the first of these conditions implies 
that $\hat{\bm\theta}\in\mathcal P(\lambda_0)$.
Now fix a $\lambda> 0$ and notice that $\mathcal P(\lambda)$ is not empty,
given that its objective function is convex, continuous and coercive;
see Proposition \ref{app:prop:existence minimizers}.
We can thus take some $\hat{\bm\theta}\in\mathcal P(\lambda)$.
Then, $\hat{\bm\theta}$ satisfies the KKT conditions of the constrained problem for
$c_0=\lVert\hat{\bm\theta}\rVert$, which implies
$\hat{\bm\theta}\in\mathcal C(c_0)$.
\end{proof}
Note that the link between the penalty parameter $\lambda$ 
and the constraint parameter $c$ is not explicit.

\subsection{Geometric interpretation}\label{sec:geometric interpretation}

We illustrate the geometry of the least squares, ridge, and lasso solutions through a simple example.
Consider the linear model (\ref{linear model}),
with $p=2$, $\varepsilon_{0i}\sim iiN(0, 1)$, $\bm\theta_0=[1.5,0.5]'$, $\E[\bm x_i\varepsilon_0]=\bm 0$, and
$$\bm x_i\sim iiN\left(\begin{bmatrix}
    0\\0
\end{bmatrix}, \begin{bmatrix}
    2&0\\0&1
\end{bmatrix}\right).$$

\begin{figure}[H]
    \centering
    \includegraphics[scale=0.3]{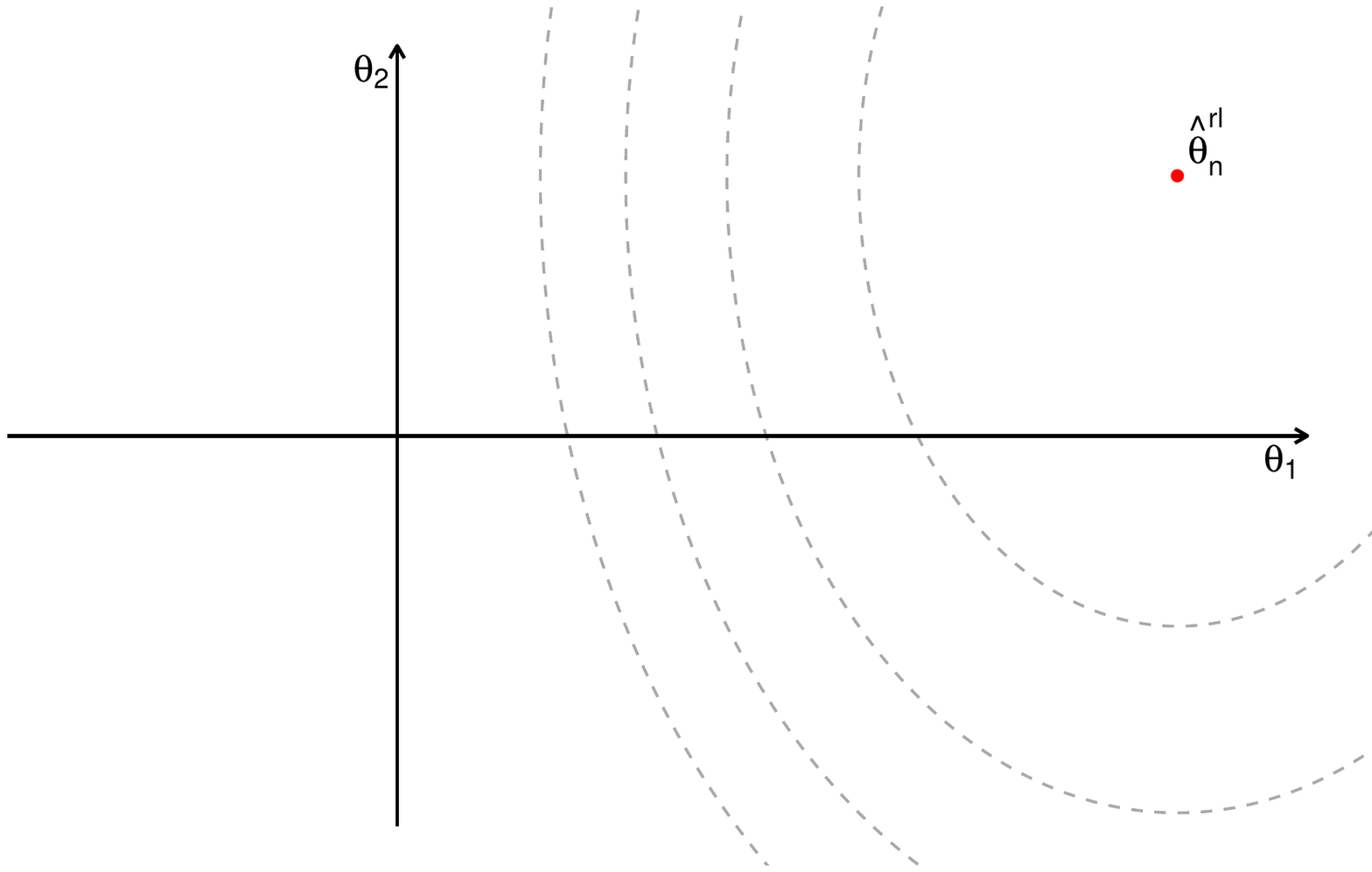}
    \caption{Geometry of the least squares solution.}
    \label{fig:ls geometry}
\end{figure}

Figure \ref{fig:ls geometry} shows the level curves of the least squares loss function $f(\bm\theta):=\norm{\bm y-\bm X\bm\theta}_2^2/2$, 
corresponding to values $f_1<f_2<f_3<f_4$.
Its minimizer, or least squares solution $\hat{\bm{\theta}}_n^{\text{ls}}$,
which coincides with the ridgeless solution $\hat{\bm\theta}_n^{rl}$, is highlighted in the figure.

\begin{figure}[H]
    \centering
    \includegraphics[scale=0.3]{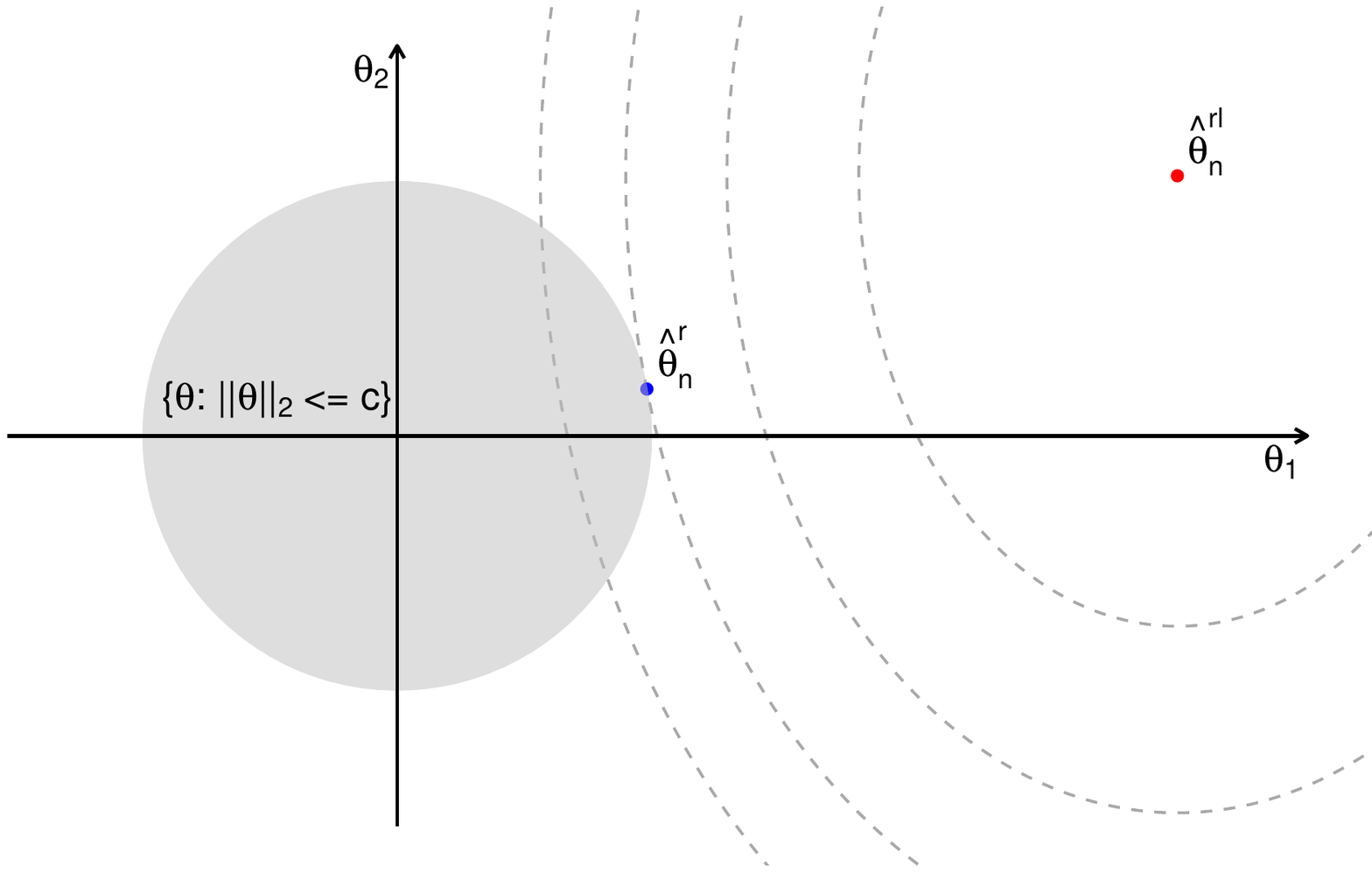}
    \caption{Geometry of the ridge solution.}
    \label{fig:ridge geometry}
\end{figure}

To illustrate the geometry of the ridge solution, we consider the constrained formulation of the ridge problem; see Proposition \ref{prop:equivalence penalized constrained}.
Figure \ref{fig:ridge geometry} demonstrates the impact of imposing the ridge constraint, represented by the sphere $\{\bm\theta\in\R^2\ :\ \norm{\bm\theta}_2\le c\}$ with $c=0.5$, on the least squares problem.
The ridge solution $\hat{\bm{\theta}}_n^{r}$ is located at the intersection between the ridge constraint and the lower level set of the least squares loss at the lowest height (see Appendix \ref{app sec:convex analysis} Definition \ref{app:def:lower level set}) for which the intersection is non-empty. If the ridgeless solution $\hat{\bm{\theta}}_n^{rl}$ lies within the constraint boundary, then $\hat{\bm{\theta}}_n^{r}$ coincides with $\hat{\bm{\theta}}_n^{rl}$.
Otherwise, the ridge solution $\hat{\bm{\theta}}_n^{r}$, by construction, is closer to the origin than $\hat{\bm{\theta}}_n^{rl}$, demonstrating the shrinkage effect of the ridge penalty.
In general, $\hat{\bm{\theta}}_n^{r}$ is dense (i.e., contains no zero elements) with probability one.

\begin{figure}[H]
    \centering
    \includegraphics[scale=0.3]{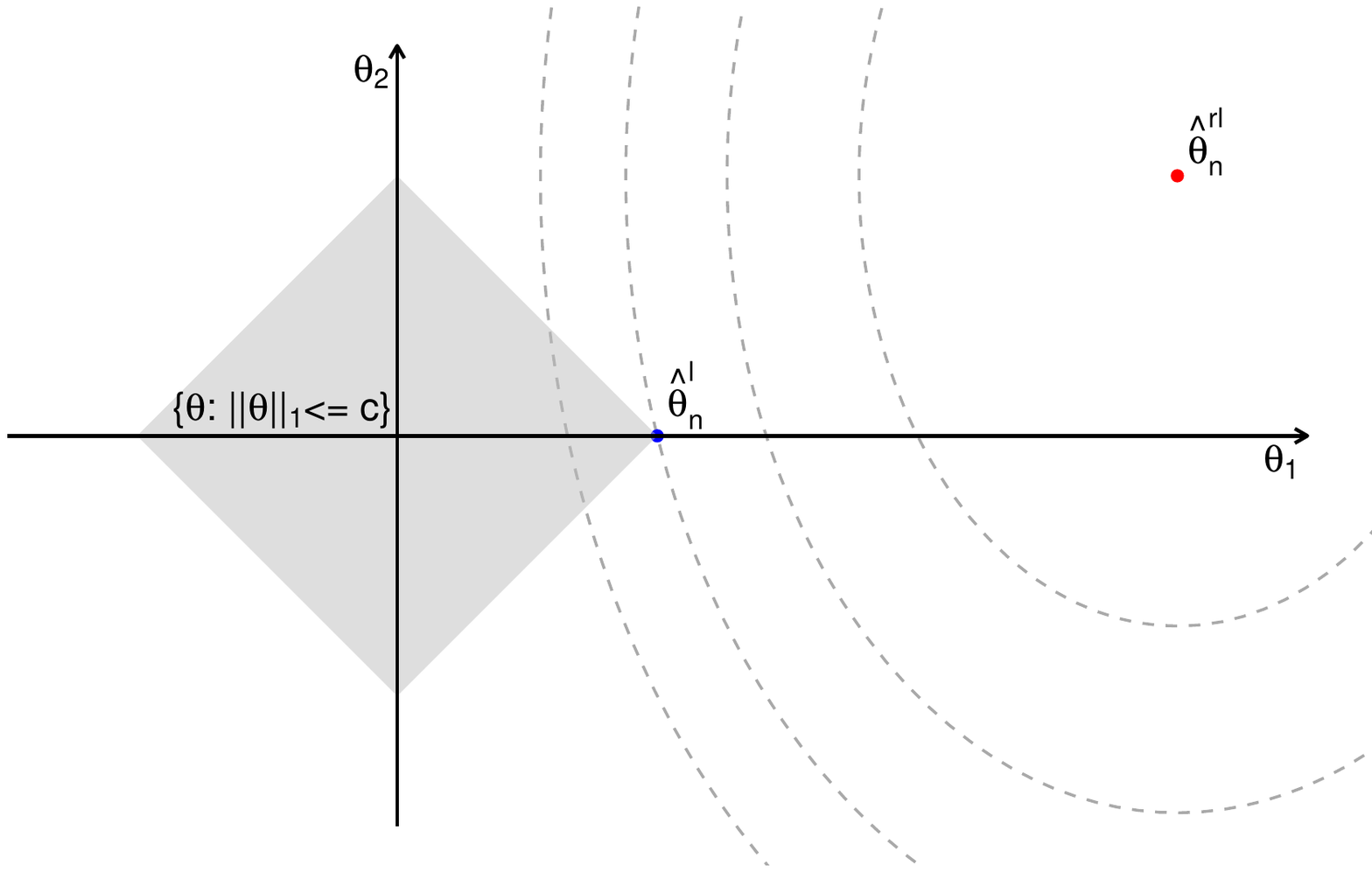}
    \caption{Geometry of the lasso solution.}
    \label{fig:lasso geometry}
\end{figure}

Figure \ref{fig:lasso geometry} illustrates the effect of the lasso constraint, represented by the rotated square $\{\bm{\theta}\in\R^2\ :\ \norm{\bm{\theta}}_1 \leq c\}$ with $c = 0.5$, on the least squares solution. 
Like the ridge solution, the lasso solution $\hat{\bm{\theta}}_n^l$ is located at the intersection of the lasso constraint and the lower level set of the least squares loss
at the lowest height for which the intersection is non-empty.
For small values of $c$, this intersection is more likely to occur along one of the coordinate axes. As a result, the lasso solution tends to be sparse, meaning that some components of $\hat{\bm{\theta}}_n^{l}$ are exactly zero.

\begin{figure}[H]
    \centering
    \includegraphics[scale=0.3]{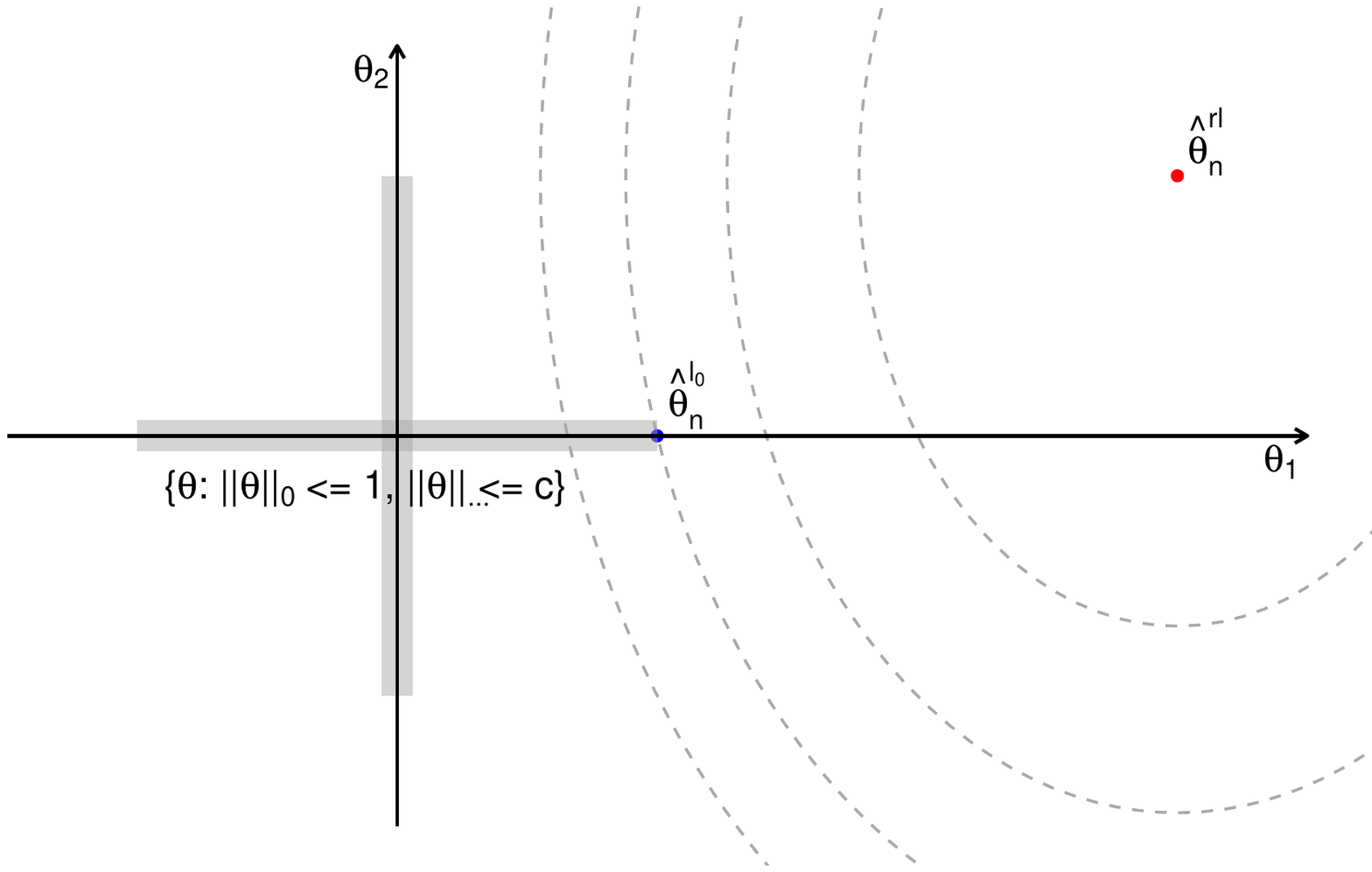}
    \caption{Geometry of the $l_0$ solution.}
    \label{fig:l0 geometry}
\end{figure}

As discussed in Section \ref{sec:least squares}, the lasso estimator serves as an approximation to the $l_0-$estimator (\ref{l0 estimator}).
This relationship becomes evident through visual comparison of Figure \ref{fig:lasso geometry} and Figure \ref{fig:l0 geometry}. 
The lasso constraint set $\{\bm\theta\ :\ \norm{\bm\theta}_1\le c\}$ 
is the convex hull (i.e., the smallest convex superset) of the constraint set underlying the $l_0-$estimator, which is given by:
$\{\bm\theta\ :\ \norm{\bm\theta}_0\le c,\ \norm{\bm\theta}_\infty\le 1\}$.
Further details on this approximation can be found in \citet{argyriou2012sparse}.

To illustrate the geometry of the ridgeless solution, consider the linear model (\ref{linear model}) with $p=2$, $\varepsilon_{0i} \sim \text{iid} N(0, 1)$, $x_{i1} \sim \text{iid} N(0, 1)$, $\E[x_{i1} \varepsilon_{0i}] = 0$, and $x_{i2} = 2x_{i1}$. In this case, the predictors are linearly dependent. As a result, the second-moment matrix of the predictors is reduced-rank:
$$\E[\bm x_i\bm x_i']=\begin{bmatrix}
    1&2\\2&4
\end{bmatrix},$$
with $\text{Rank}(\E[\bm{x}_i \bm{x}_i']) = 1$. Let $\E[x_{i1} y_i] = 1$. The identifying condition $\E[\bm x_i \varepsilon_{0i}] = \bm{0}$ holds if and only if the population coefficient $\bm{\theta}_0$ satisfies
$$\E[\bm x_i\bm x_i']\bm\theta_0=\E[\bm x_iy_i]\quad\Longleftrightarrow\quad \begin{bmatrix}
    1&2\\2&4
\end{bmatrix}\bm\theta_0=\begin{bmatrix}
    1\\2
\end{bmatrix}.$$
Thus, any coefficient in the set $\bm{\theta}_0^{\text{rl}} + \text{Kernel}(\E[\bm{x}_i \bm{x}_i'])$ satisfies this condition, where 
$$\bm\theta_0^{rl}:=\E[\bm x_i\bm x_i']^+\E[\bm x_iy_i]=[0.2,0.4]'.$$
If the sample size $n>\Rank(\E[\bm x_i\bm x_i'])$, 
then $\Ker(\bm X)\supset\Ker(\E[\bm x_i\bm x_i'])$, and
the same issue arises in the finite-sample least squares problem, where the objective function $f(\bm{\theta})$ is minimized at any point on the affine set
$$\hat{\bm\theta}_n^{rl}+\Ker(\bm X).$$

\begin{figure}[H]
    \centering
    \includegraphics[scale=0.3]{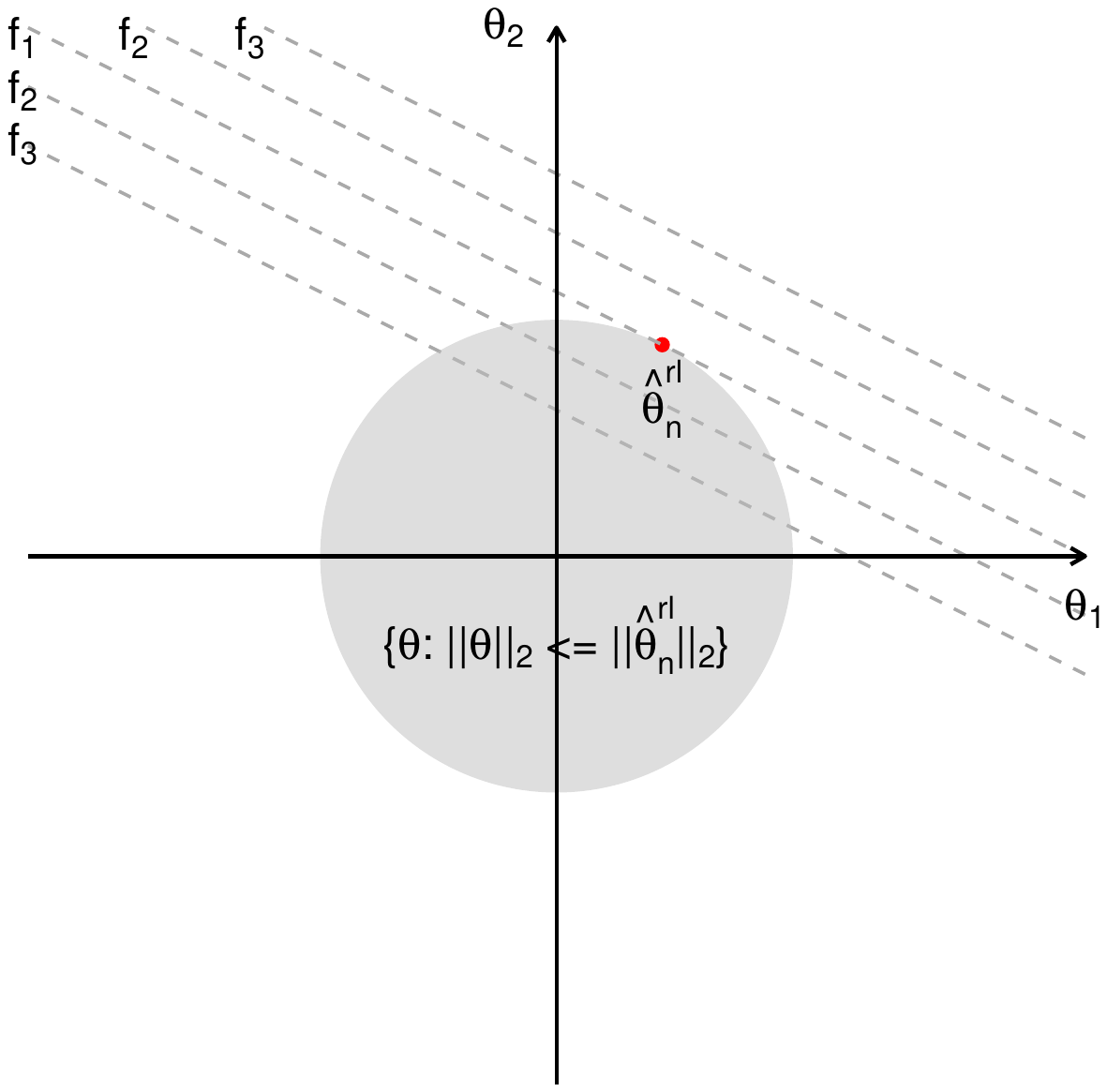}
    \caption{Geometry of the ridgeless solution.}
    \label{fig:ridgeless geometry}
\end{figure}

Figure \ref{fig:ridgeless geometry} depicts the level curves of $f(\bm{\theta})$ at $f(\hat{\bm{\theta}}_n^{\text{rl}}) = f_1 < f_2 < f_3$. These curves are parallel lines, unlike the typical ellipses seen in full-rank cases. The ridgeless estimator is the minimum $l_2$-norm solution to the least squares problem, as expected from its construction.

\subsection{Computation of lasso}

By Theorem \ref{thm: fermat's rule}, a function $f:\R^p\to\R$
is minimized at $\bm\theta^*\in\R^p$ if and only if $\bm 0\in\partial f(\bm\theta^*)$.
In the lasso problem, the objective function
\begin{equation}\label{lasso objective function}
    f^l:\R^p\to\R;\bm\theta\mapsto\frac{1}{2n}\norm{\bm y -\bm X\bm\theta}_2^2+\lambda\norm{\bm\theta}_1
\end{equation}
contains the $l_1-$norm, which makes $f^l$ non-smooth.\footnote{
  Notice that $f^l$ is the objective function in (\ref{lasso estimator}), multiplied by $1/n$.
  This term does not show up in the penalization term as it is absorbed by $\lambda$.
}
As a result, the subdifferential of $f^l$ at a minimizer $\hat{\bm\theta}_n^l(\lambda)$ is not necessarily a singleton, implying that the lasso estimator may not be unique.
Moreover, due to the complexity of $\partial f^l(\hat{\bm\theta}_n^l(\lambda))$, 
no closed-form solution exists in general for the lasso estimator.

Fortunately, Proposition \ref{prop:equivalence penalized constrained}
implies that the lasso problem is a quadratic program with a convex constraint, which allows for the computation of the lasso estimator using various quadratic programming algorithms. 
One particularly simple and effective method is the {\it cyclical coordinate descent algorithm}, which minimizes the convex objective function by iterating through each coordinate independently. This approach provides insight into how the lasso solution is obtained.

\begin{figure}[H]
    \centering
    \includegraphics[scale=0.3]{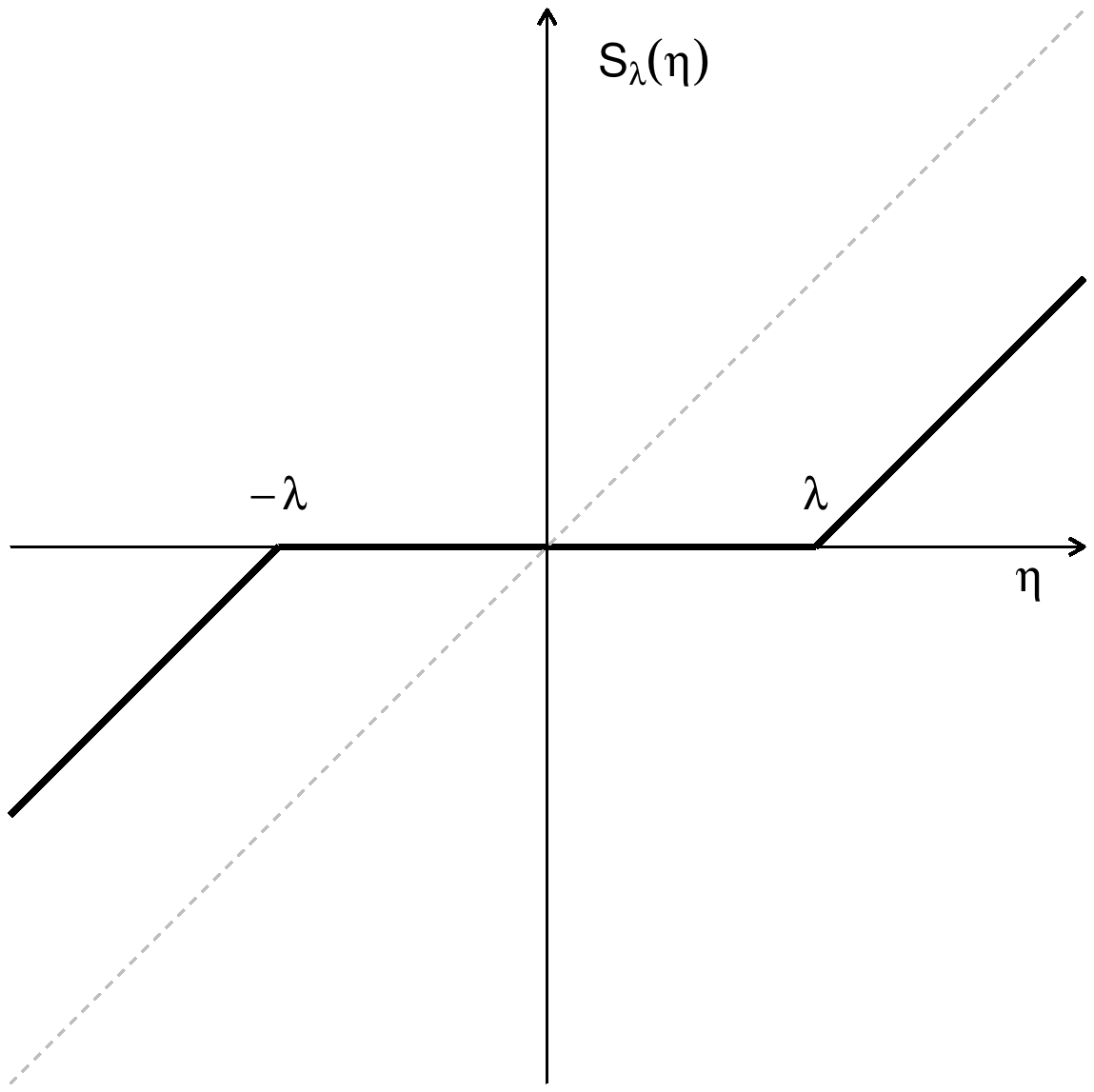}
    \caption{Soft-thresholding operator.}
    \label{fig:soft thresholding}
\end{figure}

Consider the {\it soft-thresholding operator}\index{Soft-thresholding operator} for a given $\lambda>0$, which is defined as the function 
$$\Scal_\lambda:\R\to\R; \eta\mapsto\begin{cases}
    \eta-\lambda,&\eta>\lambda\\
    0,&\eta\in[-\lambda,\lambda]\\
    \eta+\lambda,&\eta< -\lambda
\end{cases}.$$
This operator is illustrated in Figure \ref{fig:soft thresholding}.

The soft-thresholding operator provides a direct way to compute the lasso estimator in a univariate regression model, i.e.,
when there is only one predictor.
\begin{proposition}[Lasso solution for univariate regression]\label{prop:univariate lasso condition}
    Given $\lambda>0$ and $\bm X_1\in\R^n$ such that $\bm X_1'\bm X_1>0$, we have
    $$\hat{\theta}_n^l(\lambda):=\argmin_{\theta\in\R}\left\{\frac{1}{2n}\norm{\bm y -\bm X_1\theta}_2^2+\lambda|\theta|\right\}=
    \frac{\Scal_\lambda(\bm X_1'\bm y/n)}{\bm X_1'\bm X_1/n}.$$
\end{proposition}
\begin{proof}
    The subdifferential of 
    $f^l:\theta\mapsto\frac{1}{2n}\norm{\bm y -\bm X_1\theta}_2^2+\lambda|\theta|$ at $\hat{\theta}\in\R$ reads
    \begin{align*}
        \partial f^l(\hat{\theta})=b\hat{\theta}-a+\lambda\partial|\hat{\theta}|,
    \end{align*}
    where $a:=\bm X_1'\bm y/n$ and $b:=\bm X_1'\bm X_1/n$.
    From Theorem \ref{thm: fermat's rule}, and the subdifferential of the absolute value function 
    (Appendix \ref{app sec:convex analysis}, Example \ref{subdifferetnial absolute value}), 
    $\hat{\theta}^l$ is a minimizer of $f^l$ if and only if
    $$0\in\partial f^l(\hat{\theta}^l)\quad\Longleftrightarrow\quad a\in b\hat{\theta}^l+\begin{cases}
    \{\lambda\},&\hat{\theta}^l>0\\
    [-\lambda,\lambda],&\hat{\theta}^l=0\\
    \{-\lambda\},&\hat{\theta}^l<0
\end{cases}.$$
This condition reads: (i) if $\hat{\theta}^l>0$, then $\hat{\theta}^l=(a-\lambda)/b$, implying $a>\lambda$;
(ii) if $\hat{\theta}^l=0$, then $-\lambda\le a\le \lambda$; and (iii)
if $\hat{\theta}^l<0$, then $\hat{\theta}^l=(a+\lambda)/b$, implying $a< -\lambda$.
These cases are summarized by $\hat{\theta}^l=\Scal_\lambda(a)/b$.
\end{proof}
Proposition \ref{prop:univariate lasso condition} can be used to show that the $j-$th coordinate of the lasso solution 
in a multivariate regression model, i.e., when there is more than just one predictor, satisfies an expression based on the soft-thresholding operator
applied to the residual of a lasso regression of $\bm y$ onto the predictors $\bm X_k$ at position $k\neq j$. 
\begin{theorem}[Lasso solution]\label{prop:lasso solution}
Let $\bm X_j$ denote the $j-$th column of $\bm X$ and
$\bm X_{(-j)}$ denote $\bm X$ without the $j-$th column.
Assume that $\bm X_j'\bm X_j>0$ for all $j=1,\ldots,p$.
Then, given $\lambda>0$, any lasso solution $\hat{\bm\theta}_n^l(\lambda)$ is such that for all $j=1,\ldots,p$:
\begin{align}\label{lasso update rule}
    \hat{\theta}_{n,j}^l(\lambda)=\argmin_{\theta\in\R}\left\{\frac{1}{2n}\norm{\bm e_j -\bm X_j\theta}_2^2+\lambda|\theta|\right\}=
    \frac{\Scal_\lambda(\bm X_j'\bm e_j/n)}{\bm X_j'\bm X_j/n},
\end{align}
    where $\hat{\theta}_{n,j}^l(\lambda)$ is the $j-$th element of $\hat{\bm\theta}_n^l(\lambda)$, 
    $\hat{\bm\theta}_{n,(-j)}^l(\lambda)$ is $\hat{\bm\theta}_n^l(\lambda)$
    without the $j-$th element, and $\bm e_j:=\bm y-\bm X_{(-j)}\hat{\bm\theta}_{n,(-j)}^l(\lambda)$.
\end{theorem}
\begin{proof}
    The subdifferential of the lasso objective function $f^l$ 
    defined in (\ref{lasso objective function}) 
     at $\hat{\bm\theta}\in\R^p$ is
    $$\partial f^l(\hat{\bm\theta})=(\bm X'\bm X/n)\hat{\bm\theta}-\bm X'\bm y/n+\lambda\partial\lVert\hat{\bm\theta}\rVert_1,$$
    where
    $$\partial\lVert\hat{\bm\theta}\rVert_1=\{\bm v\in\R^p\ :\ v_j\in\partial |\hat{\theta}_j|\ \text{for all}\ j=1,\ldots,p\}.$$
    and the subdifferential of the absolute value function $|\cdot|$ is given in Appendix \ref{app sec:convex analysis}, Example \ref{subdifferetnial absolute value}.
    From Theorem \ref{thm: fermat's rule}, a minimizer $\hat{\bm\theta}_n^l(\lambda)$ of $f^l$ satisfies 
    $$\bm 0\in\partial f^l(\hat{\bm\theta}_n^l(\lambda))\quad\Longleftrightarrow\quad \bm X'\bm y/n\in\bm X'\bm X/n\hat{\bm\theta}_n^l(\lambda)
    +\lambda\partial\lVert\hat{\bm\theta}_n^l(\lambda)\rVert_1.$$
    This condition holds if and only if for all $j=1,\ldots,p$:
    \begin{align*}
        \bm X_j'\bm y/n&\in\bm X_j'\bm X/n\hat{\bm\theta}_n^l(\lambda)+\lambda\partial \lvert\hat{\theta}_{n,j}^l(\lambda)\rvert\quad\Longleftrightarrow\\
        \bm X_j'\bm e_j/n&\in\bm X_j'\bm X_j/n\hat{\theta}_{n,j}^l(\lambda)+\lambda\partial \lvert\hat{\theta}_{n,j}^l(\lambda)\rvert\quad\Longleftrightarrow\\
        \hat{\theta}_{n,j}^l(\lambda)&=\frac{\Scal_\lambda(\bm X_j'\bm e_j/n)}{\bm X_j'\bm X_j/n},
    \end{align*}
    where the first double implication follows from 
    \begin{align*}
        \bm X_j'\bm X\hat{\bm\theta}_n^l(\lambda)=
        \sum_{k=1}^p\bm X_j'\bm X_k\hat{\theta}_{n,k}^l(\lambda)
        =\bm X_j'\bm X_{(-j)}\hat{\bm\theta}_{n,(-j)}^l(\lambda)+\bm X_j'\bm X_j\hat{\theta}_{n,j}^l(\lambda),
    \end{align*}
    and the last double implication follows from Proposition \ref{prop:univariate lasso condition} since, by Theorem \ref{thm: fermat's rule},
    \begin{align*}
        \bm X_j'\bm e_j/n&\in\bm X_j'\bm X_j/n\hat{\theta}_{n,j}^l(\lambda)+\lambda\partial \lvert\hat{\theta}_{n,j}^l(\lambda)\rvert\quad\Longleftrightarrow\\
        \hat{\theta}_{n,j}^l(\lambda)&=\argmin_{\theta\in\R}\left\{\frac{1}{2n}\norm{\bm e_j -\bm X_j\theta}_2^2+\lambda|\theta|\right\}.
    \end{align*}
\end{proof}
Theorem \ref{prop:lasso solution} suggests that the lasso solution can be computed by a cyclical coordinate minimization
algorithm. This method is an iterative algorithm that, 
given a candidate solution $\hat{\bm\theta}^{(t)}$ at iteration $t$,
updates a single coordinate $j$ as
\[
\hat{\theta}_j^{(t+1)}
=\argmin_{\theta\in\R} 
f(\hat{\theta}_1^{(t)},\ldots,\hat{\theta}_{j-1}^{(t)},\theta,
\hat{\theta}_{j+1}^{(t)},\ldots,\hat{\theta}_p^{(t)}),
\]
while keeping all other coordinates fixed, i.e., 
$\hat{\theta}_k^{(t+1)}=\hat{\theta}_k^{(t)}$ for $k\neq j$.
A typical choice for the lasso solution would be to cycle through the
coordinates in their natural order: from $1$ to $p$.
The coordinate
descent algorithm\index{Coordinate
descent algorithm} is guaranteed to converge to a global
minimizer of any convex cost function $f:\R^p\to\R$ satisfying the
additive decomposition:
$$f:\bm\theta\mapsto g(\bm\theta)+\sum_{j=1}^ph_j(\theta_j),$$
where $g:\R^p\to\R$ is differentiable and convex, and the univariate functions $h_j:\R^p\to\R$ 
are convex (but not necessarily differentiable); see \citet{tseng2001convergence}.
What makes this algorithm work for the lasso problem is the fact that objective function (\ref{lasso objective function}) satisfies this separable structure.
\begin{remark}
    If the predictors are measured in different units, it is recommended to standardize them so that $\bm X_j'\bm X_j=1$ for all $j$.
    In this case, the lasso update
    (\ref{lasso update rule}) has the simpler form:
    $$\hat{\theta}_{n,j}^l(\lambda)=\Scal_\lambda(\bm X_j'\bm e_j/n).$$
\end{remark}

Algorithm \ref{alg:coordinate descent lasso} summarizes the pseudo-code of the cyclical coordinate descent algorithm 
for computing the lasso estimator. 
This algorithm proceeds by cyclically applying the soft-thresholding update in (\ref{lasso update rule}) for each coordinate, 
simultaneously updating the residuals 
$\bm e_j:=\bm y-\bm X_{(-j)}\hat{\bm\theta}_{n,(-j)}^l(\lambda)$.
The ridgeless or the ridge estimators can be used to initialize the procedure. 
\begin{algorithm}
\caption{Cyclical coordinate descent method for the lasso estimator.}\label{alg:coordinate descent lasso}
\begin{algorithmic}
\Require $\bm y\in\R^n$, $\bm X\in\R^{n\times p}$ with $\bm X_j'\bm X_j>0$ for all $j=1,\ldots,p$
\Require Penalty parameter $\lambda>0$
\Require Initial estimator $\hat{\bm\theta}_n^i$ (e.g., ridgeless or ridge)
\Require Maximum number of iterations $T$ and tolerance $\varepsilon>0$
\State \textbf{Standardize:} $\bm y'\bm 1=0$, $\bm X'\bm 1=\bm 0$, and $\bm X_j'\bm X_j/n=1$ for all $j$
\State $\hat{\bm\theta}^{(0)} \gets \hat{\bm\theta}_n^i$
\For{$t=1,\ldots,T$}
  \For{$j=1,\ldots,p$}
    \State $\bm e_j\gets \bm y-\bm X_{(-j)}\,\hat{\bm\theta}_{(-j)}^{(t-1)}$
    \State $\hat{\theta}_j^{(t)} \gets \mathcal S_\lambda\!\left(\bm X_j'\bm e_j/n\right)$
  \EndFor
  \If{$\|\hat{\bm\theta}^{(t)}-\hat{\bm\theta}^{(t-1)}\|_\infty<\varepsilon$}
    \State \textbf{break}
  \EndIf
\EndFor
\State \textbf{Output:} $\hat{\bm\theta}^{(t)}$
\end{algorithmic}
\end{algorithm}

In practice, it is often desirable to compute the lasso solution not for a single fixed value of $\lambda$, 
but for the entire solution path over a range of $\lambda$ values. A common approach begins by selecting 
a value of $\lambda$ just large enough that the only optimal solution is the zero vector. This value is denoted as
$\lambda_{\max}=\max_j\{|\bm X_j'\bm y|/n\}$. Starting from $\lambda_{\max}$, we progressively decrease $\lambda$ and, at each step, run coordinate descent to convergence, initializing the algorithm with the solution obtained for the previous value of $\lambda$ (a warm start).
In this
way, we can efficiently compute the solutions over a grid of $\lambda$ values. This approach is known as {\it pathwise coordinate descent}\index{Pathwise coordinate descent}.

Coordinate descent is particularly efficient for the lasso because the update rule (\ref{lasso update rule}) is available in closed form, eliminating the need for iterative searches along each coordinate. Additionally, the algorithm exploits the inherent sparsity of the problem: for sufficiently large values of 
$\lambda$, most coefficients will be zero and will remain unchanged. There are also computational strategies that can predict the active set of variables, significantly speeding up the algorithm.
More details on the pathwise coordinate descent algorithm for lasso can be found in \citet{friedman2007pathwise}.

Homotopy methods are another class of techniques for solving the lasso estimator.
They produce the entire path of solutions in a sequential fashion, starting at
zero. A homotopy
method that is particularly efficient at computing the entire lasso path is the {\it least angle regression} (LARS) algorithm; see \citet{efron2004least}.

\subsection{Finite-sample properties of ridgeless and ridge}\label{sec:properties ridgeless ridge}

This section presents finite-sample expressions and bounds for the bias, variance, MSE and MPR of the LSE, ridgeless and ridge estimators.
The main underlying assumption is that linear model $y=\bm x'\bm\theta_0+\varepsilon_0$ satisfies the typical regression condition $\E[\bm x\varepsilon_0]=\bm 0$.
Furthermore, we work with a fixed design matrix $\bm X$ (or equivalently we work conditionally on $\bm X$).

The next proposition derives the bias, MSE and MPR of the LSE when it is well-defined,
that is, when $\Rank(\bm X)=p$,
  which implies $p\le n$.
\begin{proposition}[Finite-sample properties of LSE (fixed design)]\label{prop:finite-sample properties lse}
  Assume that the linear model $y=\bm x'\bm\theta_0+\varepsilon_0$ holds with $\E[\varepsilon_0\mid\bm x]=\bm 0$.  
  Fix a realization \(\bm X\in R^{n\times p}\) with $\Rank(\bm X)=p$.
  Then (throughout, expectations are understood conditional on the fixed design $X$.):
  \begin{enumerate}[label=(\roman*)]
  \item The LSE is unbiased:
  $\E[\hat{\bm\theta}_n^{ls}]=\bm\theta_0.$
  \item The variance of the LSE is given by 
  $$\Var[\hat{\bm\theta}_n^{ls}]=
  (\bm X'\bm X)^{-1}\bm X'\Var[\bm\varepsilon_0]\bm X(\bm X'\bm X)^{-1}.$$
  \end{enumerate}
  Further let $\Var[\bm\varepsilon_0]=\sigma^2\bm I$ with $\sigma>0$.
  Then:
  \begin{enumerate}
  \item[(iii)] $\Var[\hat{\bm\theta}_n^{ls}]=\sigma^2(\bm X'\bm X)^{-1}$.
  \item[(iv)] The LSE is the best linear unbiased estimator, in the sense that
  $\Var[\tilde{\bm\theta}_n]-\Var[\hat{\bm\theta}_n^{ls}]$ 
  is positive semi-definite for any
  other unbiased linear estimator $\tilde{\bm\theta}_n$, i.e., 
  $\tilde{\bm\theta}_n=\bm A\bm y$ for some
  $\bm A\in\R^{p\times n}$.
  \item[(v)] The MSE of the LSE is given by:
  $\MSE(\hat{\bm\theta}_n^{ls},\bm\theta_0)=\frac{\sigma^2}{n}
  \sum_{j=1}^p\frac{1}{\lambda_j}$,
  where $\lambda_1\ge\ldots\ge\lambda_p>0$ are the eigenvalues
  of $\bm X'\bm X/n$.  Therefore,
  $$\MSE(\hat{\bm\theta}_n^{ls},\bm\theta_0)\le
  \frac{\sigma^2p}{\lambda_pn}.$$
  \item[(vi)] The mean predictive risk of the LSE is given by: 
  $$\MPR(\hat{\bm\theta}_n^{ls},\bm\theta_0)=p\sigma^2/n.$$
  \end{enumerate}
\end{proposition}
\begin{proof}
\begin{enumerate}[label=(\roman*)]
\item From the closed form expression (\ref{least squares estimator}):
$$\hat{\bm\theta}_n^{ls}=
(\bm X'\bm X)^{-1}\bm X'\bm\varepsilon_0+\bm\theta_0.$$
Thus, unbiasedness follows directly since:
$$\E[\hat{\bm\theta}_n^{ls}]=\bm\theta_0+
(\bm X'\bm X)^{-1}\E[\bm X'\bm\varepsilon_0]=\bm\theta_0.$$
\item The closed form expression of $\hat{\bm\theta}_n^{ls}$
immediately implies the expression
\begin{align*}
    \Var[\hat{\bm\theta}_n^{ls}]=&
    \Var[(\bm X'\bm X)^{-1}\bm X'\bm\varepsilon_0]\\=&
  (\bm X'\bm X)^{-1}\bm X'\Var[\bm\varepsilon_0]\bm X(\bm X'\bm X)^{-1}.
\end{align*}
\item Das ist trivial.
\item A linear estimator $\tilde{\bm\theta}_n=\bm A\bm y$
is unbiased if and only if $\bm A\bm X=\bm I$.
Let $\bm M:=\bm X(\bm X'\bm X)^{-1}\bm X'$, and notice that
$(\bm I-\bm M)(\bm I-\bm M)=(\bm I-\bm M)$, i.e., 
$(\bm I-\bm M)$ is idempotent.
It follows that:
\begin{align*}
  \Var[\tilde{\bm\theta}_n]-\Var[\hat{\bm\theta}_n^{ls}]=&
  \sigma^2(\bm A\bm A'-(\bm X'\bm X)^{-1})\\
  =&\sigma^2(\bm A\bm A'-\bm A\bm X(\bm X'\bm X)^{-1}\bm X'\bm A')\\
  =&\sigma^2\bm A(\bm I-\bm M)\bm A'\\
  =&\sigma^2[\bm A(\bm I-\bm M)][\bm A(\bm I-\bm M)]',
\end{align*}
which is positive semi definite.
\item Using the linearity of the $\Trace$ operator and the SVD decomposition of $\bm X$ in Definition \ref{def:spectral X}:
\begin{align*}
    \E[\lVert\hat{\bm\theta}_n^{ls}-\bm\theta_0\rVert_2^2]
    =&\E[\Trace((\hat{\bm\theta}_n^{ls}-\bm\theta_0)(\hat{\bm\theta}_n^{ls}-\bm\theta_0)')]\\
    =&\Trace(\Var[\hat{\bm\theta}_n^{ls}])\\
    =&\frac{\sigma^2}{n}\Trace((\bm X'\bm X/n)^{-1})\\
    =&\frac{\sigma^2}{n}\Trace(\bm V(\bm S'\bm S/n)^{-1}\bm V')
    =\frac{\sigma^2}{n}\sum_{j=1}^p\frac{1}{\lambda_j}.
\end{align*}
\item Simple computations give
    \begin{align*}
    \E[\lVert\Lm(\bm X,\hat{\bm\theta}_n^{ls})&-\Lm(\bm X,\bm\theta_0)\rVert_2^2/n]\\=&
    \E[\lVert\bm X(\hat{\bm\theta}_n^{ls}-\bm\theta_0)\rVert_2^2/n]\\
    =&\E[\Trace((\hat{\bm\theta}_n^{ls}-\bm\theta_0)'\bm X'
    \bm X/n(\hat{\bm\theta}_n^{ls}-\bm\theta_0))]\\
    =&\E[\Trace((\hat{\bm\theta}_n^{ls}-\bm\theta_0)(\hat{\bm\theta}_n^{ls}-\bm\theta_0)'\bm X'\bm X/n)]\\
    =&\Trace(\Var[\hat{\bm\theta}_n^{ls}]\bm X'\bm X/n)=
    \sigma^2p/n.
    \end{align*}
\end{enumerate} 
\end{proof}
This proposition shows that the LSE's accuracy
decreases:
\begin{itemize}
    \item as the variance $\sigma^2$ of the error term increases;
    \item as the number of predictors per observation $p/n$ increases;
    \item as the "degree of singularity" of the design matrix $1/\lambda_p$ increases.
\end{itemize}

\begin{remark}[Why eigenvalues of $\bm X'\bm X/n$]
We express the results in Proposition \ref{prop:finite-sample properties lse} in terms of the eigenvalues 
$\lambda_1 \ge \cdots \ge \lambda_p > 0$ 
of the normalized Gram matrix $\bm X'\bm X/n$, rather than those of $\bm X'\bm X$,
because the eigenvalues of $\bm X'\bm X/n$ remain on an $O(1)$ scale as $n$ grows; 
see, e.g., \citet{vershynin2018high}.
This keeps the $1/n$ convergence rate of the MSE explicit. 

In random design settings with i.i.d.\ rows $\bm x_i$ satisfying 
$\E[\|\bm x_i\|_2^2]<\infty$ and $\Var(\bm x_i)=\bm\Sigma$ positive definite, 
the Law of Large Numbers implies
$\bm X'\bm X/n \xrightarrow{p} \bm\Sigma$.
By continuity of the eigenvalue map
(see, e.g., Weyl's theorem; \citealp{horn2012matrix}),
it follows that
$\lambda_j(\bm X'\bm X/n) \xrightarrow{p} \lambda_j(\bm\Sigma)$
for each $j=1,\ldots,p$.
Hence, the eigenvalues of $\bm X'\bm X/n$ remain $O_p(1)$,
while those of $\bm X'\bm X$ grow linearly with $n$.
In fixed design analyses one typically assumes uniform spectral bounds,
\[
0 < \underline{\lambda}
\le
\lambda_p\big(\bm X'\bm X/n\big)
\le
\lambda_1\big(\bm X'\bm X/n\big)
\le
\overline{\lambda}
< \infty,
\]
which plays the same role.
\end{remark}

We now aim to drop the requirement that $\Rank(\bm X)=p$, to allow for high-dimensional settings where:
\begin{itemize}
    \item $n<p$, or even
    \item $\Rank(\E[\bm x\bm x'])=r_0$, not necessarily equals to $p$.
\end{itemize}
Notice that the requirement $\Rank(\E[\bm x\bm x'])=r_0$ implies \(\Rank(\bm X)\ \le\ \min\{n,r_0\}\):
\begin{proposition}[Rank bound under low-rank second moments]\label{lem:rank-bound}
Let $\bm x\in\mathbb R^p$ satisfy $\E\|\bm x\|_2^2<\infty$ and let 
$\bm\Sigma:=\E[\bm x\bm x']$ be positive semi-definite with $\Rank(\bm\Sigma)=r_0$.
Fix a realization $\bm X \in \mathbb R^{n\times p}$ with rows $\bm x_1',\ldots,\bm x_n'$, where each $\bm x_i$ has the same distribution as $\bm x$.
Then, almost surely,
\[
\Rank(\bm X)\ \le\ \min\{n,r_0\}.
\]
\end{proposition}

\begin{proof}
Since $\E\|\bm x\|_2^2<\infty$, $\bm\Sigma$ is well-defined and positive semi-definite.
If $\bm a\in\Ker(\bm\Sigma)$, then
\[
0 = \bm a'\bm\Sigma \bm a = \E\big[(\bm a'\bm x)^2\big]
\quad\Longrightarrow\quad \bm a'\bm x=0
\]
almost surely.
Hence $\bm x\in\Range(\bm\Sigma)$ almost surely, so for each $i\in\{1,\ldots,n\}$ we also have $\bm x_i\in\Range(\bm\Sigma)$ almost surely (they share the same distribution), 
i.e., all rows of $\bm X$ lie in $\Range(\bm\Sigma)$ almost surely. Therefore,
\[
\Rank(\bm X)\ \le\ \dim(\Range(\bm\Sigma))\ =\ r_0
\]
almost surely.
Combining with the trivial bound $\Rank(\bm X)\le n$ (notice that \(r_0\le p\) by construction) yields $\Rank(\bm X)\le \min\{n,r_0\}$ almost surely.
\end{proof}

We next show that, when $\Rank(\E[\bm x\bm x'])=r_0<p$, the typical linear regression condition $\E[\bm x\varepsilon_0]=\bm 0$
no longer identifies a unique estimand $\bm\theta_0$.
\begin{proposition}\label{prop:ridgeless population}
Let $y\in\R$ and $\bm x\in\R^p$ be random variables with $\E\|\bm x\|_2^2<\infty$ and $\E[y^2]<\infty$.
For any $\bm\theta\in\R^p$ let $\varepsilon(\bm\theta):=y-\bm x'\bm\theta$.
Consider the set
\[
S:=\{\bm\theta_0\in\R^p:\ y=\bm x'\bm\theta_0+\varepsilon(\bm\theta_0),\ \E[\bm x\varepsilon(\bm\theta_0)]=\bm 0\}.
\]
Then $S$ is nonempty if and only if $\E[\bm x y]\in\Range(\E[\bm x\bm x'])$, 
in which case
\[
S=(\E[\bm x\bm x'])^+\E[\bm x y]+\Ker(\bm\Sigma).
\]
\end{proposition}

\begin{proof}
By definition, $y=\bm x'\bm\theta+\varepsilon(\bm\theta)$ holds for every $\bm\theta$.
Thus the restriction in $S$ is the moment condition
\[
\E[\bm x\varepsilon(\bm\theta)] = \E[\bm x(y-\bm x'\bm\theta)] = \bm 0
\quad\Longleftrightarrow\quad
\E[\bm x\bm x']\bm\theta=\E[\bm x y].
\]
With $\bm \Sigma:=\E[\bm x\bm x']$ (symmetric positive semi-definite) and $\bm b:=\E[\bm x y]$ (well-defined by the moment assumptions), this is the linear system
$\bm \Sigma\bm\theta=\bm b$.
Such a system admits a solution if and only if $\bm b\in\Range(\bm \Sigma)$; when this holds, the set of all solutions is
\[
\{\bm\theta\in\R^p\ :\ \bm \Sigma\bm\theta=\bm b\}
= \bm \Sigma^+\bm b + \Ker(\bm \Sigma).
\]
\end{proof}
However, when $S$ is nonempty, the unique element in 
$S\cap\Range(\E[\bm x\bm x'])$ remains well defined even when 
$\Rank(\E[\bm x\bm x'])<p$. 
Moreover, when $\Rank(\E[\bm x\bm x'])=p$, this element coincides with the population parameter 
$\bm\theta_0$ by construction.
We define it as follows.
\begin{definition}[Ridgeless estimand]\label{ridgeless estimand}
The {\it ridgeless estimand}\index{Ridgeless estimand} is
the parameter vector
\[
\bm\theta_0^{rl}:=\E[\bm x\bm x']^+\E[\bm x y],
\]
which, by construction, satisfies 
$\bm\theta_0^{rl}\in\Range(\E[\bm x\bm x'])$.\footnote{
The fact that $\bm\theta_0^{rl}\in\Range(\E[\bm x\bm x'])$ follows from the identity
$\E[\bm x\bm x']^+=\Proj_{\Range(\E[\bm x\bm x'])}\E[\bm x\bm x']^+$.
The ridgeless estimand is also used in Section~\ref{sec:geometric interpretation}.
}
\end{definition}
We can now extend Proposition \ref{prop:finite-sample properties lse} to the fixed design setting where \(\Rank(\bm X)\le p\).
\begin{proposition}[Finite-sample properties of ridgeless (fixed design)]
\label{prop:finite sample properties ridgeless}
Assume that the linear model $y=\bm x'\bm\theta_0+\varepsilon_0$ holds for some \(\bm\theta_0\in\R^p\)
with $\mathbb E[\varepsilon_0\mid\bm x]=0$, and the finite second moments $\mathbb E[\|\bm x\|_2^2]<\infty$ and $\mathbb E[y^2]<\infty$ are met.
Denote $\bm\Sigma:=\mathbb E[\bm x\bm x']$, and let $r_0:=\Rank(\bm\Sigma)$. 
Fix a realization $\bm X \in \mathbb R^{n\times p}$ with rows $\bm x_1',\ldots,\bm x_n'$, where each $\bm x_i$ has the same distribution as $\bm x$.
Then (throughout, expectations and variances are understood conditionally on the fixed design $\bm X$):
\begin{enumerate}[label=(\roman*)]
\item $\mathbb E[\hat{\bm\theta}_n^{rl}]=\bm P_{\Range(\bm X')}\bm\theta_0^{rl}$. 
If $\Range(\bm X')=\Range(\bm\Sigma)$ (which implies $n\ge r_0$), then the ridgeless estimator is unbiased: $\mathbb E[\hat{\bm\theta}_n^{rl}]=\bm\theta_0^{rl}$.

\item $\Var[\hat{\bm\theta}_n^{rl}]=\bm X^{+}\Var[\bm\varepsilon(\bm\theta_0^{rl})]
(\bm X^{+})'$, where \(\Var[\bm\varepsilon(\bm\theta_0^{rl})]=\Var[\bm\varepsilon_0]\).
\end{enumerate}

Further suppose $\Var[\bm\varepsilon(\bm\theta_0^{rl})]=\sigma^2\bm I$ with $\sigma>0$, and let $r:=\Rank(\bm X)$. Then:
\begin{enumerate}
\item[(iii)] $\Var[\hat{\bm\theta}_n^{rl}]=\sigma^2(\bm X'\bm X)^{+}$.

\item[(iv)] The mean squared error satisfies
\[
\MSE(\hat{\bm\theta}_n^{rl},\bm\theta_0^{rl})
=\frac{\sigma^2}{n}\sum_{j=1}^{r}\frac{1}{\lambda_j}
+\|\bm P_{\Ker(\bm X)}\bm\theta_0^{rl}\|_2^2,
\]
where $\lambda_1\ge\cdots\ge\lambda_r>0$ are the positive eigenvalues of $(\bm X'\bm X)/n$.

\item[(v)] The mean predictive risk is
\[
\MPR(\hat{\bm\theta}_n^{rl},\bm\theta_0^{rl})=\frac{r\sigma^2}{n}.
\]

\item[(vi)] If $\Range(\bm X')=\Range(\bm\Sigma)$, then $r=r_0$ and $\Ker(\bm X)=\Ker(\bm\Sigma)$, hence
\[
\MSE(\hat{\bm\theta}_n^{rl},\bm\theta_0^{rl})
=\frac{\sigma^2}{n}\sum_{j=1}^{r_0}\frac{1}{\lambda_j}
\le \frac{\sigma^2 r_0}{\lambda_{r_0} n},
\qquad
\MPR(\hat{\bm\theta}_n^{rl},\bm\theta_0^{rl})=\frac{r_0\sigma^2}{n}.
\]
\end{enumerate}
\end{proposition}
\begin{proof}
\begin{enumerate}[label=(\roman*)]
\item The model $y=\bm x'\bm\theta_0+\varepsilon_0$ with $\mathbb E[\varepsilon_0\mid\bm x]=0$
can equivalently be written as
\[
y=\bm x'\bm\theta_0^{rl}+\varepsilon(\bm\theta_0^{rl}),
\qquad
\mathbb E[\varepsilon(\bm\theta_0^{rl})\mid\bm x]=0;
\]
see Proposition \ref{prop:ridgeless population}.
Since $\hat{\bm\theta}^{rl}=\bm X^{+}\bm y$, we have
\[
\mathbb E[\hat{\bm\theta}^{rl}]
=\bm X^{+}\mathbb E[\bm y]
=\bm X^{+}\bm X\bm\theta_0^{rl}
=\bm P_{\Range(\bm X')}\bm\theta_0^{rl},
\]
which proves the first claim. 
Unbiasedness under $\Range(\bm X')=\Range(\bm\Sigma)$ then follows because
$\bm\theta_0^{rl}\in\Range(\bm\Sigma)=\Range(\bm X')$.

\item The closed-form expression of $\hat{\bm\theta}_n^{rl}$
immediately gives
\[
\Var[\hat{\bm\theta}_n^{rl}]
=\bm X^+\Var[\bm\varepsilon(\bm\theta_0^{rl})]
(\bm X^+)'. 
\]

Since $\bm\varepsilon(\bm\theta_0^{rl})=\bm X(\bm\theta_0-\bm\theta_0^{rl})+\bm\varepsilon_0$ 
and the first term is deterministic given $\bm X$, it follows that 
$\Var[\bm\varepsilon(\bm\theta_0^{rl})]=\Var[\bm\varepsilon_0]$.

\item It follows from (ii) since $\bm X^+(\bm X^+)'=(\bm X'\bm X)^{+}$.

\item Using $\Rank(\bm X)=r$, we obtain
\[
\operatorname{tr}(\Var[\hat{\bm\theta}_n^{rl}])
    =\frac{\sigma^2}{n}\operatorname{tr}\bigl((\bm X'\bm X/n)^{+}\bigr)
    =\frac{\sigma^2}{n}\sum_{j=1}^{r}\frac{1}{\lambda_j}.
\]
Moreover,
\[
\operatorname{Bias}(\hat{\bm\theta}_n^{rl},\bm\theta_0^{rl})
=(\bm P_{\Range(\bm X')}-\bm I)\bm\theta_0^{rl}
=-\bm P_{\Ker(\bm X)}\bm\theta_0^{rl}.
\]
The result then follows from the bias–variance decomposition in Proposition~\ref{prop:MSE bias-var decomp}.

\item By Proposition~\ref{prop:prediction in range X'} and 
$\mathbb E[\hat{\bm\theta}_n^{rl}]=\bm P_{\Range(\bm X')}\bm\theta_0^{rl}$, we have 
$\Lm(\bm X,\bm\theta_0^{rl})=\bm X\E[\hat{\bm\theta}_n^{rl}]$.
Therefore: 
\begin{align*}
    \E[\lVert\Lm(\bm X,\hat{\bm\theta}_n^{rl})&-\Lm(\bm X,\bm\theta_0^{rl})\rVert_2^2/n]\\=&
    \E[\lVert\bm X(\hat{\bm\theta}_n^{rl}-\E[\hat{\bm\theta}_n^{rl}])\rVert_2^2/n]\\
    =&\E[\Trace\{(\hat{\bm\theta}_n^{rl}-\E[\hat{\bm\theta}_n^{rl}])'\bm X'
    \bm X/n(\hat{\bm\theta}_n^{rl}-\E[\hat{\bm\theta}_n^{rl}])\}]\\
    =&\E[\Trace\{(\hat{\bm\theta}_n^{rl}-\E[\hat{\bm\theta}_n^{rl}])(\hat{\bm\theta}_n^{rl}-\E[\hat{\bm\theta}_n^{rl}])'\bm X'\bm X/n\}]\\
    =&\Trace(\Var[\hat{\bm\theta}_n^{rl}]\bm X'\bm X/n)\\
    =&\sigma^2/n\Trace[(\bm X'\bm X)^+\bm X'\bm X]=\sigma^2/n\Trace(\bm X^+\bm X),
    \end{align*}
    where the last equality follows from the identity $(\bm X'\bm X)^+\bm X'=\bm X^+$.
    
    With the spectral decomposition $\bm X=\bm U\bm S\bm V'$ (Definition~\ref{def:spectral X}),
\begin{align*}
    \sigma^2/n\Trace(\bm X^+\bm X)=&\sigma^2/n\Trace(\bm V(\bm S')^+\bm S\bm V')\\
    =&\sigma^2/n\Trace\begin{bmatrix}
        \bm I_r&\bm 0_{r\times(p-r)}\\
        \bm 0_{(p-r)\times r} & \bm 0_{(p-r)\times(p-r)}
    \end{bmatrix}\\
    =&\sigma^2r/n.
\end{align*}

\item If $\Range(\bm X')=\Range(\bm\Sigma)$, then $\Rank(\bm X)=r_0$
and $\Ker(\bm X)=\Ker(\bm\Sigma)$. Therefore
$\operatorname{Bias}(\hat{\bm\theta}_n^{rl},\bm\theta_0^{rl})=\bm 0$ since $\bm\theta_0^{rl}\in\Range(\bm\Sigma)$,
and thus
\[
\MSE(\hat{\bm\theta}_n^{rl},\bm\theta_0^{rl})
=\frac{\sigma^2}{n}\sum_{j=1}^{r_0}\frac{1}{\lambda_j}.
\]
\end{enumerate}
\end{proof}
We now consider the ridge estimator, which introduces an explicit regularization parameter $\lambda>0$ to stabilize the inversion of $\bm X'\bm X$. 
Unlike the ridgeless case, ridge estimation trades bias for reduced variance, yielding a well-defined solution for any $\lambda>0$ and any $(n,p)$ configuration.
\begin{proposition}[Finite-sample properties of ridge (fixed design)]
\label{prop:finite sample properties ridge}
Assume that the linear model $y=\bm x'\bm\theta_0+\varepsilon_0$ holds for some \(\bm\theta_0\in\R^p\)
with $\mathbb E[\varepsilon_0\mid\bm x]=0$, and the finite second moments $\mathbb E[\|\bm x\|_2^2]<\infty$ and $\mathbb E[y^2]<\infty$ are met.
Let $\bm\Sigma:=\E[\bm x\bm x']$ and $r_0:=\Rank(\bm\Sigma)$.
Fix a realization $\bm X\in\R^{n\times p}$ with rows $\bm x_1',\ldots,\bm x_n'$, where each $\bm x_i$ has the same distribution as $\bm x$,
and define \(\bm Q(\lambda):=(\bm X'\bm X+\lambda\bm I)^{-1}\bm X'\bm X\) for some $\lambda>0$.
Then (throughout, expectations and variances are understood conditionally on the fixed design $\bm X$):
\begin{enumerate}[label=(\roman*)]
\item The ridge estimator is biased:
\[
\E[\hat{\bm\theta}_n^r(\lambda)]=\bm Q(\lambda)\bm\theta_0=\bm Q(\lambda)\bm\theta_0^{rl}.
\]

\item The variance of the ridge estimator is
\[
\Var[\hat{\bm\theta}_n^r(\lambda)]
=(\bm X'\bm X+\lambda\bm I)^{-1}\bm X'\Var[\bm\varepsilon(\bm\theta_0^{rl})]\bm X(\bm X'\bm X+\lambda\bm I)^{-1}.
\]
and \(\Var[\bm\varepsilon(\bm\theta_0^{rl})]=\Var[\bm\varepsilon_0]\).
\end{enumerate}

Further assume $\Var[\bm\varepsilon(\bm\theta_0^{rl})]=\sigma^2\bm I$ with $\sigma>0$, and let $r:=\Rank(\bm X)\le\min\{n,p\}$. Then:
\begin{enumerate}
\item[(iii)] $\Var[\hat{\bm\theta}_n^r(\lambda)]
=\sigma^2(\bm X'\bm X+\lambda\bm I)^{-1}\bm X'\bm X(\bm X'\bm X+\lambda\bm I)^{-1}$.
Moreover, $\Var[\hat{\bm\theta}_n^{rl}]-\Var[\hat{\bm\theta}_n^r(\lambda)]$ is positive semidefinite, and it is positive definite on $\Range(\bm X')$.

\end{enumerate}

For the remaining statements, assume $\Range(\bm X')=\Range(\bm\Sigma)$.
\begin{enumerate}
\item[(iv)] The mean squared error is
\[
\MSE(\hat{\bm\theta}_n^r(\lambda),\bm\theta_0^{rl})
=\frac{\sigma^2}{n}\sum_{j=1}^{r}\frac{\lambda_j}{(\lambda_j+\lambda/n)^2}
+\|[\bm I-\bm Q(\lambda)]\bm\theta_0^{rl}\|_2^2,
\]
where $\lambda_1\ge\cdots\ge\lambda_r>0$ are the positive eigenvalues of $(\bm X'\bm X)/n$.

\item[(v)] The mean predictive risk is
\[
\MPR(\hat{\bm\theta}_n^r(\lambda),\bm\theta_0^{rl})
=\frac{\sigma^2}{n}\sum_{j=1}^{r}\frac{\lambda_j^2}{(\lambda_j+\lambda/n)^2}
+\frac{1}{n}\lVert \bm X[\bm Q(\lambda)-\bm I]\bm\theta_0^{rl}\rVert_2^2.
\]

\item[(vi)] Following limits apply:
\[
\lim_{\lambda\to 0}\MSE(\hat{\bm\theta}_n^r(\lambda),\bm\theta_0^{rl})
=\frac{\sigma^2}{n}\sum_{j=1}^{r_0}\frac{1}{\lambda_j}
\le\frac{\sigma^2 r_0}{\lambda_{r_0}n},
\qquad
\lim_{\lambda\to 0}\MPR(\hat{\bm\theta}_n^r(\lambda),\bm\theta_0^{rl})
=\frac{r_0\sigma^2}{n}.
\]
\end{enumerate}
\end{proposition}
\begin{proof}
\begin{enumerate}[label=(\roman*)]
\item 
The model $y=\bm x'\bm\theta_0+\varepsilon_0$ with $\mathbb E[\varepsilon_0\mid\bm x]=0$
can equivalently be written as
\[
y=\bm x'\bm\theta_0^{rl}+\varepsilon(\bm\theta_0^{rl}),
\qquad
\mathbb E[\varepsilon(\bm\theta_0^{rl})\mid\bm x]=0;
\]
see Proposition \ref{prop:ridgeless population}.
Since $\hat{\bm\theta}_n^{r}(\lambda)=(\bm X'\bm X+\lambda\bm I)^{-1}\bm X'\bm y$
and $\mathbb E[\bm y]=\bm X\bm\theta_0=\bm X\bm\theta_0^{rl}$, we have
\[
\mathbb E[\hat{\bm\theta}_n^{r}(\lambda)]
= (\bm X'\bm X+\lambda\bm I)^{-1}\bm X'\bm X\bm\theta_0
=\bm Q(\lambda)\bm\theta_0,
\]
or
\[
\mathbb E[\hat{\bm\theta}_n^{r}(\lambda)]
= (\bm X'\bm X+\lambda\bm I)^{-1}\bm X'\bm X\bm\theta_0^{rl}
=\bm Q(\lambda)\bm\theta_0^{rl},
\]

\item From the closed form of $\hat{\bm\theta}_n^{r}(\lambda)$,
\[
\Var[\hat{\bm\theta}_n^{r}(\lambda)]
=(\bm X'\bm X+\lambda\bm I)^{-1}\bm X'\Var[\bm\varepsilon_0]\bm X
(\bm X'\bm X+\lambda\bm I)^{-1}.
\]

\item Under $\Var[\bm\varepsilon_0]=\sigma^2\bm I$,
\[
\Var[\hat{\bm\theta}_n^{r}(\lambda)]
=\sigma^2(\bm X'\bm X+\lambda\bm I)^{-1}\bm X'\bm X(\bm X'\bm X+\lambda\bm I)^{-1}.
\]
Let $\bm X=\bm U\bm S\bm V'$ with singular values $s_1\ge\cdots\ge s_r>0$
and write $\lambda_j=s_j^2/n$ (the positive eigenvalues of $(\bm X'\bm X)/n$).
Since $\Rank(\bm X)=r$, we have
$\bm X'\bm X/n=\sum_{j=1}^r\lambda_j\bm v_j\bm v_j'$,
where $\lambda_j=s_j^2/n$ for $j=1,\ldots,r$.
It follows $\Var[\hat{\bm\theta}_n^{rl}]=
\frac{\sigma^2}{n}\sum_{j=1}^r\frac{1}{\lambda_j}\bm v_j\bm v_j'$.
Instead, 
\begin{align*}
   \Var[\hat{\bm\theta}_n^r(\lambda)]=&
   \sigma^2\bm V(\bm S'\bm S+\lambda\bm I)^{-1}\bm S'\bm S
   (\bm S'\bm S+\lambda\bm I)^{-1}\bm V'\\
   =&\frac{\sigma^2}{n}\sum_{j=1}^r\frac{\lambda_j}{(\lambda_j+\lambda/n)^2}\bm v_j\bm v_j'.
\end{align*}
For each $j\in\{1,\ldots,r\}$: 
$$\frac{1}{\lambda_j}-\frac{\lambda_j}{(\lambda_j+\lambda/n)^2}
=\frac{(\lambda_j+\lambda/n)^2-\lambda_j^2}{\lambda_j(\lambda_j+\lambda/n)^2}
=\frac{2\lambda_j(\lambda/n)+(\lambda/n)^2}{\lambda_j(\lambda_j+\lambda/n)^2}>0.$$
Therefore the difference is positive semidefinite, and since the coefficients are strictly positive on the span of $\{\bm v_1,\ldots,\bm v_r\}=\Range(\bm X')$, it is positive definite on $\Range(\bm X')$.

\item Using the linearity of the $\Trace$ and the fact that $\bm V$ is orthogonal:
\begin{align*}
\Trace(\Var[\hat{\bm\theta}_n^r(\lambda)])
    =&\Trace\left(\frac{\sigma^2}{n}\sum_{j=1}^r\frac{\lambda_j}{(\lambda_j+\lambda/n)^2}\bm v_j\bm v_j'\right)\\
    =&\frac{\sigma^2}{n}\sum_{j=1}^r\frac{\lambda_j}{(\lambda_j+\lambda/n)^2}.
\end{align*}
Moreover, if $\Range(\bm X')=\Range(\bm\Sigma)$, then $$\Bias(\hat{\bm\theta}_n^r(\lambda),\bm\theta_0^{rl})=[\bm Q(\lambda)-\bm I]\bm\theta_0^{rl}.$$
The result then follows using Proposition \ref{prop:MSE bias-var decomp}.

\item Under $\Range(\bm X')=\Range(\bm\Sigma)$, we have $\E[\hat{\bm\theta}_n^{rl}]=\bm\theta_0^{rl}$ and $\Var[\hat{\bm\theta}_n^{rl}]=\sigma^2(\bm X'\bm X)^+$.
By definition,
\[
\MPR(\hat{\bm\theta}_n^r(\lambda),\bm\theta_0^{rl})
=\frac{1}{n}\E\lVert \bm X(\hat{\bm\theta}_n^r(\lambda)-\bm\theta_0^{rl})\rVert_2^2.
\]
Write
\[
\hat{\bm\theta}_n^r(\lambda)-\bm\theta_0^{rl}
=\bm Q(\lambda)\hat{\bm\theta}_n^{rl}-\bm\theta_0^{rl}
=[\bm Q(\lambda)-\bm I]\bm\theta_0^{rl}
+\bm Q(\lambda)\big(\hat{\bm\theta}_n^{rl}-\bm\theta_0^{rl}\big).
\]
Therefore,
\begin{align*}
\frac{1}{n}\E\lVert \bm X(\hat{\bm\theta}_n^r(\lambda)-\bm\theta_0^{rl})\rVert_2^2
&=\frac{1}{n}\E\lVert \bm X[\bm Q(\lambda)-\bm I]\bm\theta_0^{rl}
+\bm X\bm Q(\lambda)(\hat{\bm\theta}_n^{rl}-\bm\theta_0^{rl})\rVert_2^2\\
&=\frac{1}{n}\lVert \bm X[\bm Q(\lambda)-\bm I]\bm\theta_0^{rl}\rVert_2^2
+\frac{1}{n}\E\lVert \bm X\bm Q(\lambda)(\hat{\bm\theta}_n^{rl}-\bm\theta_0^{rl})\rVert_2^2\\
&\quad+\frac{2}{n}\E\Big\{ [\bm X[\bm Q(\lambda)-\bm I]\bm\theta_0^{rl}]' [\bm X\bm Q(\lambda)(\hat{\bm\theta}_n^{rl}-\bm\theta_0^{rl})]\Big\}.
\end{align*}
Since $\E[\hat{\bm\theta}_n^{rl}-\bm\theta_0^{rl}]=\bm 0$, the cross term vanishes. Hence
\[
\MPR(\hat{\bm\theta}_n^r(\lambda),\bm\theta_0^{rl})
=\frac{1}{n}\lVert \bm X[\bm Q(\lambda)-\bm I]\bm\theta_0^{rl}\rVert_2^2
+\frac{1}{n}\E\lVert \bm X\bm Q(\lambda)(\hat{\bm\theta}_n^{rl}-\bm\theta_0^{rl})\rVert_2^2.
\]
For the second term, use $\E\lVert \bm A\bm z\rVert_2^2=\Trace(\bm A\Var[\bm z]\bm A')$ with $\bm z=\hat{\bm\theta}_n^{rl}-\bm\theta_0^{rl}$, $\bm A=\bm X\bm Q(\lambda)$:
\[
\frac{1}{n}\E\lVert \bm X\bm Q(\lambda)(\hat{\bm\theta}_n^{rl}-\bm\theta_0^{rl})\rVert_2^2
=\frac{1}{n}\Trace\big(\bm X\bm Q(\lambda)\Var[\hat{\bm\theta}_n^{rl}]\bm Q(\lambda)'\bm X'\big).
\]
Since $\Var[\hat{\bm\theta}_n^{rl}]=\sigma^2(\bm X'\bm X)^+$, this equals
\[
\frac{\sigma^2}{n}\Trace\big(\bm X'\bm X\ \bm Q(\lambda)(\bm X'\bm X)^+\bm Q(\lambda)'\big).
\]
Use $\bm Q(\lambda)=(\bm X'\bm X+\lambda\bm I)^{-1}\bm X'\bm X$ and the SVD $\bm X=\bm U\bm S\bm V'$ with $\lambda_j=s_j^2/n>0$ for $j=1,\ldots,r$ to diagonalize:
\begin{align*}
\frac{\sigma^2}{n}&\Trace\big(\bm X'\bm X\ \bm Q(\lambda)(\bm X'\bm X)^+\bm Q(\lambda)'\big)\\
&=\frac{\sigma^2}{n}\Trace\Big\{\bm V\Big(\frac{\bm S'\bm S}{n}\Big)\Big[\Big(\frac{\bm S'\bm S}{n}+\frac{\lambda}{n}\bm I\Big)^{-1}\frac{\bm S'\bm S}{n}\Big]\Big(\frac{\bm S'\bm S}{n}\Big)^{+}\Big(\frac{\bm S'\bm S}{n}\Big)\Big(\frac{\bm S'\bm S}{n}+\frac{\lambda}{n}\bm I\Big)^{-1}\bm V'\Big\}\\
&=\frac{\sigma^2}{n}\sum_{j=1}^r\frac{\lambda_j^2}{(\lambda_j+\lambda/n)^2}.
\end{align*}
Combining the variance and bias contributions yields
\[
\MPR(\hat{\bm\theta}_n^r(\lambda),\bm\theta_0^{rl})
=\frac{\sigma^2}{n}\sum_{j=1}^{r}\frac{\lambda_j^2}{(\lambda_j+\lambda/n)^2}
+\frac{1}{n}\lVert \bm X[\bm Q(\lambda)-\bm I]\bm\theta_0^{rl}\rVert_2^2,
\]
as claimed.

    \item If $\Range(\bm X')=\Range(\bm\Sigma)$, then $r=r_0$ and $\Ker(\bm X)=\Ker(\bm\Sigma)$.
Moreover, $\bm Q(\lambda)=(\bm X'\bm X+\lambda\bm I)^{-1}\bm X'\bm X \to \bm\Proj_{\Range(\bm X')}$
as $\lambda\to 0$. Since $\bm\theta_0^{rl}\in\Range(\bm\Sigma)=\Range(\bm X')$, the bias term
$\|[\bm I-\bm Q(\lambda)]\bm\theta_0^{rl}\|_2^2 \to 0$. Using the spectral representation
$(\bm X'\bm X)/n=\sum_{j=1}^{r_0}\lambda_j\bm v_j\bm v_j'$ with $\lambda_j>0$, we also have
\[
\frac{\sigma^2}{n}\sum_{j=1}^{r_0}\frac{\lambda_j}{(\lambda_j+\lambda/n)^2}\;\xrightarrow[\lambda\to 0]{}\;
\frac{\sigma^2}{n}\sum_{j=1}^{r_0}\frac{1}{\lambda_j}.
\]
Hence
\[
\lim_{\lambda\to 0}\MSE(\hat{\bm\theta}_n^r(\lambda),\bm\theta_0^{rl})
=\frac{\sigma^2}{n}\sum_{j=1}^{r_0}\frac{1}{\lambda_j}
\le \frac{\sigma^2 r_0}{\lambda_{r_0}n},
\qquad
\lim_{\lambda\to 0}\MPR(\hat{\bm\theta}_n^r(\lambda),\bm\theta_0^{rl})
=\frac{r_0\sigma^2}{n}.
\]
\end{enumerate}
\end{proof}

The next proposition shows that there are penalty parameter values
for which the MSE of ridge is lower than the MSE of ridgeless.
\begin{proposition}
    Assume that the linear model $y=\bm x'\bm\theta_0+\varepsilon_0$ holds with 
  $\E[\bm x\varepsilon_0]=\bm 0$ and $\Var[\bm\varepsilon_0]=\sigma^2\bm I$ for $\sigma>0$.
  Then, for a fixed design matrix $\bm X$, there exists $\lambda^*>0$ such that 
  $$\MSE(\hat{\bm\theta}_n^r(\lambda^*),\bm\theta_0^{rl})<\MSE(\hat{\bm\theta}_n^{rl},\bm\theta_0^{rl}).$$
\end{proposition}
\begin{proof}
    See \citet{farebrother1976further}.
\end{proof}
\subsection{Finite sample properties of lasso}\label{sec:properties lasso}

In this section, we study the finite sample properties of the lasso estimator
under a fixed design matrix $\bm X$.
Since the lasso lacks a closed-form expression, its bias, variance, MSE, and MPR cannot be computed in closed-form as for least squares, ridge, or ridgeless estimators.
For this reason, rather than attempting explicit moment calculations, we derive non-asymptotic concentration inequalities for its estimation risk $\lVert\hat{\bm\theta}_n^l(\lambda)-\bm\theta_0\rVert_2^2$ and predictive risk $ \lVert\bm X(\hat{\bm\theta}_n^l(\lambda)-\bm\theta_0)\rVert_2^2/n$.

To proceed, we first show some auxiliary properties satisfied by any lasso solution.
Given a set $C$, let $|C|$ denote its cardinality, i.e., the number of elements in $C$. 
For an index set $S\subset\{1,\ldots,p\}$ with complement $S^c=\{1,\ldots,p\}\setminus S$, we write
$\bm v_S=[v_i]_{i\in S}\in\R^{|S|}$ for the subvector of $\bm v\in\R^p$ containing the entries indexed by $S$.
For a constant $\alpha\ge 1$, define the set
\begin{equation*}
    \mathcal C_\alpha(S):=\{\bm v\in\R^p\ :\ \norm{\bm v_{S^c}}_1\le\alpha\norm{\bm v_S}_1\},
\end{equation*}
i.e., the set of vectors in $\R^p$ whose coordinates outside $S$ have a total \(l_1-\)size at most \(\alpha\) times the \(l_1-\)size of the coordinates within $S$.
Finally, consider the following definition.
\begin{definition}[Support of a vector]
    The {\it support} of vector $\bm\theta\in\R^p$ is defined as $$\supp(\bm\theta):=\{j\in\{1,\ldots,p\}\ :\ \theta_j\neq 0\}.$$
\end{definition}
The next lemma shows that, for a suitable choice of the penalty parameter,
the lasso estimator satisfies two key inequalities and that its estimation error lies in $\mathcal C_\alpha(S)$ for some $\alpha\ge 1$ and an appropriate index set $S$.
\begin{lemma}[Auxiliary properties of lasso]\label{lemma:lasso properties}
    Suppose that the linear model (\ref{linear model}) holds, and let
    $s_0:=|\supp(\bm\theta_0)|$.    
    If $\lambda\ge 2\norm{\bm X'\bm\varepsilon_0/n}_\infty>0$, then
    any lasso solution $\hat{\bm\theta}_n^l(\lambda)$ satisfies: 
    \begin{enumerate}[label=(\roman*)]
        \item The predictive risk bound
    \begin{equation}\label{lem:lasso properties:predictive risk bound}
    \PR(\hat{\bm\theta}_n^l(\lambda),\bm\theta_0)\le
    12\lambda\norm{\bm\theta_0}_1.
    \end{equation} 
    \item The estimation error 
    $\hat{\bm\eta}:=\hat{\bm\theta}_n^l(\lambda)-\bm\theta_0$ belongs to $\mathcal C_3(\supp(\bm\theta_0))$ and satisfies
    \begin{equation}\label{lem:lasso properties:ineq 2}
        \lVert\bm X\hat{\bm\eta}\rVert_2^2/n\le 
        3\sqrt{s_0}\lambda\lVert\hat{\bm\eta}\rVert_2.
    \end{equation}
    \end{enumerate}
\end{lemma}
\begin{proof}
    Under the linear model (\ref{linear model}), we have for any $\bm\theta\in\R^p$:
$$\norm{\bm y-\bm X\bm\theta}_2^2=\bm y'\bm y+\bm\theta'\bm X'\bm X\bm\theta-2\bm\theta_0'\bm X'\bm X\bm\theta-2\bm\varepsilon_0'\bm X\bm\theta.$$
Since $\hat{\bm\theta}_n^l(\lambda)$ is a lasso solution,
\begin{align*}
    0\le\frac{1}{2n}\lVert\bm y-\bm X\hat{\bm\theta}_n^l(\lambda)\rVert_2^2+\lambda\lVert\hat{\bm\theta}_n^l(\lambda)\rVert_1\le
    \frac{1}{2n}\norm{\bm y-\bm X\bm\theta_0}_2^2+\lambda\norm{\bm\theta_0}_1,
\end{align*}
which holds if and only if 
    \begin{equation}\label{lem:lasso properties:basic inequality}
    \frac{1}{2n}\lVert\bm X\hat{\bm\eta}\rVert_2^2\le\bm\varepsilon_0'\bm X\hat{\bm\eta}/n+\lambda(\norm{\bm\theta_0}_1-\lVert\hat{\bm\theta}_n^l(\lambda)\rVert_1).
    \end{equation}
\begin{enumerate}[label=(\roman*)]
    \item By Hölder inequality, 
    \begin{equation*}\label{proof:properties lasso:holder}
    \bm\varepsilon_0'\bm X\hat{\bm\eta}/n\le\lvert\bm\varepsilon_0'\bm X\hat{\bm\eta}/n\rvert\le
    \norm{\bm X'\bm\varepsilon_0/n}_\infty\lVert\hat{\bm\eta}\rVert_1.
    \end{equation*}
Thus, using the choice $\lambda\ge 2\norm{\bm X'\bm\varepsilon_0/n}_\infty$ in (\ref{lem:lasso properties:basic inequality}) yields 
\begin{equation}\label{proof:lasso properties:basic inequality 2}
   0\le\frac{1}{2n}\lVert\bm X\hat{\bm\eta}\rVert_2^2\le\lambda/2\lVert\hat{\bm\eta}\rVert_1+\lambda(\norm{\bm\theta_0}_1-\lVert\hat{\bm\theta}_n^l(\lambda)\rVert_1).
\end{equation}
Using the triangle inequality 
\begin{equation}\label{proof:lasso properties:triangle eta}
\lVert\hat{\bm\eta}\rVert_1\le\lVert\hat{\bm\theta}_n^l(\lambda)\rVert_1+\lVert\bm\theta_0\rVert_1,
\end{equation}
we further obtain
$$0\le \lambda/2(\lVert\hat{\bm\theta}_n^l(\lambda)\rVert_1+\lVert\bm\theta_0\rVert_1)+\lambda(\lVert\bm\theta_0\rVert_1-\lVert\hat{\bm\theta}_n^l(\lambda)\rVert_1),$$
which, using that $\lambda>0$, implies $\lVert\hat{\bm\theta}_n^l(\lambda)\rVert_1\le3\lVert\bm\theta_0\rVert_1$.
Substituting this result into the triangle inequality (\ref{proof:lasso properties:triangle eta}) yields
\begin{equation}\label{proof:lasso:bound eta hat}
    \lVert\hat{\bm\eta}\rVert_1\le4 \lVert\bm\theta_0\rVert_1.
\end{equation}
Moreover, again by the triangle inequality,
$$\norm{\bm\theta_0}_1=\lVert(\bm\theta_0+\hat{\bm\eta})-\hat{\bm\eta}\rVert_1\le
\lVert\bm\theta_0+\hat{\bm\eta}\rVert_1+\lVert\hat{\bm\eta}\rVert_1,$$
which implies:
\begin{equation}\label{proof:lasso:reverse triangle inequality}
    \lVert\bm\theta_0+\hat{\bm\eta}\rVert_1\ge\norm{\bm\theta_0}_1-\lVert\hat{\bm\eta}\rVert_1.
\end{equation}
Using (\ref{proof:lasso:bound eta hat}) and (\ref{proof:lasso:reverse triangle inequality})
in the basic inequality (\ref{proof:lasso properties:basic inequality 2}),
we obtain
\begin{align*}
    \lVert\bm X\hat{\bm\eta}\rVert_2^2/n\le&
    \lambda\lVert\hat{\bm\eta}\rVert_1+2\lambda(\norm{\bm\theta_0}_1-\lVert\bm\theta_0+\hat{\bm\eta}\rVert_1)\\
    \le&3\lambda \lVert\hat{\bm\eta}\rVert_1\le12\lambda \lVert\bm\theta_0\rVert_1.
\end{align*}
\item Let $S_0:=\supp(\bm\theta_0)$. Using that $\bm\theta_{0S_0^c}=\bm 0$, we have
    \begin{equation*}
    \norm{\bm\theta_0}_1-\lVert\hat{\bm\theta}_n^l(\lambda)\rVert_1=
    \norm{\bm\theta_{0S_0}}_1-\lVert\bm\theta_{0S_0}+\hat{\bm\eta}_{S_0}\rVert_1-\lVert\hat{\bm\eta}_{S_0^c}\rVert_1.
    \end{equation*}
    Substituting this result into the basic inequality (\ref{proof:lasso properties:basic inequality 2}) yields:
    \begin{equation}\label{proof:properties lasso:inequality}
        0\le\lVert\bm X\hat{\bm\eta}\rVert_2^2/n\le
        \lambda\lVert\hat{\bm\eta}\rVert_1+2\lambda(\norm{\bm\theta_{0S_0}}_1-\lVert\bm\theta_{0S_0}+\hat{\bm\eta}_{S_0}\rVert_1-\lVert\hat{\bm\eta}_{S_0^c}\rVert_1).
    \end{equation}  
    By the triangle inequality,
    \begin{equation*}\label{proof:properties lasso:triangle}
    \norm{\bm\theta_{0S_0}}_1=\lVert\bm\theta_{0S_0}+\hat{\bm\eta}_{S_0}-\hat{\bm\eta}_{S_0}\rVert_1\le
    \lVert\bm\theta_{0S_0}+\hat{\bm\eta}_{S_0}\rVert_1+\lVert\hat{\bm\eta}_{S_0}\rVert_1.
    \end{equation*}
    Therefore, using the decomposition $\lVert\hat{\bm\eta}\rVert_1=\lVert\hat{\bm\eta}_{S_0}\rVert_1+\lVert\hat{\bm\eta}_{S_0^c}\rVert_1$, (\ref{proof:properties lasso:inequality}) reads
    \begin{align*}
        \nonumber0\le\lVert\bm X\hat{\bm\eta}\rVert_2^2/n\le& \lambda \lVert\hat{\bm\eta}\rVert_1+2\lambda(\lVert\hat{\bm\eta}_{S_0}\rVert_1-\lVert\hat{\bm\eta}_{S_0^c}\rVert_1)\\
        =& \lambda(3\lVert\hat{\bm\eta}_{S_0}\rVert_1-\lVert\hat{\bm\eta}_{S_0^c}\rVert_1),\label{proof:lem:lasso properties:final ineq}
    \end{align*}
    which implies that $\hat{\bm\eta}\in\mathcal C_3(\supp(\bm\theta_0))$.
    Finally, using the relation 
    between the $l_1-$ and the $l_2-$norm ($\norm{\bm v}_1\le\sqrt{s}\norm{\bm v}_2$ for every $\bm v\in\R^s$),
    and the notation $s_0:=|\supp(\bm\theta_0)|$,
    we conclude that
    \begin{align*}
        \lVert\bm X\hat{\bm\eta}\rVert_2^2/n\le&
    \lambda(3\lVert\hat{\bm\eta}_{S_0}\rVert_1-\lVert\hat{\bm\eta}_{S_0^c}\rVert_1)\le 3\lambda\lVert\hat{\bm\eta}_{S_0}\rVert_1\le 
    3\sqrt{s_0}\lambda\lVert\hat{\bm\eta}_{S_0}\rVert_2\\
    \le&
    3\sqrt{s_0}\lambda\lVert\hat{\bm\eta}\rVert_2.
    \end{align*}
\end{enumerate}
\end{proof}
We now derive the main finite-sample properties of lasso under a {\it restricted eigenvalue} condition on the design matrix.
This condition is tailored to the fact---established in Lemma \ref{lemma:lasso properties}---that 
for any penalty level $\lambda\ge 2\norm{\bm X'\bm\varepsilon_0/n}_\infty$, the
estimation error \(\hat{\bm\theta}_n^l(\lambda)-\bm\theta_0\) lies in the cone 
$\mathcal C_3(\supp(\bm\theta_0))$.
We therefore impose a lower eigenvalue bound only on vectors in this cone.
%
\begin{assumption}[Restricted eigenvalue condition]\label{ass:restricted eigenvalue condition}
    \index{Restricted eigenvalue condition}The design matrix $\bm X\in\R^{n\times p}$ 
    is such that for all $\bm\eta\in\mathcal C_3(\supp(\bm\theta_0))$ there exists $\kappa>0$ for which:
    $$\norm{\bm X\bm\eta}_2^2/n\ge\kappa\norm{\bm\eta}_2^2,$$
    where $\bm\theta_0\in\R^p$ is the coefficient of linear model (\ref{linear model}).
\end{assumption}
%
%
In the next proposition we derive finite-sample bounds on the squared $l_2$ estimation risk and predictive risk.
These bounds rely on the restricted eigenvalue condition, which controls the behavior of the design matrix on the cone where the estimation error is known to lie.
\begin{theorem}\label{prop:lasso estimation and predictive risk}
    Suppose that the linear model (\ref{linear model}) holds
    and that Assumption \ref{ass:restricted eigenvalue condition} is satisfied.
    Let $s_0:=|\supp(\bm\theta_0)|\le p$.
    Then, any lasso solution $\hat{\bm\theta}_n^l(\lambda)$
    with $\lambda\ge 2\norm{\bm X'\bm\varepsilon_0/n}_\infty>0$
    satisfies:
    \begin{enumerate}[label=(\roman*)]
        \item The estimation risk bound
        \begin{equation}\label{lasso:estimation risk bound rec}
            \lVert\hat{\bm\theta}_n^l(\lambda)-\bm\theta_0\rVert_2^2\le\frac{9}{\kappa^2}s_0\lambda^2.
        \end{equation}
        \item The predictive risk bound
        \begin{equation}\label{lasso:predictive risk bound rec}
            \PR(\hat{\bm\theta}_n^l(\lambda),\bm\theta_0)\le\frac{9}{\kappa}s_0\lambda^2.
        \end{equation}
    \end{enumerate}
\end{theorem}
\begin{proof}
In Lemma \ref{lemma:lasso properties} we obtained Inequality \eqref{lem:lasso properties:ineq 2}, namely
\[
\lVert \bm X \hat{\bm\eta} \rVert_2^2 / n
\le
3 \sqrt{s_0} \lambda \lVert \hat{\bm\eta} \rVert_2,
\]
where $\hat{\bm\eta} := \hat{\bm\theta}_n^l(\lambda) - \bm\theta_0 \in \mathcal C_3(\supp(\bm\theta_0))$.

\begin{enumerate}[label=(\roman*)]

\item
Applying Assumption \ref{ass:restricted eigenvalue condition} to the left-hand side of the inequality above yields
\[
\kappa \lVert \hat{\bm\eta} \rVert_2^2
\le
3 \sqrt{s_0} \lambda \lVert \hat{\bm\eta} \rVert_2.
\]
If $\lVert \hat{\bm\eta} \rVert_2 > 0$, dividing both sides by $\lVert \hat{\bm\eta} \rVert_2$ gives
\[
\lVert \hat{\bm\eta} \rVert_2
\le
3 \sqrt{s_0} \lambda / \kappa,
\]
and squaring produces the stated bound.
If instead $\lVert \hat{\bm\eta} \rVert_2 = 0$, the inequality holds trivially.

\item
Applying Assumption \ref{ass:restricted eigenvalue condition} to the factor $\lVert \hat{\bm\eta} \rVert_2$ on the right-hand side of \eqref{lem:lasso properties:ineq 2} yields
\[
\lVert \bm X \hat{\bm\eta} \rVert_2^2 / n
\le
3 \sqrt{s_0} \lambda
\lVert \bm X \hat{\bm\eta} \rVert_2 / \sqrt{n \kappa}.
\]
If $\lVert \bm X \hat{\bm\eta} \rVert_2 > 0$, dividing both sides by $\lVert \bm X \hat{\bm\eta} \rVert_2 / \sqrt{n}$ gives
\[
\lVert \bm X \hat{\bm\eta} \rVert_2 / \sqrt{n}
\le
3 \sqrt{s_0} \lambda / \sqrt{\kappa},
\]
and squaring yields the stated predictive risk bound.
If $\lVert \bm X \hat{\bm\eta} \rVert_2 = 0$, the claim again holds trivially.

\end{enumerate}
\end{proof}
The sparsity parameter $s_0=|\supp(\bm\theta_0)|$
plays a central role in the bounds of Theorem 
\ref{prop:lasso estimation and predictive risk}.
We say that $\bm\theta_0$ is hard sparse if it contains zero entries. More formally:
\begin{definition}[Hard sparsity]\label{def:strong sparsity}
Coefficient $\bm\theta_0\in\R^p$ is {\it hard sparse}\index{Hard sparsity} if $s_0:=|\supp(\bm\theta_0)|<p$.
\end{definition}
In high-dimensional settings, 
hard sparsity is typically imposed as an identifying restriction for $\bm\theta_0$.
Consider an asymptotic regime in which
$$\lim_{n\to\infty}p/n= K>0,\quad\text{and}\quad 
\lim_{p\to\infty}s_0/p= s_{\infty}\in(0,1].$$
Under this scaling, $s_0\to\infty$ as $n\to\infty$.
Thus, the lasso converges to $\bm\theta_0$
only if the restricted eigenvalue $\kappa$ and the penalty level $\lambda$ compensate for the 
divergence of $s_0$, i.e.,  
$\lim_{n\to\infty}s_0\lambda^2/\kappa^2=0$.
Notice however that $2\norm{\bm X'\bm\varepsilon_0/n}_\infty$,
the lower bound for $\lambda$ in Theorem 
\ref{prop:lasso estimation and predictive risk},
is monotone non-decreasing as additional columns are added to $\bm X$.
Moreover, intuition suggests that Assumption \ref{ass:restricted eigenvalue condition} with a large $\kappa$ becomes increasingly more restrictive as $p\to\infty$. 

\begin{figure}[t]
\centering
\includegraphics[scale=1]{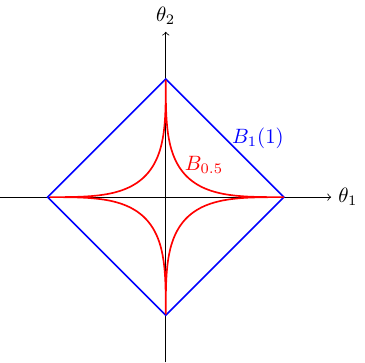}
\caption{Unit $q$-balls $B_q(1)=\{\theta\in\mathbb{R}^2: \|\theta\|_q^q\le 1\}$ for $q=1$ (blue diamond) and $q=0.5$ (red curve).  }
\label{fig:qballs}
\end{figure}

\begin{remark}
The results in Lemma \ref{lemma:lasso properties} and Theorem \ref{prop:lasso estimation and predictive risk} can be extended under a milder assumption known as weak sparsity, which formalizes the idea that $\bm\theta_0$ may be well approximated by a hard sparse vector.
\begin{definition}[Weak sparsity]\label{def:weak sparsity}
    Coefficient $\bm\theta_0\in\R^p$ is {\it weak sparse}\index{Weak sparsity} if $\bm\theta_0\in B_q(r)$ where, for some $q\in[0,1]$ and radius $r>0$,
    $$B_q(r):=\{\bm\theta\in\R^p\ :\ \norm{\bm\theta}_q^q\le r\}.$$
\end{definition}
\noindent Setting $q=0$ in Definition \ref{def:weak sparsity} recovers hard sparsity with $s=r$.
Figure \ref{fig:qballs} represents the borders of set $B_q(r)$ for $r=1$ and $q=0.5$ and $q=1$.
For $q\in(0,1]$, weak sparsity restricts the decays of the ordered coefficients:\footnote{For $q\in(0,1]$, the map $t\mapsto t^q$ is concave, so small coefficients are penalized less than large ones. As a consequence, the quantity $\sum_{j=1}^p |\theta_{0j}|^q$ is dominated by the largest entries. For $\bm\theta_0$ to satisfy $\sum_j |\theta_{0j}|^q \le r$, the ordered coefficients $|\theta_{0(1)}|\ge|\theta_{0(2)}|\ge\cdots$ must therefore decay sufficiently quickly.}
$$\max_{j=1,\ldots,p}|\theta_{0j}|=\theta_{0(1)}\ge\theta_{0(2)}\ge\ldots,\ge\theta_{0(p-1)}\ge\theta_{0(p)}=\min_{j=1,\ldots,p}|\theta_{0j}|.$$
More precisely, if the coefficients satisfy the bound $|\theta_{0j}|\le Cj^{-\alpha}$ for some $C\ge 0$ and $\alpha>0$,
then $\bm\theta_0\in B_q(r)$ for a radius $r$ depending on $C$ and $\alpha$.
\end{remark}

\subsubsection{Restricted eigenvalue condition}

Inequality (\ref{lem:lasso properties:ineq 2}) of Lemma \ref{lemma:lasso properties} provides an upper bound on the predictive risk of the lasso estimator in terms of its estimation error.  
Conversely, for an appropriate choice of the penalty parameter, the restricted eigenvalue condition in Assumption \ref{ass:restricted eigenvalue condition} yields an upper bound on the estimation error in terms of predictive risk.  
Combining these two ingredients leads to the estimation and predictive risk bounds in Theorem \ref{prop:lasso estimation and predictive risk}.

To provide intuition for why the restricted eigenvalue condition is needed, consider the constrained form of the lasso estimator
\[
\hat{\bm\theta}_n^{lc}(R)
:= \argmin_{\bm\theta\in\mathbb{R}^p}
\Bigl\{ L_n(\bm\theta) : \|\bm\theta\|_1 \le R \Bigr\},
\qquad
R := \|\bm\theta_0\|_1,
\]
where 
\[
L_n(\bm\theta)
:= \frac{1}{2n} \| \bm y - \bm X \bm\theta \|_2^2
\]
is the least squares loss.  
With the choice $R=\|\bm\theta_0\|_1$, the true parameter is feasible, and we have  
$L_n(\hat{\bm\theta}_n^{lc}(R)) \le L_n(\bm\theta_0)$.  
Under mild conditions, the loss difference $L_n(\bm\theta_0)-L_n(\hat{\bm\theta}_n^{lc}(R))$ decreases with $n$.  
A natural question is when this decrease in the loss translates into a decrease in the estimation error  
$\|\hat{\bm\eta}\|_2^2$, where $\hat{\bm\eta}:=\hat{\bm\theta}_n^{lc}(R)-\bm\theta_0$.

Since $L_n$ is quadratic with Hessian $\nabla_{\bm\theta}^2 L_n = \bm X'\bm X / n$, the estimation error will shrink when the loss has sufficient curvature in the directions of $\hat{\bm\eta}$.  
In a low-dimensional setting ($p \le n$), a simple sufficient condition ensuring sufficient curvature is that all eigenvalues of $\bm X'\bm X/n$ are uniformly bounded below by a positive constant, that is, for all $\bm\eta\in\R^p$:
\[
\|\bm X \bm\eta\|_2^2/n
\ge
\kappa \|\bm\eta\|_2^2.
\]
This global eigenvalue condition is stronger than what is required for the restricted eigenvalue assumption, but it guarantees the desired curvature of the loss function in every direction.

In the high-dimensional regime ($p>n$), $\bm X'\bm X/n$ is singular and this uniform curvature condition cannot hold globally.  
However, Lemma \ref{lemma:lasso properties} shows that the lasso estimation error lies in the cone \(\mathcal C_3(\supp(\bm\theta_0))\).
The restricted eigenvalue condition requires curvature only along these directions.
Note that this condition does not require $\bm X_{S_0}$, the matrix formed by the columns of \(\bm X\) with index in \(S_0\), to have full column rank, nor does it imply $|S_0| \le \operatorname{Rank}(\bm X)$.  
It simply demands that the design has sufficient curvature within the cone of directions in which the lasso error can lie.  
Under this restricted curvature condition, a small loss difference still implies a small estimation error, even when $p \gg n$.

Verifying that a given deterministic design satisfies the restricted eigenvalue condition is challenging.  
Developing probabilistic conditions ensuring that random designs satisfy it with high probability is an active area of research.


\subsubsection{Slow rates and fast rates}

Consider the case in which the error term in the linear model \eqref{linear model} is sub-Gaussian with mean zero and variance proxy $\sigma^2$. Under this assumption, one can select a tuning parameter $\lambda$ that depends only on the unknown $\sigma^2$
and guarantees that the estimation and prediction risks of the lasso estimator are upper bounded with high probability.

\begin{figure}[t]
\centering
\includegraphics[scale=1]{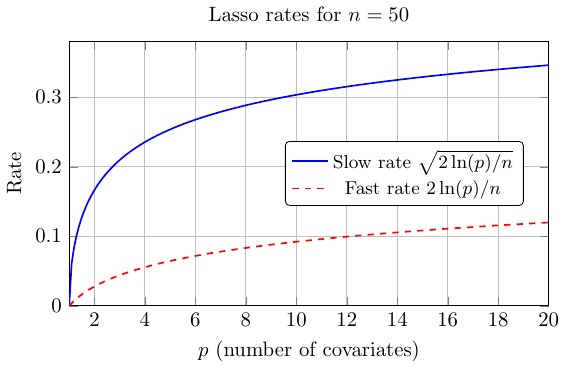}

\caption{Comparison of slow and fast lasso rates for $n=50$ and $p \in [1,20]$.}
\label{fig:slow_fast_rates_small}
\end{figure}

\begin{theorem}
    Suppose that the linear model (\ref{linear model}) holds
    and that $\bm\varepsilon_0$ is a vector of independent 
    random variables
    $\varepsilon_{0i}\sim\subG(\sigma^2)$ with variance proxy $\sigma^2>0$.
    Further suppose that the columns of $\bm X$ are standardized so that 
    $\max_{j=1,\ldots,p}\norm{\bm X_j}_2/\sqrt{n}\le C$
    for some constant $C>0$.
    Then, for all $\delta>0$, 
    any lasso solution $\hat{\bm\theta}_n^l(\lambda)$
    with regularization parameter 
    \begin{equation}\label{thm:lasso:lambda slow}
        \lambda=2C\sigma\left(\sqrt{2\ln(p)/n}+\delta\right)
    \end{equation}
    satisfies with probability at least $1-2e^{-n\delta^2/2}$:
    \begin{equation}\label{thm:lasso:predictive risk slow bound}
    \PR(\hat{\bm\theta}_n^l(\lambda),\bm\theta_0)\le
    24C\norm{\bm\theta_0}_1\sigma(\sqrt{2\ln(p)/n}+\delta).
    \end{equation} 
    Further suppose that Assumption \ref{ass:restricted eigenvalue condition} holds and let $s_0:=|\supp(\bm\theta_0)|\le p$.
    Then, with probability at least $1-2e^{-n\delta^2/2}$:
    \begin{enumerate}[label=(\roman*)]
        \item The estimation risk bound
        \begin{equation}\label{lasso:estimation risk bound rec fast}
            \lVert\hat{\bm\theta}_n^l(\lambda)-\bm\theta_0\rVert_2^2\le\frac{72C^2\sigma^2s_0}{\kappa^2}
            (2\ln(p)/n+\delta^2).
        \end{equation}
        \item The predictive risk bound
        \begin{equation}\label{lasso:predictive risk bound rec fast}
            \PR(\hat{\bm\theta}_n^l(\lambda),\bm\theta_0)\le\frac{72C^2\sigma^2s_0}{\kappa}(2\ln(p)/n+\delta^2).
        \end{equation}
    \end{enumerate}
\end{theorem}
\begin{proof}
    For any $t \ge 0$, by the union bound:
\begin{align*}
    \P\left[\norm{\bm X'\bm\varepsilon_0/n}_\infty \ge t\right]
    &= \P\left[\max_{j=1,\ldots,p}\lvert\bm X_j'\bm\varepsilon_0/n\rvert \ge t\right] \\
    &= \P\left[\bigcup_{j=1}^p \left\{\,\lvert\bm X_j'\bm\varepsilon_0/n\rvert \ge t\,\right\}\right] \\
    &\le \sum_{j=1}^p \P\left[\lvert\bm X_j'\bm\varepsilon_0/n\rvert \ge t\right],
\end{align*}
where the second equality follows from the fact that the maximum of a set of
numbers exceeds $t$ if and only if at least one element in the set exceeds $t$.
    Since $\varepsilon_{01},\ldots,\varepsilon_{on}$ are independent
    random variables with $\subG(\sigma^2)$ distribution, 
    from Proposition \ref{app:prop:sum indep sub gauss}
    we have that for any $t\in\R$:
    \begin{align*}
    \sum_{j=1}^p\P\left[\lvert\bm X_j'\bm\varepsilon_0/n\rvert\ge t\right]&\le
    \sum_{j=1}^p2\exp\left(-\frac{t^2}{2\sigma^2\norm{\bm X_j/n}_2^2}\right)\\
    &\le2p\exp\left(-\frac{t^2n}{2\sigma^2C^2}\right),
    \end{align*}
    where the last inequality follows from the standardization of the columns of $\bm X$, namely
\[
\left\|\bm X_j/n\right\|_2^2
= \|\bm X_j\|_2^2/n^2
\le C^2 n/n^2
= C^2/n.
\]
    Substituting $t=C\sigma\left(\sqrt{2\ln(p)/n}+\delta\right)$ we get
    \begin{align*}
        2p\exp\left(-\frac{t^2n}{2\sigma^2C^2}\right)&=
        2\exp(\ln(p))\exp\left(-n\delta^2/2-\ln(p)-\delta \sqrt{2n\ln(p)}\right)\\
        &=2\exp\left(-n\delta^2/2\right)\exp\left(-\delta\sqrt{2n\ln(p)}\right)\\
        &\le 2\exp\left(-n\delta^2/2\right),
    \end{align*}
    since $-\delta\sqrt{2n\ln(p)}<0$.
    We conclude that, for all $\delta>0$:
    $$\P[2\norm{\bm X'\bm\varepsilon_0/n}_\infty\le 
    2C\sigma(\sqrt{2\ln(p)/n}+\delta)]\ge 1- 2e^{-n\delta^2/2}.$$
    Consequently, if we set $\lambda=2C\sigma(\sqrt{2\ln(p)/n}+\delta)$,
    we obtain from (\ref{lem:lasso properties:predictive risk bound}) of Lemma \ref{lemma:lasso properties}
    that (\ref{thm:lasso:predictive risk slow bound}) holds with probability at least $1- 2e^{-n\delta^2/2}$.
    Moreover, under Assumption \ref{ass:restricted eigenvalue condition}, 
    we obtain from (\ref{lasso:estimation risk bound rec}) and (\ref{lasso:predictive risk bound rec}) of Theorem \ref{prop:lasso estimation and predictive risk}
    that (\ref{lasso:estimation risk bound rec fast}) and (\ref{lasso:predictive risk bound rec fast}) hold
    with probability at least $1- 2e^{-n\delta^2/2}$,
    by using the inequality:\footnote{
      This inequality follows from $2ab\le a^2+b^2$ for any two real numbers $a$ and $b$.
    }
    $$2\ln(p)/n+\delta^2+2\sqrt{2\ln(p)/n}\delta\le 2(2\ln(p)/n+\delta^2).$$
\end{proof}
As long as $n \ge 2\ln(p)$, the ratio $2\ln(p)/n$ can be much smaller than $\sqrt{2\ln(p)/n}$.  
This distinction motivates the terminology commonly used in the high-dimensional statistics literature.  
The bound \eqref{thm:lasso:predictive risk slow bound}, which scales on the order of
\[
\sqrt{2\ln(p)/n},
\]
is referred to as the {\it slow rate} for the predictive risk of the lasso, whereas  
the bound \eqref{lasso:predictive risk bound rec fast}, which scales on the order of
\[
2\ln(p)/n,
\]
is called the {\it fast rate}.
These rates are depicted in Figure \ref{fig:slow_fast_rates_small}.

The slow rate arises from ensuring that the tuning parameter satisfies  
$\lambda \ge 2\|\bm X'\bm\varepsilon_0/n\|_\infty$, a condition that can be enforced with high probability under sub-Gaussian errors and column normalization of the design.  
This argument does not rely on any structural assumptions on~$\bm X$ beyond these probabilistic and scaling considerations, and therefore the resulting bound holds without the restricted eigenvalue condition.  
The slow rate thus provides a general upper bound on the predictive risk that depends $\|\bm\theta_0\|_1$, $n$ and $p$.

By contrast, the fast rate requires additional geometric structure on the design, as formalized by the restricted eigenvalue condition.  
This assumption guarantees sufficient curvature of the least squares loss along the cone in which the lasso estimation error lies.  
Under this condition, the lasso achieves sharper bounds of order $s_0 \ln(p)/n$ up to multiplicative constants.  
In cases where $s_0$ is much smaller than $p$, this represents a substantial improvement over the slow rate.

\bibliographystyle{plainnat}
\bibliography{articles}

\begin{thebibliography}{23}
\providecommand{\natexlab}[1]{#1}
\providecommand{\url}[1]{\texttt{#1}}
\expandafter\ifx\csname urlstyle\endcsname\relax
  \providecommand{\doi}[1]{doi: #1}\else
  \providecommand{\doi}{doi: \begingroup \urlstyle{rm}\Url}\fi

\bibitem[Albert(1972)]{albert1972regression}
Arthur Albert.
\newblock Regression and the moore-penrose pseudoinverse.
\newblock 1972.

\bibitem[Argyriou et~al.(2012)Argyriou, Foygel, and Srebro]{argyriou2012sparse}
Andreas Argyriou, Rina Foygel, and Nathan Srebro.
\newblock Sparse prediction with the $ k $-support norm.
\newblock \emph{Advances in Neural Information Processing Systems}, 25, 2012.

\bibitem[Axler(2024)]{axler2024linear}
Sheldon Axler.
\newblock \emph{Linear algebra done right}.
\newblock Springer Nature, 2024.

\bibitem[Bauschke et~al.(2017)Bauschke, Combettes, Bauschke, and Combettes]{bauschke2017correction}
Heinz~H Bauschke, Patrick~L Combettes, Heinz~H Bauschke, and Patrick~L Combettes.
\newblock \emph{Correction to: convex analysis and monotone operator theory in Hilbert spaces}.
\newblock Springer, 2017.

\bibitem[Billingsley(2017)]{billingsley2017probability}
Patrick Billingsley.
\newblock \emph{Probability and measure}.
\newblock John Wiley \& Sons, 2017.

\bibitem[B{\"u}hlmann and Van De~Geer(2011)]{buhlmann2011statistics}
Peter B{\"u}hlmann and Sara Van De~Geer.
\newblock \emph{Statistics for high-dimensional data: methods, theory and applications}.
\newblock Springer Science \& Business Media, 2011.

\bibitem[Efron et~al.(2004)Efron, Hastie, Johnstone, and Tibshirani]{efron2004least}
Bradley Efron, Trevor Hastie, Iain Johnstone, and Robert Tibshirani.
\newblock Least angle regression.
\newblock \emph{The Annals of Statistics}, 32\penalty0 (2):\penalty0 407--499, 2004.

\bibitem[Farebrother(1976)]{farebrother1976further}
Richard~William Farebrother.
\newblock Further results on the mean square error of ridge regression.
\newblock \emph{Journal of the Royal Statistical Society. Series B (Methodological)}, pages 248--250, 1976.

\bibitem[Friedman et~al.(2007)Friedman, Hastie, H{\"o}fling, and Tibshirani]{friedman2007pathwise}
Jerome Friedman, Trevor Hastie, Holger H{\"o}fling, and Robert Tibshirani.
\newblock Pathwise coordinate optimization.
\newblock \emph{The annals of applied statistics}, 1\penalty0 (2):\penalty0 302--332, 2007.

\bibitem[Gauss(1809)]{gauss1809theoria}
Carl~F Gauss.
\newblock Theoria motus corporum coelestium in sectionibus conicis solem ambientium.
\newblock \emph{sumtibus Frid. Perthes et I. H. Besser}, 1809.

\bibitem[Hastie et~al.(2009)Hastie, Tibshirani, Friedman, and Friedman]{hastie2009elements}
Trevor Hastie, Robert Tibshirani, Jerome~H Friedman, and Jerome~H Friedman.
\newblock \emph{The elements of statistical learning: data mining, inference, and prediction}, volume~2.
\newblock Springer, 2009.

\bibitem[Hastie et~al.(2015)Hastie, Tibshirani, and Wainwright]{hastie2015statistical}
Trevor Hastie, Robert Tibshirani, and Martin Wainwright.
\newblock Statistical learning with sparsity.
\newblock \emph{Monographs on statistics and applied probability}, 143\penalty0 (143):\penalty0 8, 2015.

\bibitem[Hoerl and Kennard(1970)]{hoerl1970ridge}
Arthur~E Hoerl and Robert~W Kennard.
\newblock Ridge regression: Biased estimation for nonorthogonal problems.
\newblock \emph{Technometrics}, 12\penalty0 (1):\penalty0 55--67, 1970.

\bibitem[Horn and Johnson(2012)]{horn2012matrix}
Roger~A Horn and Charles~R Johnson.
\newblock \emph{Matrix analysis}.
\newblock Cambridge university press, 2012.

\bibitem[Legendre(1805)]{legendre1805nouvelles}
Adrien-Marie Legendre.
\newblock Nouvelles méthodes pour la détermination des orbites des comètes.
\newblock \emph{F. Didot}, 1805.

\bibitem[Moore(1920)]{moore1920reciprocal}
Eliakim~H Moore.
\newblock On the reciprocal of the general algebraic matrix.
\newblock \emph{Bulletin of the american mathematical society}, 26:\penalty0 294--295, 1920.

\bibitem[Nerlove et~al.(1961)]{nerlove1961returns}
Marc Nerlove et~al.
\newblock \emph{Returns to scale in electricity supply}.
\newblock Institute for mathematical studies in the social sciences, 1961.

\bibitem[Penrose(1955)]{penrose1955generalized}
Roger Penrose.
\newblock A generalized inverse for matrices.
\newblock In \emph{Mathematical proceedings of the Cambridge philosophical society}, volume~51, pages 406--413. Cambridge University Press, 1955.

\bibitem[Tibshirani(1996)]{tibshirani1996regression}
Robert Tibshirani.
\newblock Regression shrinkage and selection via the lasso.
\newblock \emph{Journal of the Royal Statistical Society Series B: Statistical Methodology}, 58\penalty0 (1):\penalty0 267--288, 1996.

\bibitem[Tibshirani(2013)]{tibshirani2013lasso}
Ryan~J Tibshirani.
\newblock The lasso problem and uniqueness.
\newblock \emph{The Electronic Journal of Statistics}, 7:\penalty0 1456--1490, 2013.

\bibitem[Tseng(2001)]{tseng2001convergence}
Paul Tseng.
\newblock Convergence of a block coordinate descent method for nondifferentiable minimization.
\newblock \emph{Journal of optimization theory and applications}, 109:\penalty0 475--494, 2001.

\bibitem[Vershynin(2018)]{vershynin2018high}
Roman Vershynin.
\newblock \emph{High-dimensional probability: An introduction with applications in data science}, volume~47.
\newblock Cambridge university press, 2018.

\bibitem[Wainwright(2019)]{wainwright2019high}
Martin~J Wainwright.
\newblock \emph{High-dimensional statistics: A non-asymptotic viewpoint}, volume~48.
\newblock Cambridge university press, 2019.

\end{thebibliography}

\chapter{Appendix}

\section{Linear algebra}\label{app:sec:linear algebra}

This section introduces a selection of definitions and results 
from linear algebra that are used in these lecture notes.
A book-length exposition of linear algebra can be found in \citet{axler2024linear}, among others.

\subsection{Vector space}
We introduce useful definitions and results for real vector spaces\index{Vector space}.
\begin{definition}[Vector space]
    A (real) {\it vector space} is a set $V$ along with an addition on $V$
    and a scalar multiplication on $V$ with the following properties:
    \begin{enumerate}
        \item (commutativity) $u+v=v+u$ for all $u,v\in V$;
        \item (associativity) $(u+v)+w=u+(v+w)$ and $(ab)v=a(bv)$ for all $u,v,w\in V$ and all $a,b\in\R$;
        \item (additive identity) there exists an element $0\in V$ such that $v+0=v$ for all $v\in V$;
        \item (additive inverse) for every $v\in V$, there exists $w\in V$
        such that $v+w=0$;
        \item (multiplicative identity) $1v=v$ for all $v\in V$;
        \item (distributive properties) $a(u+v)=au+av$ and $(a+b)v=av+bv$ for all $a,b\in\R$ and all $u,v\in V$.
    \end{enumerate}
\end{definition}
\begin{definition}[Subspace]
    A subset $U$ of a vector space $V$ is a {\it subspace}\index{Subspace}
    of $V$ if $U$ is a vector space,
    (using the same addition and scalar multiplication as on $V$).
\end{definition}

\begin{proposition}
    A subset $U$ of a vector space $V$ is a subspace of $V$
    if and only if it satisfies these conditions:
    \begin{enumerate}[label=(\roman*)]
        \item (additive identity) $0\in U$;
        \item (closed under addition) $u,v\in U$ implies $u+v\in U$;
        \item (closed under scalar multiplication) $a\in\R$ and $u\in U$
        implies $au\in U$.
    \end{enumerate}
\end{proposition}

\begin{definition}[Linear combination]
    A {\it linear combination}\index{Linear combination} of 
    vectors $v_1,\ldots,v_n$ in vector space $V$
    with coefficients $a_1,\ldots,a_n\in\R$ is:
    $$a_1v_1+\ldots+a_nv_n.$$
\end{definition}

\begin{definition}[Span]
    The {\it span}\index{Span} of vectors $v_1,\ldots,v_n$ in vector space $V$ 
    is defined as
    $$\Span(v_1,\ldots,v_n):=\{a_1v_1+\ldots+a_nv_n\ :\ 
    a_1,\ldots,a_n\in\R\}.$$
\end{definition}

\begin{definition}[Linear independence]
    The vectors $v_1,\ldots,v_n$ in vector space $V$
    are {\it linearly independent}\index{Linear independence} if 
    $$\{a_1,\ldots,a_n\in\R\ :\ a_1v_1+\ldots+a_nv_n=0\}=
    \{a_1=\ldots=a_n=0\}.$$
\end{definition}

\begin{definition}[Linear dependence]
    The vectors $v_1,\ldots,v_n$ in vector space $V$
    are {\it linearly dependent}\index{Linear dependence} if they are not linearly independent.
\end{definition}

\begin{definition}[Basis]
    A {\it basis}\index{Basis} of a vector space $V$ is a set of vectors in $V$
    that are linearly independent and span $V$.
\end{definition}

\subsection{Inner products and norms}

\begin{definition}[Inner product]
    An {\it inner product}\index{Inner product} on a vector space $V$ is a function
    that takes each ordered pair $(v,u)$ of elements of $V$
    to a number $\langle v,u\rangle\in\R$ and satisfies:
    \begin{enumerate}
        \item (positivity) $\langle v,v\rangle\ge 0$ for all $v\in V$;
        \item (definiteness) $\langle v,v\rangle= 0$ if and only if $v=0$;
        \item (additivity in first slot) 
        $\langle v+u,w\rangle=\langle v,w\rangle+\langle u,w\rangle$ for all $v,u,w\in V$;
        \item (homogeneity in first slot) 
        $\langle av,u\rangle=a\langle v,u\rangle$ for all $a\in\R$ and all $v,u\in V$;
        \item (conjugate symmetry) $\langle v,u\rangle=\langle u,v\rangle$ for all $v,u\in V$.
    \end{enumerate}
\end{definition}
\begin{proposition}[Basic properties of an inner product]
    An inner product $\langle\cdot,\cdot\rangle$ on vector space $V$
    satisfies:
    \begin{enumerate}
        \item $\langle0,v\rangle=\langle v,0\rangle$
        for every $v\in V$.
        \item $\langle v,u+w\rangle=\langle v,u\rangle+\langle v,w\rangle$
        for every $v,u,w\in V$.
        \item $\langle v,au\rangle=a\langle v,u\rangle$ for all
        $a\in\R$ and all $v,u\in V$.
    \end{enumerate}
\end{proposition}
\begin{definition}[Orthogonal vectors]
    Two vectors $v$ and $u$ in vector space $V$ are {\it orthogonal}\index{Orthogonal vectors}
    if $\langle v,u\rangle=0$.
\end{definition}
\begin{definition}[Orthogonal subspace]
    $U$ and $W$ are {\it orthogonal subspaces}\index{Orthogonal subspaces} of vector space $V$ if
    $\langle u,w\rangle=0$ for all $u\in U$ and all $w\in W$.
\end{definition}
\begin{definition}[Orthonormal basis]
    The set of vectors $\{v_1,\ldots,v_n\}$ in vector space $V$ 
    is an {\it orthonormal basis}\index{Orthonormal basis}
    of $V$ if it is a basis of $V$ such that $\langle v_i,v_j\rangle=0$
    and $\norm{v_i}=1$ for all $i,j=1,\ldots,n$ with $i\neq j$.
\end{definition}
\begin{definition}[Norms]
    Given inner product $\langle\cdot,\cdot\rangle$ on vector space $V$,
    the {\it norm}\index{Norm} of $v\in V$ is defined by $\norm{v}:=\sqrt{\langle v, v\rangle}$. 
\end{definition}
\begin{proposition}[Properties of norms]
    For $v$ in vector space $V$:
    \begin{enumerate}
        \item $\norm{v}=0$ if and only if $v=0$.
        \item $\norm{av}=|a|\norm{v}$ for all $a\in\R$.
    \end{enumerate}
\end{proposition}

\begin{definition}[Normed vector space]\index{Normed vector space}\label{def:normed_space}
    A \emph{normed vector space} is a pair $(V,\norm{\cdot})$ where $V$ is a real vector space and $\norm{\cdot}:V\to[0,+\infty)$ is a norm on $V$, that is, a mapping satisfying the properties in the previous proposition together with the triangle inequality
    \[
        \norm{v+u} \le \norm{v} + \norm{u}
        \qquad\text{for all }v,u\in V.
    \]
\end{definition}

\begin{example}[Canonical examples of normed vector spaces]
\label{ex:normed_space_examples}
\begin{enumerate}[label=(\roman*)]

\item The Euclidean space\index{Euclidean space} $\R^d$ with the Euclidean norm
\[
\norm{h}_2 := \Bigg( \sum_{j=1}^d h_j^2 \Bigg)^{1/2},\qquad
h\in\R^d,
\]
is a normed vector space.

\item Let $(\mathcal{X},\mathcal A,\P)$ be a \emph{probability space}\index{Probability space} and let
$L^2(\mathcal{X})$ denote the collection of (equivalence classes of)
square-integrable real-valued functions $f:\mathcal{X}\to\R$ with
$\int_{\mathcal{X}} f(x)^2d\P(x) < \infty$.
Then $L^2(\mathcal{X})$ is a normed vector space under the norm
\[
\norm{f}_{L^2}
:=
\Bigg( \int_{\mathcal{X}} f(x)^2d\P(x) \Bigg)^{1/2}.
\]

\end{enumerate}
\end{example}

\begin{definition}[Linear function]
    $L:V\to W$ from a vector space $V$ to 
    another vector space $W$ is a {\it linear function}\index{Linear function} if:
    \begin{enumerate}[label=(\roman*)]
        \item $L(v+u)=L(v)+L(u)$ for all $v,u\in V$;
        \item $L(av)=aL(v)$ for all $a\in\R$ and $v\in V$.
    \end{enumerate}
\end{definition}
\begin{theorem}[Cauchy–Schwarz inequality]\index{Cauchy–Schwarz inequality}
    Suppose $v$ and $u$ are two vectors in vector space $V$.
    Then, $$|\langle v,u\rangle|\le\norm{v}\norm{u}.$$
    This inequality is an equality if and only if there is $a\in\R$
    such that $v=au$.
\end{theorem}
\begin{theorem}[Triangle inequality]\index{Triangle inequality}
    Suppose $v$ and $u$ are two vectors in vector space $V$.
    Then, $$\norm{v+u}\le\norm{v}+\norm{u}.$$
    This inequality is an equality if and only if there is $a\ge 0$
    such that $v=au$.
\end{theorem}

\begin{theorem}[Parallelogram equality]
    Suppose $v$ and $u$ are two vectors in vector space $V$.
    Then, $$\norm{v+u}^2+\norm{v-u}^2=2(\norm{v}^2+\norm{u}^2).$$
\end{theorem}

\subsection{The Euclidean space}

\begin{definition}[(Real) $n-$tuple]\index{n-tuple}
    A {\it (real) $n-$tuple} is an ordered list of $n$ real numbers.
\end{definition}

With a slight abuse of terminology, we sometimes use the term {\it vector} to mean a (real) $n-$tuple.

\begin{definition}[Real Euclidean space]
    The {\it real Euclidean space}\index{Euclidean space} of dimension $n$,
    denoted $\R^n$, is the set of all $n-$tuples.
\end{definition}

Elements of a real Euclidean space are written in bold. 
For example, $ a\in\R^n$, which means
$ a=(a_1,\ldots,a_n)$ with $a_1,\ldots,a_n\in\R$.


\begin{definition}[Euclidean inner product]
    The {\it Euclidean inner product}\index{Euclidean inner product} of $ v, u\in\R^n$ 
    is defined as $\langle v, u\rangle_e:=\sum_{i=1}^nv_iu_i$.
\end{definition}

\begin{definition}[$l_p-$norm]
    The {\it $l_p-$norm}\index{l$_p-$norm} $\norm{\cdot}_p$ on $\R^n$
    is defined for all $ v\in\R^n$ as
    $\norm{ v}_p:=\left(\sum_{i=1}^{n}|v_i|^p\right)^{1/p}$ when $p\in[1,+\infty)$,
    and $\norm{ v}_p:=\max_{i=1}^{n}\vert v_i\vert$ when $p=+\infty$. 
\end{definition}

\begin{proposition}[Equivalence of norms in $\R^d$]\index{Equivalence of norms}
\label{prop:equivalence_norms}
All norms on $\R^d$ are equivalent.

In particular, for any $1 \le p, q \le \infty$, there exist constants
$0 < C_{p,q} \le D_{p,q} < \infty$
such that, for all $v \in \R^d$,
\[
C_{p,q} \|v\|_q \le \|v\|_p \le D_{p,q} \|v\|_q.
\]
Equivalently, $\|v_n\|_p \to 0$ if and only if $\|v_n\|_q \to 0$.
\end{proposition}

\begin{remark}
The equivalence of norms implies that the choice of norm does not affect
notions of convergence, continuity, or differentiability on $\R^d$.
\end{remark}

\begin{theorem}[Hölder inequality]\index{Hölder inequality}
    Let $p,q\in[1,+\infty]$ satisfy $\tfrac{1}{p}+\tfrac{1}{q}=1$, and
    let $ v, u\in\R^n$.
    Then
    \[
        |\langle  v, u\rangle|
        \le\norm{ v}_p\norm{ u}_q.
    \]
\end{theorem}

\subsection{Matrices}

\begin{definition}[Matrix]\index{Matrix}
    An $n\times p$ {\it matrix} is a collection of $p$ 
    $n-$tuples. 
\end{definition}

The collection of all $n\times p$ matrices is denoted $\R^{n\times p}$.
For a matrix $ A\in\R^{n\times p}$, we write
$ A=[ A_1,\ldots, A_p]$ where 
$ A_1,\ldots, A_p\in\R^n$ are $p$ $n-$tuples.
Written more explicitly,
$$ A=\begin{bmatrix}
    A_{1,1}& \ldots& A_{1,p}\\
    \vdots& \ddots& \vdots\\
    A_{n,1}& \ldots& A_{n,p}
\end{bmatrix},$$
that is, the elements of $ A$, the $n-$tuples, 
    are organized in columns.
    We denote:
\begin{itemize}
    \item the $i,j-$th element of $ A$ by $A_{i,j}$;
    \item the $j-$th column $ A$ by $ A_j$;
    \item the $i-$th row $ A$ by $ A_{(i)}$.
\end{itemize}
Notice that a matrix in $\R^{n\times p}$ can be equivalently 
seen as a collection of $n$ $p-$tuples, where the $p-$tuples
represent the rows of the matrix.

\begin{definition}[Column and row vector]
    A {\it $n-$column vector}\index{Column vector} is a $n-$tuple seen as a 
    matrix in $\R^{n\times 1}$. A 
    {\it $n-$row vector}\index{Row vector} is a $n-$tuple seen as a 
    matrix in $\R^{1\times n}$.
\end{definition}
Throughout these lecture notes, we denote $n-$tuples as column vectors,
and use the simple notation $ v\in\R^n$ instead of $ v\in\R^{n\times 1}$.

\begin{definition}[Matrix addition]\index{Matrix addition}
    The sum of two matrices 
    of the same size 
    is the matrix obtained by adding
    corresponding entries in the matrices.
    That is, for $ A, B\in\R^{n\times p}$, 
    we define $ A+ B= C$ where $ C\in\R^{n\times p}$
    and $C_{i,j}=A_{i,j}+B_{i,j}$ for $i=1,\ldots,n$ and
    $j=1,\ldots,p$.
\end{definition}

\begin{definition}[Matrix-scalar multiplication]\index{Matrix-scalar multiplication}
    The product of a scalar and a matrix is the matrix obtained by multiplying
each entry in the matrix by the scalar
    That is, for $ A\in\R^{n\times p}$ and $a\in\R$, 
    we define $a A= B$ where $ B\in\R^{n\times p}$
    and $B_{i,j}=aA_{i,j}$ for $i=1,\ldots,n$ and
    $j=1,\ldots,p$.
\end{definition}

\begin{definition}[Matrix multiplication]\index{Matrix multiplication}
    Given $ A\in\R^{n\times p}$ and $ B\in\R^{p\times m}$,
    the product $ A B= C$ where $ C\in\R^{n\times m}$
    and $C_{i,j}=\sum_{r=1}^pA_{i,r}B_{r,j}$ for $i=1,\ldots,n$ and
    $j=1,\ldots,m$.
\end{definition}

Note that we define the product of
two matrices only when the number of
columns of the first matrix equals the
number of rows of the second matrix.

\begin{definition}[Transpose of a matrix]\index{Matrix transpose}
    The {\it transpose} of a matrix 
    $ A\in\R^{n\times p}$ is the matrix $ B\in\R^{p\times n}$ 
    with $j,i-$entry given by $B_{j,i}=A_{i,j}$ for $i=1,\ldots,n$
    and $j=1,\ldots,p$.
    We denote it by $ A'$.
\end{definition}

It follows that the Euclidean inner product between
$ v, u\in\R^n$ is 
$$\langle v, u\rangle_e= v' u.$$

\begin{definition}[Range of a matrix]\index{Matrix Range}
    The {\it range} of a matrix $ A\in\R^{n\times p}$
    is defined as
    $$\Range( A):=\{ u\in\R^n\ :\  u= A v\ 
    \text{for some}\  v\in\R^p\}.$$
\end{definition}
The range of a matrix is also called the {\it column space}, i.e.,
the space spanned by the matrix's columns, since:
\begin{proposition}
    Let $ A=[ A_1,\ldots, A_p]\in\R^{n\times p}$.
    Then, $\Range( A)=\Span( A_1,\ldots, A_p).$
\end{proposition}

\begin{definition}[Kernel of a matrix]\index{Matrix kernel}
    The {\it kernel}, or {\it null space}, 
    of a matrix $ A\in\R^{n\times p}$ is defined as
    $$\Ker( A):=\{ v\in\R^p\ :\  A v= 0\}.$$
\end{definition}

\begin{proposition}
    Let $ A\in\R^{n\times p}$.
    Then, $\Range( A)$ and $\Ker( A')$ are orthogonal subspaces
    of $\R^n$ such that $\R^n=\Range( A)+\Ker( A')$.
\end{proposition}

\begin{definition}[Rank of a matrix]\index{Matrix rank}
    The {\it rank} of a matrix $ A\in\R^{n\times p}$,
    denoted $\Rank( A)$, is
    the maximum number of linearly independent columns of $ A$.
\end{definition}

\begin{proposition}
    Let $ A\in\R^{n\times p}$. Then, $\Rank( A)\le\min\{n,p\}$.
\end{proposition}

\begin{definition}[Eigenvalue]\index{Eigenvalue}
    Let $ A\in\R^{n\times n}$. A scalar $\lambda\in\R$ is an {\it eigenvalue} of $ A$
    if there exists $ v\in\R^n$ with $ v\neq 0$ such that \( A v=\lambda v\).
\end{definition}
\begin{definition}[Eigenvector]\index{Eigenvector}
    Given $ A\in\R^{n\times n}$ and an eigenvalue $\lambda$ of $ A$,
    a vector $ v\in\R^n$ with $ v\neq 0$ satisfying $ A v=\lambda v$
    is an {\it eigenvector} corresponding to $\lambda$.
\end{definition}
\begin{proposition}
    Let $ A\in\R^{n\times n}$. Then, $ A$ has at most
    $\Rank( A)$ distinct nonzero eigenvalues.
\end{proposition}

\begin{proposition}
    Suppose $\lambda_1,\ldots,\lambda_r\in\R$ are distinct eigenvalues
    of $ A\in\R^{n\times n}$ and $ v_1,\ldots, v_r\in\R^n$ 
    are corresponding eigenvectors. Then, $ v_1,\ldots, v_r$
    are linearly independent.
\end{proposition}

\begin{definition}[Singular values]\index{Singular values}
    The {\it singular values} of $ A\in\R^{n\times p}$
    are the nonnegative square roots of the eigenvalues of $ A' A$.
\end{definition}

\begin{definition}[Symmetric matrix]\index{Symmetric matrix}
    A square matrix $ A\in\R^{n\times n}$ is {\it symmetric}
    if $ A'= A$.
\end{definition}

\begin{definition}[Positive definite matrix]\index{Positive definite matrix}
    A square symmetric matrix $ A\in\R^{n\times n}$
    is {\it positive definite} if $ v' A v>0$
    for all $ v\in\R^n$ such that $ v\neq 0$.
\end{definition}

\begin{definition}[Positive semi-definite matrix]\index{Positive semi-definite matrix}
    A square symmetric matrix $ A\in\R^{n\times n}$
    is {\it positive semi-definite} if $ v' A v\ge0$
    for all $ v\in\R^n$.
\end{definition}

\begin{proposition}
    A square symmetric matrix $ A\in\R^{n\times n}$
    is positive definite (positive semi-definite) if and only if 
    all of its eigenvalues are positive (nonnegative).
\end{proposition}

\begin{definition}[Identity matrix]\index{Identity matrix}
    The {\it identity} matrix on $\R^n$ is defined as
    $$ I:=\begin{bmatrix}
        1&&0\\&\ddots&\\0&&1
    \end{bmatrix}\in\R^{n\times n}.$$
\end{definition}

\begin{definition}[Diagonal of a matrix]\index{Matrix diagonal}
    The {\it diagonal} of a square matrix $ A\in\R^{n\times n}$
    indicates the elements "on the diagonal": $A_{1,1},\ldots,A_{n,n}$.
\end{definition}

\begin{definition}[Diagonal matrix]\index{Diagonal matrix}
    A square matrix $ A\in\R^{n\times n}$ is a {\it diagonal}
    matrix if all its elements outside of the diagonal are zero.
    We can write $ A=\diag(A_{1,1},\ldots,A_{n,n})$.
\end{definition}

\begin{definition}[Invertible matrix, matrix inverse]\index{Invertible matrix}
    A square matrix $ A\in\R^{n\times n}$ is {\it invertible} if 
    there is a matrix $ B\in\R^{n\times n}$ such that 
    $ A B= B A= I$. We call $ B$ the {\it inverse}
    of $ A$ and denote it by $ A^{-1}$.
\end{definition}

\begin{proposition}
    A square matrix $ A\in\R^{n\times n}$ is invertible
    if and only if $\Rank( A)=n$, or equivalently,
    if and only if $\Ker( A)=\{0\}$.
\end{proposition}
\begin{proposition}
    If a square symmetric matrix $ A\in\R^{n\times n}$ is positive definite,
    then it is invertible.
\end{proposition}
\begin{definition}[Orthogonal matrix]\index{Orthogonal matrix}
    A square matrix $ P\in\R^{p\times p}$ is {\it orthogonal},
    or orthonormal, if $ P' P= P P'= I$.
\end{definition}
\begin{definition}[Projection matrix]\index{Projection matrix}
    A square matrix $ P\in\R^{p\times p}$ is a 
    {\it projection} matrix if
    $ P= P^2$.
\end{definition}
\begin{definition}[Orthogonal projection matrix]\index{Orthogonal projection}
    A square matrix $ P\in\R^{p\times p}$ is an 
    {\it orthogonal projection} matrix if it is a projection
    matrix and 
    $ P= P'$.
\end{definition}
Projections and orthogonal projections have the following properties.
\begin{proposition}
    For any projection matrix $ P\in\R^{p\times p}$ and vector
    $ b\in\R^p$, we have
    $$ b= P b+( I- P) b.$$
    If $ P$ is an orthogonal projection matrix, then
    $$( P b)'( I- P) b=0.$$
\end{proposition}

\begin{definition}[Trace]\index{Trace}
    The {\it trace} of a square matrix $ A\in\R^{n\times n}$,
    denoted $\Trace( A)$, is the sum of its diagonal elements:
    $$\Trace( A)=A_{11}+\ldots+A_{n,n}.$$
\end{definition}

\begin{proposition}
    The $\Trace$ is a linear function.
\end{proposition}

\begin{proposition}[Properties of the trace]
    \begin{enumerate}
        \item $\Trace( A)=\lambda_1+\ldots+\lambda_n$
        for all $ A\in\R^{n\times n}$ with
        $\lambda_1,\ldots,\lambda_n$ denoting the 
        (not necessarily distinct) eigenvalues of $ A$.
        \item $\Trace( A)=\Trace( A')$ for all $ A\in\R^{n\times n}$.
        \item $\Trace( A B)=\Trace( B A)$ for all for all $ A, B\in\R^{n\times n}$.
        \item $\Trace( A' B)=\Trace( A B')=\Trace( B' A)
        =\Trace( B A')$ for all $ A, B\in\R^{n\times p}$.
    \end{enumerate}
\end{proposition}

\subsection{Moore-Penrose inverse}

The Moore-Penrose inverse, or matrix pseudoinverse, 
is a generalization of the inverse of a matrix that was independently introduced by \citet{moore1920reciprocal} and \citet{penrose1955generalized}.
\begin{definition}[Moore-Penrose inverse]\index{Moore-Penrose inverse}
   The matrix $ A^+\in\R^{p\times n}$
   is a {\it Moore-Penrose inverse} of $ A\in\R^{n\times p}$ if
   \begin{enumerate}[label=(\roman*)]
       \item $ A A^+ A= A$;
       \item $ A^+ A A^+= A^+$;
       \item $( A A^+)'= A A^+$;
       \item $( A^+ A)'= A^+ A$.
   \end{enumerate}
\end{definition}
\subsubsection*{Properties and examples of the Moore-Penrose inverse}
We now list the main properties of the Moore-Penrose inverse.
\begin{proposition}
    For any matrix $ A\in\R^{n\times p}$, the 
    Moore-Penrose inverse $ A^+$ exists and is unique.
\end{proposition}
\begin{proposition}
    Let $ A\in\R^{n\times p}$ have $\Rank( A)=p$. Then,
    $ A^+=( A' A)^{-1} A'$.
\end{proposition}
\begin{proposition}
    Let the square matrix $ A\in\R^{p\times p}$ have $\Rank( A)=p$. Then, $ A^+= A^{-1}$.
\end{proposition}
\begin{proposition}
    Let $ A\in\R^{n\times p}$. Then,
    $$ A^+=\lim_{\lambda\to 0}( A' A+\lambda I)^{-1} A'
    = \lim_{\lambda\to 0} A'( A A'+\lambda I)^{-1}.$$
\end{proposition}
\begin{proof}
    See \citet{albert1972regression}.
\end{proof}
\begin{proposition}
    Let $ A\in\R^{n\times p}$. Then:
    \begin{enumerate}
        \item $ A=( A^+)^+$.
        \item $ A^+=( A' A)^+ A'= A'( A A')^+$.
        \item $( A')^+=( A^+)'$.
        \item $( A' A)^+= A^+( A')^+$.
        \item $( A A')^+=( A')^+ A^+$.
        \item $\Range( A^+)=\Range( A')=\Range( A^+ A)=\Range( A' A)$.
        \item $\Ker( A^+)=\Ker( A A^+)=
        \Ker(( A A')^+)=\Ker( A A')=\Ker( A')$.
    \end{enumerate}
\end{proposition}
\begin{proposition}
    For any matrix $ A\in\R^{n\times p}$:
    \begin{enumerate}
        \item $ A A^+$ is an orthogonal projection onto $\Range( A)$.
        \item $ I- A A^+$ is an orthogonal projection onto $\Ker( A')$.
        \item $ A^+ A$ is an orthogonal projection onto $\Range( A')$.
        \item $ I- A^+ A$ is an orthogonal projection onto $\Ker( A)$.
    \end{enumerate}
\end{proposition}
We also collect some examples of the Moore-Penrose inverse.
\begin{example}
    If $a\in\R$, then $a^+=\begin{cases}
        a^{-1}&a\neq 0\\0&a=0
    \end{cases}$.
\end{example}
\begin{example}
    If $ A=\diag(A_1,\ldots,A_{p-k},0,\ldots,0)\in\R^{p\times p}$, 
    then 
    $$ A^+=\diag(1/A_1,\ldots,1/A_{p-k},0,\ldots,0).$$
\end{example}
\begin{example}
    If $ A=\begin{bmatrix}
        1\\2
    \end{bmatrix}$, then $ A^+=[1/5, 2/5]$.
\end{example}
\begin{example}
    If $ A=\begin{bmatrix}
        1&0\\0&0
    \end{bmatrix}$, then $ A^+=\begin{bmatrix}
        1&0\\0&0
    \end{bmatrix}$.
\end{example}
\begin{example}
    If $ A=\begin{bmatrix}
        1&1\\1&1
    \end{bmatrix}$, then $ A^+=\begin{bmatrix}
        1/4&1/4\\1/4&1/4
    \end{bmatrix}$.
\end{example}

\subsubsection*{Systems of linear equations and least squares}
The Moore-Penrose inverse plays a central role
in the study of solutions to systems of linear equations\index{System of linear equations} 
and linear least squares problems.
\begin{theorem}[Solutions of systems of linear equations]\label{app:thm:solutions system lin eq}
    For $ A\in\R^{n\times p}$ and $ b\in\R^n$, let
    $L:=\{\theta\in\R^p\ :\  A\theta= b\}$. 
    The following statements hold:
    \begin{enumerate}[label=(\roman*)]
        \item If $ b\notin\Range( A)$, then $L$ is empty.
        \item If $ b\in\Range( A)$, then
        $L= A^+ b+\Ker( A)$.
    \end{enumerate}
\end{theorem}
\begin{corollary}
    Given a square matrix $ A\in\R^{p\times p}$ and $ b\in\R^p$, let
    $L:=\{\theta\in\R^p\ :\  A\theta= b\}$.
    Then, $ A^+ b$ is the unique element of $L$ if and only if $\Rank( A)=p$.
    In this case, $ A^+= A^{-1}$.
\end{corollary}
\begin{corollary}
    For $ X\in\R^{n\times p}$ and $ y\in\R^n$:
    $$\argmin_{\theta\in\R^p}\norm{ y- X\theta}_2^2=
     X^+ y+\Ker( X).$$
\end{corollary}

\subsection{Eigenvalue and Singular value decomposition}

This section introduces the eigenvalue and the singular value decompositions,
which are matrix factorizations with many applications to statistics and machine learning.
\begin{definition}[Singular value decomposition]\label{app:def:svd}
    The {\it Singular Value Decomposition}\index{Singular value decomposition} (SVD) of a matrix $ A\in\R^{n\times p}$ with rank $r:=\Rank( A)$ is given by 
    $$ A= U S V'=\sum_{i=1}^rs_i u_i v_i',$$ 
    where
    $ U\in\R^{n\times n}$ and $ V\in\R^{p\times p}$ are 
    orthogonal matrices, and 
    $$ S=\begin{bmatrix}
        \diag(s_1,\ldots,s_r)& 0\\  0& 0
    \end{bmatrix}\in\R^{n\times p},$$
    where $s_1,\ldots,s_r$ are the positive {\it singular values} of $ A$.
\end{definition}
\begin{proposition}[Existence]
    Any matrix $ A\in\R^{n\times p}$ admits a singular value decomposition.
\end{proposition}
The next proposition demonstrates the relation of the SVD to the 
four fundamental subspaces of a matrix.
\begin{proposition}
    Consider Definition \ref{app:def:svd}. Then,
    \begin{enumerate}[label=(\roman*)]
        \item $\{ u_1,\ldots, u_r\}$ is an orthonormal basis of $\Range( A)$.
        \item $\{ u_{r+1},\ldots, u_{n}\}$ is an orthonormal basis of $\Ker( A')$.
        \item $\{ v_1,\ldots, v_r\}$ is an orthonormal basis of $\Range( A')$.
        \item $\{ v_{r+1},\ldots, v_{p}\}$ is an orthonormal basis of $\Ker( A)$.
    \end{enumerate}
\end{proposition}
We thus have that the orthogonal projection onto $\Range( A)$ is
\[
    \sum_{j=1}^r  u_j  u_j',
\]
and the orthogonal projection onto $\Range( A')$ is
\[
    \sum_{j=1}^r  v_j  v_j'.
\]
Moreover, if $r<n$, then $\Ker( A')$ has orthonormal basis 
$\{ u_{r+1},\ldots, u_n\}$, and the orthogonal projection onto $\Ker( A')$ is
\[
    \sum_{j=r+1}^n  u_j  u_j'.
\]
Similarly, if $r<p$, then $\Ker( A)$ has orthonormal basis
$\{ v_{r+1},\ldots, v_p\}$, and the orthogonal projection onto $\Ker( A)$ is
\[
    \sum_{j=r+1}^p  v_j  v_j'.
\]

\begin{proposition}
    The Moore-Penrose inverse of a matrix $ A\in\R^{n\times p}$ admitting SVD decomposition
    $ A= U S V'$ is given by
    $ A^+= V S^+ U'$.
\end{proposition}

\begin{definition}[Eigenvalue decomposition]
    The {\it eigenvalue decomposition}\index{Eigenvalue decomposition} of a square matrix 
    $ A\in\R^{n\times n}$ with
    $n$ linearly independent eigenvectors $ Q_1,\ldots, Q_n$
    corresponding to eigenvalues $\lambda_1,\ldots,\lambda_n$
    is given by
    $$ A= Q\Lambda Q^{-1},$$
    where $ Q=[ Q_1,\ldots, Q_n]$ and
    $\Lambda=\diag(\lambda_1,\ldots,\lambda_n)$.
\end{definition}

\begin{proposition}[Relation between the singular value and the eigenvalue decompositions]
    Given a matrix $ A\in\R^{n\times p}$ with SVD 
    $ A= U S V'$:
    \begin{enumerate}
        \item $ A' A= V S' S V'$;
        \item $ A A'= U S S' U'$.
    \end{enumerate}
\end{proposition}


\section{Convex analysis}\label{app sec:convex analysis}
This section introduces a selection of definitions and results from convex analysis that are used in these lecture notes.
A book-length exposition of convex analysis can be found in \citet{bauschke2017correction}, among others.

\subsection{Convex sets and convex functions}

\begin{definition}[Closed set]
    A set $C\subset\R^p$ is {\it closed}\index{Closed set} if it contains all of its limit points.
\end{definition}
\begin{definition}[Bounded set]
   A set $C\subset\R^p$ is \emph{bounded} if there exists $r>0$ and a point
$\theta_0\in\R^p$ such that
\[
\|\theta - \theta_0\|_2 < r
\qquad \text{for all } \theta \in C.
\]
\end{definition}
\begin{definition}[Convex set]
    A set $C\subset\R^p$ is {\it convex}\index{Convex set} if for all $0<\alpha<1$ and all $\theta,\beta\in C$:
    $$\alpha \theta+(1-\alpha)\beta\in C.$$
    In particular, $\R^p$ and $\emptyset$ are convex.
\end{definition}
\begin{definition}[Epigraph of a function]
    The {\it epigraph}\index{Epigraph} of function $f:\R^p\to(-\infty,+\infty]$ is
    $$\epi(f):=\{(\theta,\xi)\in\R^p\times\R\ :\ f(\theta)\le\xi\}.$$
\end{definition}
\begin{definition}[Domain of a function]
    The {\it domain}\index{Domain} of function $f:\R^p\to(-\infty,+\infty]$ is 
    $$\dom(f):=\{\theta\in\R^p\ :\ f(\theta)<+\infty\}.$$
\end{definition}
\begin{definition}[Lower level set of a function]\label{app:def:lower level set}
    The {\it lower level set}\index{Lower level set} of function $f:\R^p\to(-\infty,+\infty]$ at height $\xi\in\R$ is 
    $$\text{lev}_{\le\xi}(f):=\{\theta\in\R^p\ :\ f(\theta)\le\xi\}.$$
\end{definition}
\begin{definition}[Proper function]
    function $f:\R^p\to(-\infty,+\infty]$ is {\it proper}\index{Proper function} if 
    $\dom(f)\neq\emptyset$.
\end{definition}
\begin{definition}[Convex function]
    Let $f:\R^p\to(-\infty,+\infty]$ be a proper function. Then $f$ is {\it convex}\index{Convex function} if its epigraph
    $\epi(f)$ is convex.
    Equivalently, $f$ is convex if for all $0< \alpha< 1$ and
    all $\theta,\beta\in\R^p$:
    $$f(\alpha\theta + (1-\alpha)\beta)\le \alpha f(\theta)+ (1-\alpha)f(\beta).$$
\end{definition}
\begin{definition}[Strictly convex function]
    Let $f:\R^p\to(-\infty,+\infty]$ be a proper function. Then $f$ is {\it strictly convex}\index{Strictly convex function} if for all $0< \alpha< 1$ and
    all $\theta,\beta\in\R^p$ with $\theta\neq\beta$:
    $$f(\alpha\theta + (1-\alpha)\beta)< \alpha f(\theta)+ (1-\alpha)f(\beta).$$
\end{definition}
\begin{proposition}
If $f:\R^p\to(-\infty,+\infty]$ is proper, convex, and 
$\theta$ is in the interior of $\dom(f)$, then $f$ is continuous at $\theta$.
\end{proposition}
\begin{definition}[Limit inferior]
    The {\it limit inferior}\index{Limit inferior} of $f:\R^p\to(-\infty,+\infty]$ at a point $\theta^*\in\R^p$ is
    $$\underset{\theta\to\theta^*}{\lim\inf} f(\theta)=\lim_{\varepsilon\to 0}\left(
    \inf\{f(\theta)\ :\ \theta\neq\theta^*, \norm{\theta-\theta^*}\le\varepsilon\}\right).$$
\end{definition}
\begin{definition}[Lower semicontinuous function]
    Function $f:\R^p\to(-\infty,+\infty]$ is {\it lower semicontinuous}\index{Lower semicontinuous function} at $\theta^*\in\R^p$
    if
    $$\underset{\theta\to\theta^*}{\lim\inf} f(\theta)\ge f(\theta^*).$$
\end{definition}
\begin{definition}[Coercive function]
    Function $f:\R^p\to(-\infty,+\infty]$ is {\it coercive}\index{Coercive function} if
    $$\lim_{\norm{\theta}
\to+\infty}f(\theta)=+\infty.$$
\end{definition}

\subsection{Subdifferential}

\begin{definition}[Subdifferential]
    Let $f:\R^p\to(-\infty,+\infty]$ be a proper function. The {\it subdifferential}\index{Subdifferential} of $f$
    is the set-valued operator:\footnote{Given a set $C$, the set of all subsets of $C$, including the empty set and $C$ itself, is denoted $2^C$.
    This set is called the power set of $C$.}
    $$\partial f:\R^p\to 2^{\R^p};\theta\mapsto\left\{
    \beta\in\R^p\ :\ \langle v-\theta,\beta\rangle+f(\theta)\le f( v)\ \forall  v\in\R^p\right\}.$$
    Let $\theta\in\R^p$. Then $f$ is {\it subdifferentiable} at $\theta$ if $\partial f(\theta)\neq\emptyset$;
    the elements of $\partial f(\theta)$ are the {\it subgradients}\index{Subgradient} of $f$ at $\theta$.
\end{definition}
\begin{remark}[Graphical representation of subgradient]
Graphically, a vector $\beta\in\R^p$ is a subgradient of a proper
function $f:\R^p\to(-\infty,+\infty]$ at $\theta\in\dom(f)$ if 
$$f_{\beta,\theta}: v\mapsto\langle v-\theta,\beta\rangle+f(\theta),$$
which coincides with $f$ at $\theta$, lies below $f$.
\end{remark}
\begin{theorem}[Supporting hyperplane theorem]\index{Supporting hyperplane theorem}
\label{thm:supporting_hyperplane}
Let $f:\R^p\to(-\infty,+\infty]$ be proper, convex, and lower semicontinuous.
Then, for every $\theta \in \operatorname{int}(\dom(f))$, the subdifferential is nonempty:
\[
\partial f(\theta) \neq \emptyset.
\]
\end{theorem}
\begin{example}\label{subdifferetnial absolute value}
    The subdifferential of the absolute value function $|\cdot|$ at $\theta\in\R$ is given by
$$\partial |\theta|=\begin{cases}
    \{1\},&\theta>0\\
    [-1,1],&\theta=0\\
    \{-1\},&\theta<0
\end{cases}.$$
See \citet[Example 16.15]{bauschke2017correction}.
\end{example}

\begin{proposition}[Relation between gradient and subgradient]\index{Subgradient}
\label{prop:subgradient_gradient}
Let $f:\R^p \to (-\infty,+\infty]$ be a proper convex function.

\begin{enumerate}[label=(\roman*)]
\item If $f$ is differentiable at $\theta\in\R^p$, then the subdifferential
is the singleton
\[
\partial f(\theta) = \{ \nabla f(\theta) \}.
\]

\item Equivalently, a vector $\beta\in\R^p$ is a subgradient of $f$ at $\theta$
if and only if
\[
f(v) \ge f(\theta) + \langle v - \theta, \nabla f(\theta) \rangle
\qquad \text{for all } v\in\R^p.
\]
\end{enumerate}
\end{proposition}

\begin{remark}
Thus, the subdifferential generalizes the classical gradient to the
nondifferentiable setting.  At points where $f$ is smooth,
subgradients coincide with the unique gradient.  
At nondifferentiable points (e.g.\ $\theta=0$ for $|\theta|$),
the subdifferential becomes a set of valid linear under-estimators.
\end{remark}

\subsection{Minimizers of convex optimization problems}
\begin{definition}[Global minimizer]
    A point $\theta^*\in\R^p$ is a {\it (global) minimizer}\index{Global minimizer} of a proper function $f:\R^p\to(-\infty,+\infty]$
    over $C\subset\R^p$
    if $f(\theta^*)=\inf_{\theta\in C}f(\theta)$.
    The set of minimizers of $f$ over $C$ is
\[
\argmin_{\theta\in C} f(\theta)
:= \{\theta\in C : f(\theta)=\inf_{\eta\in C} f(\eta)\}.
\]
\end{definition}

\begin{definition}[Local minimizer]\index{Local minimizer}
A point $\theta^* \in \R^p$ is a \emph{local minimizer} of $f$ if 
there exists $r>0$ such that
\[
\theta^* \in 
\operatorname*{arg\,min}_{\theta : \|\theta - \theta^*\| \le r} 
f(\theta).
\]
\end{definition}

\begin{proposition}[Convex problems: local minimizers are global minimizers]
    Let $f:\R^p\to(-\infty,+\infty]$ be proper and convex. Then every local minimizer of $f$ is a minimizer.
\end{proposition}

\begin{proposition}[Convex problems: $\argmin$ is convex]\label{app:prop:argmin convex}
    Let $f:\R^p\to(-\infty,+\infty]$ be proper and convex and $C\subset\R^p$. Then $\argmin_{\theta\in C}f(\theta)$ is convex.
\end{proposition}

\begin{proposition}[Existence of minimizers]\label{app:prop:existence minimizers}
    Let $f:\R^p\to(-\infty,+\infty]$ be proper, convex and lower semicontinuous and $C$ be a closed convex subset
    of $\R^p$ such that $C\cap\dom(f)\neq\emptyset$. Suppose that one of the following holds:
    \begin{enumerate}[label=(\roman*)]
        \item f is coercive.
        \item $C$ is bounded.
    \end{enumerate}
    Then $f$ has a minimizer over $C$.
\end{proposition}

\begin{proposition}[Uniqueness of minimizers]\label{app:prop:uniqueness minimizers}
    Let $f:\R^p\to(-\infty,+\infty]$ be proper and strictly convex. Then $f$ has at most one minimizer.
\end{proposition}

Global minimizers of proper functions can be characterized by a simple
rule which extends a seventeenth century result
due to Pierre Fermat.
\begin{theorem}[Fermat's rule]\label{thm: fermat's rule}\index{Fermat's rule}
    Let $f:\R^p\to(-\infty,+\infty]$ be proper. Then
    $$\argmin_{\theta\in\R^p} f(\theta)=\{\theta^*\in\R^p\ :\  0\in\partial f(\theta^*)\}.$$
\end{theorem}

\begin{theorem}[Hilbert projection theorem]\label{app:thm:hilber projection theorem}\index{Hilbert projection theorem}
    For every vector $\theta\in\R^p$ and every nonempty closed convex
    $C\subset\R^p$, there exists a unique vector $\beta\in\R^p$
    for which 
    $$\norm{\theta-\beta}_2^2=
    \inf_{\eta\in C}\norm{\theta-\eta}_2^2.$$
    If $C$ is a vector subspace of $\R^p$, then the minimizer $\beta$
    is the unique element in $C$ such that $\theta-\beta$ is
    orthogonal to $C$.
\end{theorem}

\section{Probability theory}

This section introduces a selection of definitions and results 
from probability theory that are used in these lecture notes.
A book-length exposition of probability theory can be found in 
\citet{billingsley2017probability} 
and \citet{vershynin2018high}, among others.

\subsection{Foundations of probability}

\begin{definition}[Measurable space]\index{Measurable space}
A \emph{measurable space} is a pair $(\Omega,\mathcal{F})$ where:
\begin{enumerate}[label=(\roman*)]
\item $\Omega$ is a nonempty set (the \emph{sample space}); and
\item $\mathcal{F}$ is a $\sigma$-algebra\index{Sigma-algebra} of subsets of $\Omega$,
i.e.\ a collection of subsets of $\Omega$ satisfying:
\begin{enumerate}
    \item $\Omega\in\mathcal{F}$;
    \item if $A\in\mathcal{F}$, then $A^c\in\mathcal{F}$;
    \item if $A_1,A_2,\dots\in\mathcal{F}$, then $\bigcup_{i=1}^\infty A_i\in\mathcal{F}$.
\end{enumerate}
\end{enumerate}
Elements of $\mathcal{F}$ are called \emph{measurable sets}.
\end{definition}

\begin{definition}[Probability measure]\index{Probability measure}
Let $(\Omega,\mathcal{F})$ be a measurable space.
A \emph{probability measure} is a function $\P:\mathcal{F}\to[0,1]$ satisfying:
\begin{enumerate}[label=(\roman*)]
    \item $\P(\Omega)=1$;
    \item (countable additivity) if $A_1,A_2,\dots$ are pairwise disjoint sets in $\mathcal{F}$, then
    \[
        \P\Big(\bigcup_{i=1}^\infty A_i\Big)
        = \sum_{i=1}^\infty \P(A_i).
    \]
\end{enumerate}
\end{definition}

\begin{definition}[Probability space]\index{Probability space}
A \emph{probability space} is a triple $(\Omega,\mathcal{F},\P)$ consisting of a measurable space $(\Omega,\mathcal{F})$ and a probability measure $\P$ on it.
\end{definition}

All random variables used in these lecture notes are real-valued and
defined on a common, complete probability space $(\Omega,\mathcal{F},\P)$.

\subsection{Random variables and distributions}

\begin{definition}[Random variable]\index{Random variable}
A \emph{random variable} is a measurable function $X:(\Omega,\mathcal{F}) \to (\R,\mathcal{B})$,  
where $\mathcal{B}$ denotes the Borel $\sigma$-algebra on $\R$.
Measurability means that $X^{-1}(B)\in\mathcal{F}$ for every $B\in\mathcal{B}$.
\end{definition}

\begin{definition}[Distribution of a random variable]\index{Distribution of a random variable}\index{Push-forward measure}
Let $X:\Omega\to\R$ be a random variable.
The \emph{distribution} (or \emph{law}) of $X$ is the probability measure $\P_X$ on $(\R,\mathcal{B})$ defined by
\[
\P_X(B) := \P(X\in B) = \P(X^{-1}(B)),
\qquad B\in\mathcal{B}.
\]
It is the push-forward measure of $\P$ under $X$.
\end{definition}

\begin{definition}[Cumulative distribution function]\index{Cumulative distribution function}
The \emph{CDF} of a random variable $X$ is the function
\[
F_X:\R\to[0,1], \qquad F_X(x) := \P[X\le x].
\]
\end{definition}

\subsection{Expectation, moments, and $L^p$ spaces}

\begin{definition}[Expected value]\index{Expected value}
Let $X$ be a random variable such that $\int_\Omega |X|d\P < \infty$.
The \emph{expectation} of $X$ is
\[
\E[X] := \int_\Omega Xd\P.
\]
\end{definition}

\begin{definition}[Moment generating function]\index{Moment generating function}
The \emph{MGF} of a random variable $X$ is the function
\[
M_X(t) := \E[\exp(tX)],\qquad t\in\R,
\]
whenever the expectation exists.
\end{definition}

\begin{definition}[Moment of order $p$]\index{Moment of order $p$}
For $p>0$, the \emph{$p$th moment} of a random variable $X$ is $\E[|X|^p]$, when finite.
\end{definition}

\begin{definition}[$L^p$-norm]\index{L$^p$-norm}
For $p>0$, the $L^p$-norm of a random variable $X$ is
\[
\norm{X}_{L^p} := \big(\E[|X|^p]\big)^{1/p}.
\]
For $p=\infty$, the $L^\infty$-norm is
\[
\norm{X}_{L^\infty}
:= \ess\sup |X|
:= \inf\{ b\in\R : \P(|X|>b)=0\}.
\]
\end{definition}

\begin{definition}[$L^p$ space]\index{L$^p$ space}
The space $L^p(\Omega,\mathcal{F},\P)$ consists of all random variables with finite $L^p$-norm:
\[
L^p := \{X : \norm{X}_{L^p} < \infty\}.
\]
\end{definition}

\begin{definition}[Conjugate exponents]\index{Conjugate exponents}
Numbers $p,q\in[1,\infty]$ are \emph{conjugate exponents} if $1/p + 1/q = 1$.
\end{definition}
For $p,q\in[1,\infty]$ conjugate exponents and $X\in L^p$, $Y\in L^q$,
we define the duality pairing
\[
\langle X,Y\rangle_{L^p-L^q} := \E[XY].
\]
In particular, for $p=q=2$ this is the usual inner product on $L^2$:
\[
\langle X,Y\rangle_{L^2}:=\E[XY].
\]
The inner product in $L^2$ is for all $X,Y\in L^2$:
$$\langle X,Y\rangle_{L^2}:=\E[XY].$$
The Variance of $X\in L^2$ is
$$\Var[X]:=\E[(X-\E[X])^2]=\norm{X-\E[X]}_{L^2}^2,$$
and the standard deviation is
$$\sigma(X):=\sqrt{\Var[X]}=\norm{X-\E[X]}_{L^2}.$$
The covariance between $X,Y\in L^2$ is
$$\Cov[X,Y]:=\langle X-\E[X], Y-\E[Y]\rangle_{L^2}=\E[(X-\E[X])(Y-\E[Y])].$$

\subsection{Classical inequalities}

\begin{theorem}[Jensen's inequality]\index{Jensen's inequality}
Let $X$ be an integrable random variable (i.e., $X\in L^1$) and 
let $f:\R\to\R$ be a convex function such that $f(X)$ is integrable.
Then
\[
    f(\E[X])\le\E[f(X)].
\]
\end{theorem}

The following proposition is a consequence of Jensen's inequality.
\begin{proposition}
Let $(\Omega,\mathcal F,\P)$ be a probability space.
If $1 \le p \le q \le \infty$ and $X \in L^q$, then
\[
    \norm{X}_{L^p} \le \norm{X}_{L^q}.
\]
\end{proposition}

Therefore, for $1 \le p \le q \le \infty$ we have
$L^q \subset L^p$.

\begin{theorem}[Minkowski's inequality]\index{Minkowski's inequality}
    For any $p\in[1,\infty]$ and any random variables $X,Y\in L^p$:
    $$\norm{X+Y}_{L^p}\le\norm{X}_{L^p}+\norm{Y}_{L^p}.$$
\end{theorem}

\begin{theorem}[Cauchy-Schwarz inequality]\index{Cauchy-Schwarz inequality}\label{thm:cauchy-schwarz}
    For any random variables $X,Y\in L^2$:
    $$|\langle X,Y\rangle|=|\E[XY]|\le
    \norm{X}_{L^2}\norm{Y}_{L^2}.$$
\end{theorem}

\begin{theorem}[Hölder's inequality]\index{Hölder's inequality}\label{app:Holder inequality}
    For any random variables $X\in L^p$ and $Y\in L^q$
    with {\it conjugate exponents} $p,q\in(1,\infty)$:
    $$|\E[XY]|\le\norm{X}_{L^p}\norm{Y}_{L^q}.$$
    This inequality also holds for $p=1$ and $q=\infty$.
\end{theorem}

The tails and the moments 
of a random variable are connected.

\begin{proposition}[Integral identity]\index{Integral identity}
    For any nonnegative random variable $X$:
    $$\E[X]=\int_0^\infty \P[X>x]dx.$$
    The two sides of this identity are either both finite 
    or both infinite.
\end{proposition}

\begin{theorem}[Markov's inequality]\index{Markov's inequality}
    For any nonnegative random variable $X$ and $x>0$:
    $$\P[X\ge x]\le\E[X]/x.$$
\end{theorem}

A consequence of Markov’s inequality is Chebyshev’s inequality,
which bounds the
concentration of a random variable about its mean.

\begin{theorem}[Chebyshev's inequality]\index{Chebyshev's inequality}
    Let $X$ be a random variable with finite mean $\mu$ and
    finite variance $\sigma^2$. Then, for any $x>0$:
    $$\P[|X-\mu|\ge x]\le\sigma^2/x^2.$$
\end{theorem}

\begin{proposition}[Generalization of Markov's inequality]\index{Markov's inequality}
    For any random variable $X$ with mean $\mu\in\R$ 
    and finite moment of order $p\ge 1$, 
    and for any $x>0$:
    $$\P[|X-\mu|\ge x]\le\E[|X-\mu|^p]/x^p.$$
\end{proposition}

\subsection{Concentration of sums of independent random variables}

Concentration inequalities quantify how a random variable
deviates around its mean.

\begin{definition}[Symmetric Bernoulli distribution]\index{Symmetric Bernoulli distribution}
    A random variable $X$ has a {\it symmetric Bernoulli} distribution
    if $$\P[X=-1]=\P[X=+1]=1/2.$$
\end{definition}

\begin{theorem}[Hoeffding's inequality]\index{Hoeffding's inequality}
    Let $X_1,\ldots,X_n$ be an independent symmetric 
    Bernoulli random variables, and $ a\in\R^n$.
    Then, for any $x\ge 0$:
    $$\P\left[\sum_{i=1}^na_iX_i\ge x\right]\le
    \exp\left(-\frac{x^2}{2\norm{ a}_2^2}\right).$$
\end{theorem}

\begin{theorem}[Two-sided Hoeffding's inequality]
    Let $X_1,\ldots,X_n$ be an independent symmetric 
    Bernoulli random variables, and $ a\in\R^n$.
    Then, for any $x> 0$:
    $$\P\left[\left\lvert\sum_{i=1}^na_iX_i\right\rvert\ge x\right]\le
    2\exp\left(-\frac{x^2}{2\norm{ a}_2^2}\right).$$
\end{theorem}

\begin{theorem}[Hoeffding's inequality for bounded random variables]
    Let $X_1,\ldots,X_n$ be an independent random variables. 
    Assume that $X_i\in[l_i,u_i]$ with $l_i,u_i\in\R$ and $l_i\le u_i$. 
    Then, for any $x> 0$:
    $$\P\left[\sum_{i=1}^n(X_i-\E[X_i])\ge x\right]\le
    \exp\left(-\frac{2x^2}{\sum_{i=1}^n(u_i-l_i)^2}\right).$$
\end{theorem}

\begin{theorem}[Chernoff's inequality]\index{Chernoff's inequality}
    Let $X_i$ be independent Bernoulli random variables
    with parameter $p_i\in[0,1]$. 
    Let $S_n:=\sum_{i=1}^nX_i$ and its mean $\mu:=\E[S_n]$.
    Then, for any $x> 0$:
    $$\P[S_n\ge x]\le\exp(-\mu)\left(\frac{e\mu}{x}\right)^x.$$
\end{theorem}

\begin{proposition}[Tails of the standard normal distribution]
    Let $Z\sim N(0,1)$. Then, for all $z>0$:
    $$\left(\frac{1}{z}-\frac{1}{z^3}\right)
    \frac{1}{\sqrt{2\pi}}e^{-z^2/2}\le
    \P[Z\ge z]\le \frac{1}{z}\frac{1}{\sqrt{2\pi}}e^{-z^2/2}.$$
    In particular, for $z\ge 1$:
    $$\P[Z\ge z]\le\frac{1}{\sqrt{2\pi}}e^{-z^2/2}.$$
\end{proposition}

\begin{proposition}[Tails of the normal distribution]
    Let $X\sim N(\mu,\sigma^2)$ with $\mu\in\R$ and $\sigma>0$.
    Then, for all $x\ge 0$:
    $$\P[X-\mu\ge x]\le\exp\left(\frac{-x^2}{2\sigma^2}\right).$$
\end{proposition}

\begin{proposition}
    Let $Z\sim N(0,1)$. Then, for all $z\ge 0$:
    $$\P[|Z|\ge z]\le 2\exp(-z^2/2).$$
\end{proposition}

\subsubsection{Sub-Gaussian random variables}

\begin{proposition}[Sub-Gaussian properties]\label{app:prop:subgaussian properties}\index{Sub-Gaussian properties}
    Let $X$ be a random variable.
    Then, there are constants $C_1,\ldots,C_5>0$ for which
    the following properties are equivalent.
    \begin{enumerate}[label=(\roman*)]
        \item The tails of $X$ satisfy for all $x\ge 0$:
        $$\P[|X|\ge x]\le 2\exp(-x^2/C_1^2).$$
        \item The moments of $X$ satisfy for all $p\ge 1$:
        $$\norm{X}_{L^p}=\E[|X|^p]^{1/p}\le C_2\sqrt{p}.$$
        \item The MGF of $X^2$ satisfies, for all $t\in\R$ such that $|t|\le 1/C_3$,
\[
    \E[\exp(t X^2)]\le\exp(C_3^2 t^2).
\]
        \item The MGF of $X^2$ is bounded at some point, namely
        $$\E[\exp(X^2/C_4^2)]\le 2.$$
    \end{enumerate}
    If further $\E[X]=0$, these properties are equivalent to:
    \begin{enumerate}
        \item[(v)] The MGF of $X$ satisfies for all $t\in\R$:
        $$\E[\exp(tX)]\le \exp(C_5^2t^2).$$
    \end{enumerate}
\end{proposition}

\begin{definition}[Sub-Gaussian random variables]\index{Sub-Gaussian random variable}
    A random variable $X$ that satisfies the equivalent conditions
    of Proposition \ref{app:prop:subgaussian properties} is
    a {\it sub-Gaussian} random variable, denoted $X\sim\subG$.
\end{definition}

Gaussian, symmetric Bernoulli, uniform, and bounded random variables are
examples of sub-Gaussian random variables.
The tails of the distribution of a sub-Gaussian random variable 
decay at least as fast as the tails
of a Gaussian distribution.
The Poisson, exponential, Pareto, and Cauchy distribution are
examples of distributions that are not sub-Gaussian.

\begin{definition}[Variance proxy]\index{Variance proxy}
    For a random variable $X\sim\subG$,
    if there is some $s>0$ such that for all $t\in\R$:
    $$\E[e^{(X-\E[X])t}]\le \exp(s^2t^2/2),$$
    then $s^2$ is called {\it variance proxy}.
\end{definition}

\begin{proposition}[Weighted sum of independent sub-Gaussian random variables]
\label{app:prop:sum indep sub gauss}
    Let $X_1,\ldots,X_n$
    be independent sub-Gaussian random variables, 
    all with variance proxy $\sigma^2$ where $\sigma>0$.
    Then, for any $ a\in\R^n$:
    $$\P\left[\sum_{i=1}^na_iX_i\ge t\right]\le\exp\left(
    -\frac{t^2}{2\sigma^2\norm{ a}_2^2}\right),$$
    and 
    $$\P\left[\sum_{i=1}^na_iX_i\le -t\right]\le\exp\left(
    -\frac{t^2}{2\sigma^2\norm{ a}_2^2}\right).$$
\end{proposition}

\begin{definition}[Sub-Gaussian norm]\index{Sub-Gaussian norm}
    The {\it sub-Gaussian norm} $\norm{X}_{\psi_2}$ of random variable $X$
    is defined as
    $$\norm{X}_{\psi_2}:=\inf\{t>0\ :\ \E[\exp(X^2/t^2)]\le 2\}.$$
\end{definition}

\begin{proposition}
    If $X$ is a sub-Gaussian random variable, then $X-\E[X]$
    is sub-Gaussian and for a constant $C>0$:
    $$\norm{X-\E[X]}_{\psi_2}\le C\norm{X}_{\psi_2}.$$
\end{proposition}

\begin{proposition}[Sums of independent sub-Gaussian]
    Let $X_1,\ldots,X_n$ be independent sub-Gaussian
    random variables with mean zero. Then
    $\sum_{i=1}^nX_i$ is also a sub-Gaussian random variable,
    and, for a constant $C>0$:
    $$\norm{\sum_{i=1}^nX_i}_{\psi_2}^2\le C\sum_{i=1}^n
    \norm{X_i}_{\psi_2}^2.$$
\end{proposition}

We can now extend the Hoeffding's inequality 
to sub-Gaussian distributions.

\begin{proposition}[General Hoeffding's inequality]\index{General Hoeffding's inequality}
    Let $X_1,\ldots,X_n$ be independent sub-Gaussian
    random variables with mean zero and $C>0$ a constant. 
    Then, for every $t\ge 0$:
    $$\P\left[\left\lvert\sum_{i=1}^nX_i\right\rvert\ge t\right]\le
    2\exp\left(-\frac{Ct^2}{\sum_{i=1}^n\norm{X_i}^2_{\psi_2}}\right).$$
\end{proposition}

\begin{proposition}
    Let $X_1,\ldots,X_n$ be independent sub-Gaussian
    random variables with mean zero, $ a\in\R^n$,
    $K=\max_{i=1}^n\norm{X_i}_{\psi_2}$ and $C>0$ a constant. 
    Then, for every $t\ge 0$:
    $$\P\left[\left\lvert\sum_{i=1}^na_iX_i\right\rvert\ge t\right]\le
    2\exp\left(-\frac{Ct^2}{K^2\norm{ a}_2^2}\right).$$
\end{proposition}

\begin{proposition}[Khintchine’s inequality]\index{Khintchine’s inequality}
    Let $X_1,\ldots,X_n$ be independent sub-Gaussian
    random variables, all with mean zero and unit variance proxy,
    $ a\in\R^n$, $K=\max_{i=1}^n\norm{X_i}_{\psi_2}$ and $C>0$
    a constant.
    Then, for every $p\in[2,\infty)$:
    $$\left(\sum_{i=1}^na_i^2\right)^{1/2}\le
    \norm{\sum_{i=1}^na_iX_i}_{L^p}\le CK\sqrt{p}\left(\sum_{i=1}^na_i^2\right)^{1/2}.$$
\end{proposition}

The sub-Gaussian distribution does not embed distributions whose tails are heavier than Gaussian. 

\subsubsection{Sub-exponential random variables}

\begin{proposition}[Sub-exponential properties]\label{app:prop:subexponential properties}\index{Sub-exponential properties}
    Let $X$ be a random variable.
    Then, there are constants $K_1,\ldots,K_5>0$ for which
    the following properties are equivalent.
    \begin{enumerate}[label=(\roman*)]
        \item The tails of $X$ satisfy for all $x\ge 0$:
        $$\P[|X|\ge x]\le 2\exp(-x/K_1).$$
        \item The moments of $X$ satisfy for all $p\ge 1$:
        $$\norm{X}_{L^p}=\E[|X|^p]^{1/p}\le K_2p.$$
        \item The MGF of $|X|$ satisfies, for all $t\in\R$ such that $0\le t\le 1/K_3$,
\[
    \E[\exp(t |X|)]\le\exp(K_3 t).
\]
        \item The MGF of $|X|$ is bounded at some point, namely
        $$\E[\exp(|X|/K_4)]\le 2.$$
    \end{enumerate}
    If further $\E[X]=0$, these properties are equivalent to:
    \begin{enumerate}
        \item The MGF of $X$ satisfies for all $t\in\R$
        such that $|t|\le 1/K_5$:
        $$\E[\exp(tX)]\le \exp(K_5^2t^2).$$
    \end{enumerate}
\end{proposition}

\begin{definition}[Sub-exponential random variables]\index{Sub-exponential random variables}
    A random variable $X$ that satisfies the equivalent conditions
    of Proposition \ref{app:prop:subexponential properties} is
    a {\it sub-exponential} random variable.
\end{definition}

Sub-Gaussian random variables, bounded random variables, and many light-tailed
distributions such as Poisson, exponential, and (appropriately parametrized)
Gamma or $\chi^2$ distributions are examples of sub-exponential random
variables. In contrast, heavy-tailed distributions with polynomial decay
(Pareto, Cauchy, certain $t$-distributions, etc.) are not sub-exponential.

\begin{definition}[Sub-exponential norm]\index{Sub-exponential norm}
    The {\it sub-exponential norm} $\norm{X}_{\psi_1}$ of random variable $X$
    is defined as
    $$\norm{X}_{\psi_1}:=\inf\{t>0\ :\ \E[\exp(|X|/t)]\le 2\}.$$
\end{definition}

\begin{proposition}
    If $X$ is a sub-exponential random variable, then $X-\E[X]$
    is sub-exponential and for a constant $C>0$:
    $$\norm{X-\E[X]}_{\psi_1}\le C\norm{X}_{\psi_1}.$$
\end{proposition}

\begin{proposition}[Square of a sub-Gaussian is sub-exponential]
Let $X$ be a sub-Gaussian random variable. Then $X^2$ is sub-exponential, and
there exists a constant $C>0$ such that
\[
    \norm{X^2}_{\psi_1} \le C \norm{X}_{\psi_2}^2.
\]
Conversely, if $X^2$ is sub-exponential, then $X$ is sub-Gaussian, with
\[
    \norm{X}_{\psi_2}^2 \le C' \norm{X^2}_{\psi_1}
\]
for some constant $C'>0$.
\end{proposition}

\begin{proposition}[Product of sub-Gaussians is sub-exponential]
    Let $X$ and $Y$ be sub-Gaussian random variables. Then 
    $XY$ is sub-exponential, and there exists a constant $C>0$ such that
    \[
        \norm{XY}_{\psi_1} \le C \norm{X}_{\psi_2}\norm{Y}_{\psi_2}.
    \]
\end{proposition}

\begin{theorem}[Bernstein's inequality]\index{Bernstein's inequality}
    Let $X_1,\ldots,X_n$ be independent 
    sub-exponential random variables with mean zero and
    let $C>0$ be a constant.
    Then, for every $t\ge 0$:
    $$\P\left[\left\lvert\sum_{i=1}^nX_i\right\rvert\ge t\right]
    \le 2\exp\left(-C\min\left\{\frac{t^2}{\sum_{i=1}^n\norm{X_i}_{\psi_1}^2},
    \frac{t}{\max_i\norm{X_i}_{\psi_1}}\right\}\right).$$
\end{theorem}

\begin{theorem}[Bernstein's inequality for weighted sums]
    Let $X_1,\ldots,X_n$ be independent 
    sub-exponential random variables with mean zero,
    $C>0$ be a constant,
    $K:=\max_{i=1}^n\norm{X_i}_{\psi_1}$ and
    $ a\in\R^n$. Then:
    $$\P\left[\left\lvert\sum_{i=1}^na_iX_i\right\rvert\ge t\right]
    \le 2\exp\left(-C\min\left\{\frac{t^2}{K^2\norm{ a}_2^2},
    \frac{t}{K\norm{ a}_\infty}\right\}\right).$$
\end{theorem}

\begin{corollary}[Bernstein's inequality for averages]
    Let $X_1,\ldots,X_n$ be independent 
    sub-exponential random variables with mean zero,
    $C>0$ be a constant
    and $K:=\max_{i=1}^n\norm{X_i}_{\psi_1}$. Then:
    $$\P\left[\left\lvert\sum_{i=1}^nX_i/n\right\rvert\ge t\right]
    \le 2\exp\left(-Cn\min\left\{\frac{t^2}{K^2},
    \frac{t}{K}\right\}\right).$$
\end{corollary}

\printindex

\end{document}